\documentclass[11pt]{article}
\usepackage{fullpage}
\usepackage{amsmath}
\usepackage{color}
\usepackage{amsfonts}
\usepackage{amssymb}
\usepackage{graphicx}
\usepackage{graphics}
\usepackage{subfig}
\usepackage{psfrag}
\usepackage{epsfig,latexsym,verbatim}
\usepackage{rotating}
\usepackage{lscape}
\usepackage{geometry}
\usepackage{mathrsfs}
\usepackage{float}
\usepackage[colon]{natbib}
\usepackage{breakcites}
\usepackage{authblk}
\usepackage{enumitem}
\usepackage{authblk}
\usepackage{multirow}
\numberwithin{equation}{section}
\usepackage{placeins}
\usepackage{rotating}
\usepackage{etoolbox}
\usepackage{algorithm, algorithmicx}
\usepackage{bm}
\usepackage{csquotes}
\usepackage{url}
\usepackage[font=small]{caption}
\usepackage[colorlinks=true,linkcolor=blue]{hyperref}
\usepackage{array}
\usepackage{booktabs}
\usepackage{rotating}
\usepackage{multicol}
\usepackage{algpseudocode}

\newtheorem{theorem}{Theorem}
\newtheorem{lemma}{Lemma}
\newtheorem{proposition}{Proposition}

\graphicspath{{fig/}}
\allowdisplaybreaks[3]
\title{\Large\textbf{A fusion learning method to subgroup analysis of Alzheimer's disease}
\footnotetext{%
Mingming Liu and Jing Yang are joint first authors. Correspondence should be addressed to Shujie Ma. Email: shujie.ma@ucr.edu.}}

\author[1]{Mingming Liu}
\author[2]{Jing Yang}
\author[3]{Yushi Liu}
\author[3]{Bochao Jia}
\author[3]{Yun-Fei Chen}
\author[3]{Luna Sun}
\author[1]{Shujie Ma}

\affil[1]{Department of Statistics, University of California at Riverside, U.S.A}
\affil[2]{Key Laboratory of Computing and Stochastic Mathematics (Ministry of Education), College of Mathematics and Statistics, Hunan Normal University, People's Republic of China}

\affil[3]{Global Statistical Science, Eli Lilly and Company, U.S.A}

\begin{document}
\date{}
\maketitle

\begin{center}
{\Large \textbf{Abstract}}
\end{center}
Uncovering the heterogeneity in the disease progression of Alzheimer's is a key factor to disease understanding and treatment development, so that interventions can be tailored to target the subgroups that will benefit most from the treatment, which is an important goal of precision medicine. However, in practice, one top methodological challenge hindering the heterogeneity investigation is that the true subgroup membership of each individual is often unknown. In this article, we aim to identify latent subgroups of individuals who share a common disorder progress over time, to predict latent subgroup memberships, and to estimate and infer the heterogeneous trajectories among the subgroups. To achieve these goals, we apply a concave fusion learning method proposed in Ma and Huang (2017) and Ma et al. (2019) to conduct subgroup analysis for longitudinal trajectories of the Alzheimer's disease data. The heterogeneous trajectories are represented by subject-specific unknown functions which are approximated by B-splines. The concave fusion method can simultaneously estimate the spline coefficients and merge them together for the subjects belonging to the same subgroup to automatically identify subgroups and recover
the heterogeneous trajectories. The resulting estimator of the disease trajectory of each subgroup is supported by an asymptotic distribution. It provides a sound theoretical basis for further conducting statistical inference in subgroup analysis.

\clearpage\pagebreak\newpage \pagenumbering{arabic} \newlength{\gnat} %
\setlength{\gnat}{23pt} \baselineskip=\gnat

\section{Introduction}
\label{s:intro}
\noindent Alzheimer's disease (AD) is the leading cause of dementia for adults. It is a progressive disease that worsens over time. Patients with AD show symptoms of memory loss, mental decline, delusion and so forth as the disease progresses.  The progression of AD varies from person to person, and patients with AD have experienced it in different ways. The lack of a good understanding of the heterogeneity in the disease progression through the population is a key reason for failures of disease-modifying treatments for AD. As a result, very little progress has been made for the AD treatment development since 2003 
\citep{Yiannopoulou2019}. To overcome this difficulty,  one has to first understand the heterogeneity in the disease trajectories, so that interventions can be tailored to target the subgroups that will benefit most from the treatment, which is an important goal of precision medicine. The progression of AD is often measured by cognitive scores at multiple time points, resulting in a collection of longitudinal data. One major methodological challenge hindering the heterogeneity investigation is that the true subgroup membership of each individual is often unknown. 

The growth mixture modeling (GMM) method \citep{fraley2002model,SLMK2007, Jung2008,mcnicholas2010model} has been popularly used for the identification and prediction of latent subpopulations for longitudinal data. This method requires to specify the underlying distribution of the data, which is often hard to obtain for longitudinal data, because of their complex structure. The k-means algorithm \citep{hartigan1979algorithm} is another popular clustering method. It divides the data into subgroups based on the distances between measurement vectors of subjects. It is difficult to apply this method to cluster  functional curves, especially arising from longitudinal data with missing measurements. Moreover, both GMM and k-means methods need to pre-specify the number of subgroups, which is often unknown in practice, and thus introduces additional complications in the estimation procedure. 

To overcome these challenges, we apply the concave fusion learning method proposed in \cite{ma2017concave,Ma2019} to conduct subgroup analysis for longitudinal trajectories of the AD data. This semi-supervised machine learning method applies concave penalty functions to pairwise differences of clinical outcomes or unknown treatment coefficients in a regression model. It can automatically identify memberships from latent subgroups and estimate the number of subgroups simultaneously without specifying the underlying distribution. Although the fusion learning method was originally considered in \cite{ma2017concave,Ma2019} for the cross-sectional data setting with independent observations, it also has a great potential for subgroup analysis of other data settings such as longitudinal data and survival data. In this article, we extend this method to the longitudinal AD data, and investigate its numerical performance through extensive simulation studies with both balanced and unbalanced correlated repeated measures designs. Moreover, we propose two different data-driven methods based on the modified Bayes Information Criterion BIC and the Calinski-Harabasz (CH) index, respectively, for selecting the optimal tuning parameter involved in the concave fusion penalization method, while the CH method was not considered in \cite{ma2017concave,Ma2019}. We also thoroughly investigate the performance of these two data-driven methods through numerical studies.

To cluster the AD patients based on their cognitive scores observed over time, we consider a subject-specific nonparametric regression model, in which the heterogeneity can be driven by observed or unobserved latent covariates. More specifically, we model each patient's cognitive scores through an unknown functional curve of time. We approximate each curve by B-splines \citep{de2001practical,LiuYang2010,XueLiang2013,Ma2014}, and then apply pairwise fusion penalties to the spline coefficients, so that patients with similar disease trajectories can be automatically clustered into the same homogeneous subgroup. As a result, patients in the same identified subgroup share the same disease progressive curve. We use an alternating direction method of multipliers (ADMM) algorithm \citep{boyd2011distributed} that has a good convergence property to  solve the optimization problem.
Different from the GMM method, our method does not require to pre-specify the number of subgroup, nor does it need to provide the underlying distribution of the data.  Instead, our estimation procedure only involves a working correlation matrix \citep{liangzeger1986,WangCarrollLin2005,Ma2012,MaSongWang2013} for the repeated measures of each subject. We show that the resulting estimator of the functional curve for each subgroup is robust to the specification of the correlation matrix, i.e., it is still a consistent estimator even if the working correlation matrix is mis-specified. Moreover, we establish point-wise asymptotic normality of the functional curve estimator for each subgroup, so that statistical inference can be further conducted based on our clustering and estimation results. 

\indent The rest of this article is organized as follows. Section \ref{SEC:model} describes the proposed model. Section \ref{SEC:estimation} introduces the model estimation procedure using concave fusion penalization method. In Section \ref{theory}, we establish the theoretical properties of the proposed estimators. Simulation studies are presented in Section \ref{SEC:simulation}. Section \ref{SEC:real} illustrates the application of the proposed method to Alzheimer's disease data. Discussions are provided in Section \ref{discussion}. The related computation procedure and technical proofs are included in the Supplementary Material.

\section{Model}
\label{SEC:model}

{\noindent{In a longitudinal study, subjects are usually measured repeatedly over a time period. Suppose the data consist of $ \left(Y_i(t_{ij}), t_{ij} \right), i=1,\dots,n, j=1, \dots, m_i$, where $\left \{ t_{ij}, j=1, \dots, m_i \right \}$ are the distinct time points that the measurements of the $i$th subject are taken, and $ Y_{i}(t_{ij})$ is the observed response for the $i$th subject at time $\ t_{ij}$. Our goal of this article is to understand how the change of trajectories may differ across individual subjects. To study the longitudinal trajectories of the $i$th subject, we consider the subject-specific nonparametric regression model:  
\begin{equation}
Y_{i}(t_{ij})=\beta _{i}(t_{ij})+\varepsilon
_{i}(t_{ij}),
\label{model1}
\end{equation}
where $\beta
_{i}(t)$'s are the unknown smooth functions of $t$, and the errors $\varepsilon _{i}(t)$'s satisfy $E (\varepsilon _{i}(t))=0$ and Cov$(\varepsilon
_{i}(t),\varepsilon _{i^{\prime }}(t^{\prime}))=\delta (t,t^{\prime})I\{i=i^{\prime}\}$ with $I\{\cdot\}$ being an indicator function. For simplicity, we denote $Y_{ij}=Y_i(t_{ij})$ and $\varepsilon _{ij}=\varepsilon _{i}(t_{ij})$. Model (\ref{model1}) can be rewritten as 
\begin{equation}
Y_{ij}=\beta _{i}(t_{ij})+\varepsilon
_{ij}.  
\label{model2}
\end{equation}%
}
}

{\indent In this model, the trajectory of the $i$th subject over time is represented by the subject-specific unknown function ${\beta_i}(t)$. Due to the heterogeneity of the trajectories, we assume $\beta
_{i}(t)$'s arise from $K$ different groups with $ K\geq 1$. To be specific, we have ${\beta}_i(t)={\alpha}_k(t)$ for all $i \in \mathcal{G}_k$, where $\mathcal{G}=(\mathcal{G}_1,...,\mathcal{G}_K)$ is a mutually exclusive partition of $\left \{1,...,n  \right \}$ and ${\alpha}_k(t)$ is the common function for all the ${\beta}_i(t)$'s from group $\mathcal{G}_k$. {In practice, the number of subgroups $K$ can be much smaller than the sample size $n$, and it is often unknown.}
}

\section{Estimation}
\label{SEC:estimation}
\noindent {In order to identify the subgroups of the heterogeneous trajectories, we first approximate the nonparametric functions ${\beta}_i(\cdot )$'s in (\ref{model2}) using B-splines. Referring to \citep{ma2016inference}, let $a_0=\zeta_0<\zeta_1<\cdots<\zeta_{J}<\zeta_{J+1}=b_0$ be a partition of $[a_0, b_0]$ into $J+1$ subintervals $I_l=[\zeta_l, \zeta_{l+1}), l=0,\cdots, J-1$ and $I_{J}=[\zeta_{J}, b_0]$, where $\left \{ \zeta_l \right \}^{J}_{l=1}$ is a sequence of interior knots. Denote the $r$th order normalized B-spline basis as $\left \{ B_{1}(t), \dots, B_{S}(t)\right \}^T$ (see \citep{de2001practical}), in which $S=J+r$ is the number of basis functions.} Then, $\beta_i(t_{ij})$ in (\ref{model2}) can be approximated by a linear combination of the B-spline functions, 
\begin{align}
\beta_{i}(t_{ij}) \approx \sum_{d=1}^{S}\gamma_{id}B_{d}(t_{ij})= \boldsymbol{B}(t_{ij})^T\boldsymbol{\gamma}_{i},~~i=1,\dots, n, ~~j=1,\dots, m_i,
\label{3.1}
\end{align}
where $\boldsymbol{B}(t_{ij})=(B_1(t_{ij}),\dots, B_{S}(t_{ij}))^{ T}$ and $ \boldsymbol{\gamma}_{i}=(\gamma_{i1},\dots,\gamma_{i{S}})^{T}$. In this case, the trajectory heterogeneity represented by $\beta_i(t)$ is reflected on the B-spline coefficient $\boldsymbol{\gamma}_i$. Therefore, our goal can be transformed into identifying the subgroups based on the $\boldsymbol{\gamma}_i$'s.

\indent Let $\boldsymbol{Y}_{i}=(Y_{i1},\dots, Y_{im_{i}})^{{T}},\boldsymbol{\varepsilon}_{i}=(\varepsilon_{i1},\dots, \varepsilon_{im_{i}})^{{T}}$ and $\boldsymbol{X}_i=(\boldsymbol{B}_{i1}, \dots, \boldsymbol{B}_{im_i})^{T}$, where $\boldsymbol{B}_{ij}=\boldsymbol{B}(t_{ij})$. Given (\ref{3.1}), for each $i$, model (\ref{model2}) can be written in matrix notation as
\begin{align}
\boldsymbol{Y}_i\approx \boldsymbol{X}_i\boldsymbol{\gamma}_i+\boldsymbol{\varepsilon}_i, ~~ i=1,\cdots,n. \label{mamodel}
\end{align}
{As in \citep{liangzeger1986, WangCarrollLin2005, MaSongWang2013}, we let $\boldsymbol{\Sigma}_i$ and $\boldsymbol{V}_i$ be the true and assumed working covariance of $\boldsymbol{Y}_i$, where $\boldsymbol{\Sigma}_i=\text{Var}(\boldsymbol{Y}_i)$ and $\boldsymbol{V}_i=\boldsymbol{A}_i^{1/2}\boldsymbol{R}_i\boldsymbol{A}_i^{1/2}$, $\boldsymbol{A}_i$ represents a $m_i \times m_i$ diagonal matrix containing the marginal variances of $Y_{ij}$, and $\boldsymbol{R}_i$ is an invertible working correlation matrix. The true covariance $\boldsymbol{\Sigma}_i$ is often unknown in practice, so we use a working covariance $\boldsymbol{V}_i$ to replace $\boldsymbol{\Sigma}_i$ in the estimation procedure. The structure of the working correlation $\boldsymbol{R}_i$ is pre-specified. Throughout, we assume that $\boldsymbol{V}_i$ depends on a nuisance finite dimensional parameter vector $\boldsymbol{\eta}$.}

\indent Following \cite{Ma2019}, we utilize a fusion learning approach with concave penalty to estimate model (\ref{mamodel}). For any vector $\boldsymbol{a}$, define its $L_2$ norm as $\left \| \boldsymbol{a}\right \|_2=(\sum a^2_i)^{1/2}$. The objective function is constructed as
\begin{equation}
Q_n\left ( \boldsymbol {\gamma}; \lambda  \right )=\frac{1}{2}\sum _{i=1}^{n}(\boldsymbol{Y}_{i} -\boldsymbol{X}_{i} \boldsymbol{\gamma }_{i})^{T}\boldsymbol{V}_{i}^{-1}(\boldsymbol{Y}_{i} - \boldsymbol{X}_{i} \boldsymbol{\gamma }_{i})+\sum _{1\leq i <j\leq n}p\left (\left \Vert \boldsymbol {\gamma}_{i}-\boldsymbol {\gamma }_{j}\right \Vert_2, \lambda \right ),
\label{obf}
\end{equation}
where $\boldsymbol{\gamma}=\left(\boldsymbol{\gamma}_{1}^{T},\dots, \boldsymbol{\gamma}_{n}^{ T}\right)^{T}$ and $p\left(\cdot ,\lambda \right)$ is a concave penalty function with a tuning parameter $\lambda \geq 0$. For a given $\lambda > 0$, define
\begin{equation}
\hat{\boldsymbol{\gamma}}(\lambda)=\arg\min_{\boldsymbol{\gamma}}Q_n(\boldsymbol{\gamma}; \lambda).
\label{3.4}
\end{equation}
\noindent  When $\lambda$ is large enough, the penalty shrinks some pairs of $\left \| \boldsymbol{\gamma}_i-\boldsymbol{\gamma}_j \right \|_2$ to zero. For two subjects with $\left \| \hat{\boldsymbol{\gamma}}_i(\lambda)-\hat{\boldsymbol{\gamma}}_j(\lambda) \right \|_2=0$, they are clustered into the same group. Based on this fact, we can partition the heterogeneous trajectories into subgroups.  For convenience, we write $\hat{\boldsymbol{\gamma}}(\hat{\lambda})\equiv\hat{\boldsymbol{\gamma}}$. Let $\left \{\hat{\boldsymbol{\theta}}_{1}, \dots, \hat{\boldsymbol{\theta}}_{\hat{K}} \right \}$ be the unique values of $\hat{\boldsymbol\gamma}$, where $\hat{K}$ is the number of these distinct values. In the $k$th subgroup, we denote the set of the corresponding indices by $\hat{\mathcal{G}}_k=\left \{i: \hat{\boldsymbol{\gamma}}_i=\hat{\boldsymbol{\theta}}_{k}, ~ 1\leq i\leq n \right \}$ with $1\leq k\leq \hat{K}$. To select the optimal tuning parameter ${\lambda}$, a data-driven procedure such as BIC or the Calinski-Harabasz index is considered. {It is noteworthy that our method can also be applied to the case that the true number of subgroups $K$ is known. In this scenario,  we will choose a $\lambda$ value that corresponds to the estimated number of subgroups $\hat{K}$ which is equal to or the closest one to the true number of subgroups $K$. If two $\hat{K}$ values are equally distant from $K$, we use the larger one to determine the $\lambda$ value.}

\indent An appropriate selection of the penalty is very critical to the model estimation. Instead of choosing lasso penalty $p_{\tau}(t, \lambda)=\lambda \left | t \right |$ \citep{tibshirani1996regression}, which results in biased estimates due to the over-shrinkage of large coefficients, we use the minimax concave penalty (MCP) \citep{zhang2010nearly} by inducing nearly unbiased estimators with the form 
\begin{align*}
p_{\tau}(t, \lambda)=\lambda \int_{0}^{\left | t \right |} (1-x/(\tau\lambda))_+dx, ~~ \tau>1,  
\end{align*}
where $\tau$ is a parameter controlling the concavity of the penalty function, {and $(a)_+=a$, if $a>0$ and $(a)_+=0$, otherwise}. Moreover, it is more aggressive in enforcing a sparser solution. Consequently, MCP is a more desirable choice. The details for computational procedure using ADMM algorithm for a given value of $\lambda$ are provided in the Supplementary Material.

\indent Another problem is how to choose the working covariance matrix $\boldsymbol{V}_i$. Here we consider an unequally spaced AR(1) structure for the working covariance matrix $\boldsymbol{V}_i$, such that $V_{i}(t,s)=\sigma^2 \rho^{\kappa\left | t-s \right |}$, where  $\kappa=\frac{1}{|t_{(1)}-t_{(2)}|}$ with $t_{(1)}$, $t_{(2)}$ being the first two time points. Note that our estimator of the functional curve for each subgroup is consistent even if the working covariance matrix is mis-specified, i.e., $\boldsymbol{V}_i\neq\boldsymbol{\Sigma}_i$. First, we estimate $\sigma^2$ by taking the mean of the estimated variance $\hat{\sigma}^2_i$, $i=1, \dots, n$, where $\hat{\sigma}^2_i$ is calculated within subject by using ordinary least squares (OLS) residuals. Due to the fact that these residuals may be small and thus underestimate the true errors, we modify these residuals by replacing $\hat{\varepsilon}_{ij}$ with $\hat{\varepsilon}^*_{ij}=\hat{\varepsilon}_{ij}/(1-h_{ij})$, where $h_{ij}$ is the $j$th diagonal element of the projection matrix $\boldsymbol{H}_i$ for subject $i$. This modification is suggested by \cite{mackinnon1985some}. Given (\ref{mamodel}), we have $\boldsymbol{H}_i=\boldsymbol{X}_i(\boldsymbol{X}_i^T\boldsymbol{X}_i)^{-1}\boldsymbol{X}_i^T$. Next, we estimate correlation $\rho$ by taking the average of the estimated correlation between the two adjacent time points, in which we only consider the adjacent time points having the scaled distance equalling 1, i.e. $\kappa\left | t-s \right |=1$. Accordingly, ${\boldsymbol{V}}_i$ can be obtained. 

\begin{figure}[tbp]
\centering
\includegraphics[width=6.5cm]{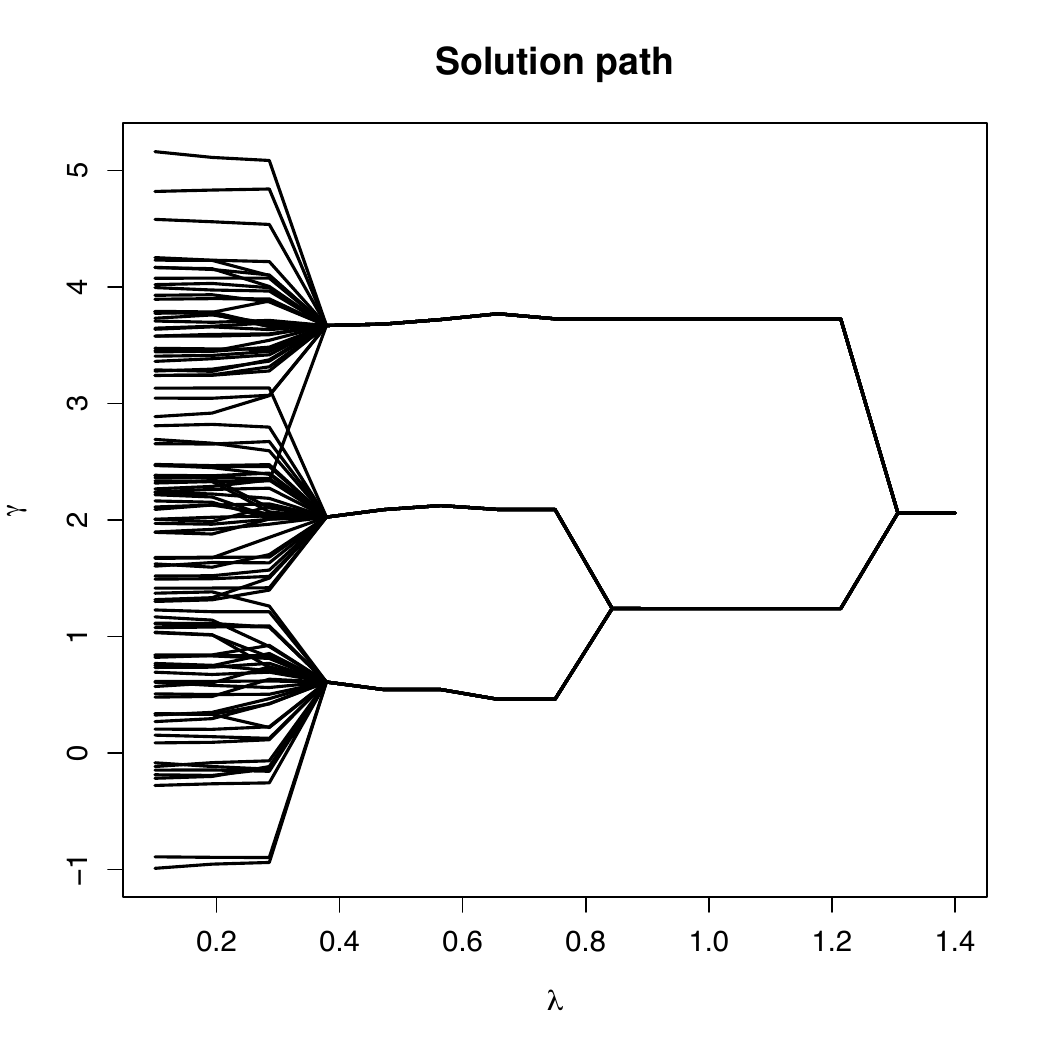}
\caption{Solution path for $(\hat{\gamma}_{31}(\lambda), \dots, \hat{\gamma}_{3n}(\lambda))$ against $\lambda$ with $n=100, \ T=20$ for balanced data of Middle case from Three Subgroup Example in Section \ref{SEC:simulation}.}
\label{path}
\end{figure}

\indent Figure \ref{path} illustrates the solution path for the estimates of B-spline coefficients $(\hat{\gamma}_{31}(\lambda), \dots, \hat{\gamma}_{3n}(\lambda))$ against $\lambda$, which is computed on a grid of $\lambda$ values in interval $[\lambda_\text{min}, \lambda_\text{max}]$. More details about the solution path are presented in  the Supplementary Material. From Figure \ref{path}, we observe that when $\lambda$ is very small, too many subgroups are identified. With $\lambda$ value increasing, the estimated number of subgroups decreases, then becomes to 1 for a large $\lambda$ value. If the actual number of subgroups is given ($K=3$), based on the solution path, we can select a $\lambda$ between 0.6 and 0.8 as the tuning parameter, where $\hat{K}$ equals the true number of subgroups; otherwise, BIC or the Calinski-Harabasz index is used to decide the optimal tuning parameter $\lambda$. 


\section{Theoretical properties}
\label{theory}
\noindent In this section, we establish the theoretical properties of the proposed estimators. We first introduce some notations. Let $\boldsymbol{\beta}(t)=(\beta_1(t), \dots, \beta_n(t))^T$ with $\beta_i(t)$ being the function of the $i$th subject, and $\boldsymbol{\alpha}(t)=(\alpha_1(t),\dots,\alpha_K(t))^T$ with $\alpha_k(t)$ being the common function for the $k$th subgroup. {For any square integrable function $g(t)$ on the compact support $\mathbb{T}$, denote its $L_2$ norm by $\|g\|_2= \left \{ \int_{\mathbb{T}} g(t)^2 dt \right \}^{1/2}$ and squared $L_2$ norm by $\|g\|_2^2=\int_{\mathbb{T}} g(t)^2 dt$. Then, for a vector valued function $\textbf{g}(t)=(g_1(t),\dots,g_L(t))^T$, its squared $L_2$ norm is defined as $\|\textbf{g}\|_2^2=\sum_{l=1}^L{\|g_l\|_2^2}$.} Let $b=\min_{k\neq k^{\prime}}\|\alpha_k -\alpha_{k^{\prime}}\|_2$ be the minimum distance between smoothing functions $\alpha_k$ and $\alpha_{k^{\prime}}$ from any two clusters.  

{ We also give the definitions for notations $O(\cdot)$ and $O_p(\cdot)$ as follows. If $\{x_n\}_1^{\infty}$ is any real sequence, $\{b_n\}_1^{\infty}$ is a sequence of positive real numbers, and there exists a constant $C_{*}<\infty$ such that $|x_n|/b_n \leq C_{*}$ for all $n$, we say that $x_n$ is at most of the order of magnitude of $b_n$, and write $x_n=O(b_n)$. If, for any $\varepsilon>0$, there exists $C_{\varepsilon}< \infty$ such that the stochastic sequence $\{X_n\}_1^{\infty}$ satisfies $\sup_{n}P(|X_n|>C_{\varepsilon})<\varepsilon$, we write $X_n=O_p(1)$. If $\{Y_n\}_1^{\infty}$ is another sequence, either stochastic or nonstochastic, and $X_n/Y_n=O_p(1)$, we say that $X_n=O_p(Y_n)$, or in words, $X_n$ is at most of order $Y_n$ in probability.}

\subsection{Asymptotic properties}
\noindent \textbf{Definition.} A random sequence $\{\xi_k, k\geq1\}$ is said to be $\alpha$-mixing if the $\alpha$-mixing coefficient
\[
\alpha(s)\mathop=\limits^{\mbox{def}}\sup_{k\geq 1}\sup \{ |P(A\cap B)-P(A)P(B)|: A\in \mathcal{F}_{s+k}^{\infty}, B\in \mathcal{F}_{1}^{k} \}
\]
converges to 0 as $s\rightarrow \infty$, where $\mathcal{F}_a^b$ is the $\sigma$ algebra generated by $\xi_a,\xi_{a+1},\ldots,\xi_b$. 

{Among various mixing conditions used in the literature, the $\alpha$-mixing is reasonably weak and is known to be fulfilled by many stochastic processes including many time series models. For instance, \cite{gorodetskii1978strong} derived the conditions under which a linear process is $\alpha$-mixing. The linear autoregressive and the bilinear time series models are strongly mixing with mixing coefficients decaying exponentially under very mild assumptions, see the page 99 of \citep{doukhan2012mixing} for more details. We refer to \citep{kitamura1997empirical, cai2000functional} and references therein for more discussions on the $\alpha$-mixing condition.}

We denote by $C^{(r)}=\left \{ \phi | \phi^{(r)} \in C(\mathbb{T}) \right \} $ the space of the $r$th order smooth functions on the compact support $\mathbb{T}$ such that their $r$th order derivatives belong to $C(\mathbb{T})$, which is the class of all continuous functions on $\mathbb{T}$.

\bigskip

\noindent \textbf{Regularity conditions:}
\begin{enumerate}
    \item[(C1)] The observation time points $t_{ij}$, $i=1,\ldots,n$, $j=1,\ldots,m_i$, are chosen independently from a distribution $F(\cdot)$ with the density $f(\cdot)$. Moreover, the density function $f(t)$ is uniformly bounded away from 0 and infinity on its compact support $\mathbb{T}$. Without loss of generality, we assume $\mathbb{T}=[a_0,b_0]$.
    
   \item [(C2)] There exists a positive constant $M$ such that $E(\varepsilon(t)^4)\leq M $ for all $t \in \mathbb{T}$. In addition, the random sequence $\{\varepsilon_{ij}\}$ for each $i$ satisfies $\alpha$-mixing condition with the $\alpha$-mixing coefficient satisfying $\alpha(s)\leq C^{*} s^{-\alpha}$ for $\alpha>\frac{2+\kappa_0}{1-\kappa_0}$, where $0<\kappa_0<1$, and $C^{*}$ is a positive constant with $0<C^{*}<\infty$.
   
   \item [(C3)] The functions $\beta_i(\cdot)\in C^{(r)}$, for $i=1,\ldots,n$.
   
   \item [(C4)] The spline knot sequences $\{\zeta_l\}_{l=0}^{J+1}$ have bounded mesh ratio. That is, for some positive constant $C_{01}$, $$
    \frac{\max_{0\leq l\leq J}|\zeta_{l+1}-\zeta_{l}|}{\min_{0\leq l\leq J}|\zeta_{l+1}-\zeta_{l}|} \leq C_{01}.$$

   \item [(C5)] {There are positive constants $0<C_1 < C_2<\infty$ such that the eigenvalues of $\boldsymbol{\Sigma}$=diag$(\boldsymbol{\Sigma}_1,\ldots,\boldsymbol{\Sigma}_n)$ and $\boldsymbol{V}$=diag$(\boldsymbol{V}_1,\ldots,\boldsymbol{V}_n)$ lie between $C_1$ and $C_2$.}
\end{enumerate}

{Condition (C1) is identical to condition (C1) in \citep{huang2004polynomial} and assumption (A1) in \citep{noh2010sparse}. This condition ensures that the observation time points are randomly scattered and it can be modified or weakened according to Remarks 3.1 and 3.2 in \citep{huang2004polynomial}. Condition (C2) is a standard requirement for moments and the mixing coefficient for an $\alpha$-mixing process as assumed in \citep{kitamura1997empirical} and \citep{cai2000functional}. This condition allows the errors to be weakly dependent. Many linear and nonlinear time series models like the linear autoregressive and the bilinear time series models are strongly mixing with the mixing coefficients decaying exponentially, see \citep{doukhan2012mixing} (page 99)  for more details. Conditions (C3)-(C4) are frequently assumed in the spline approximation literature; see for example \citep{zhu2008asymptotics, wang2011estimation, Ma2014}. The smoothness condition on $\beta_i(\cdot)$ given by Condition (C3) determines the rate of the approximation error of the spline estimator $\hat{\beta}_i(\cdot)$. Condition (C4) ensures that the knot sequence has a bounded mesh ratio; that is, the knots are quasi-uniform. Condition (C5) is commonly used in the literature related to longitudinal data, such as in \citep{huang2007efficient, MaSongWang2013} and the references therein.}


Let the nonparametric function subspace $M_\mathcal{G}^{\beta}$ corresponding to the group partition be
$M_\mathcal{G}^{\beta}= \left\{ \beta(\cdot): \beta_i(\cdot)=\alpha_k(\cdot),~\beta_i(\cdot)\in C^{(r)}, \mbox{for}~i \in \mathcal{G}_k, 1\leq k \leq K  \right\}$, while the subspace $M_\mathcal{G}^{\gamma}$ of B-spline coefficients corresponding to the group partition is denoted by $M_\mathcal{G}^{\boldsymbol{\gamma}}= \left\{ \boldsymbol{\gamma}: \boldsymbol{\gamma}_i=\boldsymbol{\theta}_k,~\boldsymbol{\gamma}_i\in R^{S},~\mbox{for}~i \in \mathcal{G}_k, 1\leq k \leq K  \right\}$, where $\boldsymbol{\theta}_k$ is the common B-spline coefficients in the $k$th subgroup. By using the proposed method, we have $\hat{\boldsymbol{\gamma}}=(\hat{\boldsymbol{\gamma}}^T_1, \dots, \hat{\boldsymbol{\gamma}}^T_n)^T$, where $\hat{\boldsymbol{\gamma}}_i$ is the estimated B-spline coefficient for subject $i$ with $\hat{\boldsymbol{\gamma}}_i=\hat{\boldsymbol{\theta}}_k$ for all $i \in \hat{\mathcal{G}}_k$. Then, the estimated function for each $i$ is 
\begin{align}
\hat{\beta}_i(t)=\boldsymbol{B}(t)^T\hat{\boldsymbol{\gamma}}_i,
\label{beta_hat}
\end{align}
for any $t \in \mathbb{T}$. Let $\hat{\boldsymbol{\alpha}}^{or}(t)=(\hat{{\alpha}}^{or}_1(t), \dots, \hat{{\alpha}}^{or}_K(t))$, where $\hat{\alpha}^{or}_k(t)$ is the estimated common function for group $\mathcal{G}_k$ by assuming that the true memberships are known. 


\begin{theorem}\label{Th1}
Suppose conditions (C1)-(C5) hold, and for any fixed $K$, if $J = O(N_0^\varsigma)$ with $0<\varsigma< 1$, the oracle estimator $\hat{\boldsymbol{\alpha}}^{or}$ satisfies $\|\hat{\boldsymbol{\alpha}}^{or}-\boldsymbol{\alpha}\|_2^2=O_p(J/N_0+J^{-2r})$, {where $N_0=\min_{1\leq k\leq K}N_k$ and $N_k=\sum_{i\in \mathcal{G}_k}m_i$.}
\end{theorem}

{ It is worth noting that the convergence rate given in Theorem \ref{Th1} consists of two parts, which are the approximation error of order $J^{-2r}$ and the estimation error of order $J/N_0$. We can see that the increase of $J$ leads to smaller approximation error but larger estimation error, whereas the decrease of $J$ leads to larger approximation error but smaller estimation error, i.e., there is a trade-off between the bias and variance. By letting $J/N_0=J^{-2r}$, we can obtain the optimal order of $J$ which is  $ N_0^{1/(2r+1)}$. Plugging it into the convergence rate, it follows that $\|\hat{\boldsymbol{\alpha}}^{or}-\boldsymbol{\alpha}\|_2^2=O_p(J/N_0+J^{-2r})=O_p\left( N_0^{-2r/(2r+1)} \right)$, which reaches the minimax convergence rate for spline regression.} 

{
The following theorem gives the convergence rate of the estimated function $\hat{\beta}_i(t)$ in (\ref{beta_hat}) for each $i$.}

\begin{theorem}\label{Th2}
Suppose conditions (C1)-(C5) hold, if there exists a constant $C>0$ such that $Cb\geq \tau \lambda$ and $J = O(m_{(n)}^\varsigma)$ with $0<\varsigma< 1$, then, for each $i$, $\|\hat{{\beta}_i}-{\beta}_i\|_2^2=O_p(J/m_{(n)}+J^{-2r})$, {where $m_{(n)}=\min_{1\leq i\leq n } m_i$.}
\end{theorem}

\begin{theorem}\label{Th3}
Assume $\hat{\mathcal{G}}$ and $\mathcal{G}_0$ respectively be the estimated and true subgroup membership. Under the same conditions in Theorem \ref{Th2}, we have $P\left( \hat{\mathcal{G}}=\mathcal{G}_0 \right) \rightarrow 1$ as $m_{(n)}\rightarrow\infty$.
\end{theorem}

{
Theorem \ref{Th3} gives the model selection consistency result for the penalized method.} Thus, given the estimated subgroup membership, we may write {$\hat{\boldsymbol{\alpha}}(t)=(\hat{{\alpha}}_1(t),\ldots,\hat{{\alpha}}_K(t))^T$ for any given $t\in \mathbb{T}$,} and the following theorem holds.

\begin{theorem}\label{Th4}
Under the same conditions in Theorem \ref{Th3}. If $J /m_{(n)}^{1/(2r+1)} \rightarrow \infty$, we have
\[
\emph{Var}\left(\hat{\boldsymbol{\alpha}}(t)\right)^{-1/2}\left(\hat{\boldsymbol{\alpha}}(t)-\boldsymbol{\alpha}(t)\right) \mathop \to \limits^d N(\boldsymbol{0},\boldsymbol{I}_K),
\]
where $\boldsymbol{I}_K$ is a K-dimensional identity matrix and $\emph{Var}\left(\hat{\boldsymbol{\alpha}}(t)\right)$ is given in (\ref{var}) of the Supplementary Material. In particular,
\[
\emph{Var}\left(\hat{\alpha}_k(t)\right)^{-1/2}\left(\hat{\alpha}_k(t)-\alpha_k(t)\right)  \mathop \to \limits^d N(0,1)
\]
for $k=1,\ldots,K$, where $\emph{Var}\left(\hat{\alpha}_k(t)\right)=\boldsymbol{e}_{k}^T\emph{Var}\left(\hat{\boldsymbol{\alpha}}(t)\right)\boldsymbol{e}_{k}$, and $\boldsymbol{e}_{k}$ is the $K$-dimensional vector with the $k$th element taken to be 1 and 0 elsewhere.
\end{theorem}

We can use the asymptotic distribution established in Theorem \ref{Th4} to construct pointwise confidence intervals of the functional curve for each subgroup. 
\section{Simulation studies}
\label{SEC:simulation}
In this section, we investigate the performance of our proposed approach by conducting simulation studies. Balanced and unbalanced data are both considered. 

Two different criteria are used to select the optimal tuning parameter. {One is the modified Bayes Information Criterion (BIC) \citep{bic} for high-dimensional data settings by minimizing}
\begin{equation}
    \text{BIC}(\lambda)=\text{log} \left [ \sum^n_{i=1}(\boldsymbol{Y}_i-\boldsymbol{X}_i\hat{\boldsymbol{\gamma}}_i(\lambda))^{T}\boldsymbol{R}^{-1}_i (\boldsymbol{Y}_i-\boldsymbol{X}_i\hat{\boldsymbol{\gamma}}_i(\lambda)) /N\right  ]+C_n\frac{\text{log}N}{N}(\hat{K}(\lambda)S),
\label{BIC}
\end{equation}
where $C_n$ is a positive number which can depend on $n$ and $N=\sum^n_{i=1}m_i$. Following \cite{ma2017concave}, we let $C_n=c \ \text{log}(\text{log}(nS))$, where $c$ is a positive constant, and we choose $c=0.6$. {The other criterion is the Calinski-Harabasz index \citep{calinski1974dendrite} by maximizing }
\begin{equation}
\text{CH}(\lambda)=\frac{B_{\hat{K}(\lambda)}/(\hat{K}(\lambda)-1)}{W_{\hat{K}(\lambda)}/(n-\hat{K}(\lambda))},
\end{equation}
where $B_{\hat{K}(\lambda)}$ and $W_{\hat{K}(\lambda)}$ are the between and within group sum of square errors of the estimated subgroups given a $\lambda$ value. We apply this index to the initial value  $\boldsymbol{\gamma}_i^0$'s presented in \ref{initial_value} of Supplementary Material, which are the ordinary least squares estimates of  (\ref{mamodel}) within each subject. Note that $\text{CH}(\lambda)$ is not defined for $\hat{K}(\lambda)=1$. Based on these criteria, we can select the optimal $\lambda$ and obtain the corresponding group membership. Here we use fixed values for $\vartheta$ and $\tau$ in ADMM algorithm: $\vartheta=1$ and $\tau=3$.

\indent To evaluate the accuracy of the clustering results, we provide three measures: Rand Index (RI) \citep{rand1971objective}, Normalized Mutual Information (NMI) \citep{vinh2010information} and accuracy percentage (\%). The accuracy percentage (\%) is defined as the proportion of subjects that are correctly identified. These three values are between 0 and 1, with higher values indicating better performance. 

\subsection{Two Subgroups Example}
We simulate data from the heterogeneous model with two subgroups
\begin{equation*}
Y_{ij}=\beta _{i}(t_{ij})+\varepsilon
_{ij},~~i=1,\dots,n,~j=1,\dots,m_i, 
\end{equation*}%
where $\beta_i(t)=\alpha_1(t)$ if $i \in \mathcal{G}_1$ and $\beta_i(t)=\alpha_2(t)$ if $i \in \mathcal{G}_2$. 

\indent We first consider balanced data. In this situation, we have $m_i=T$ for all $i$'s. The time points $t_{ij}$'s are chosen equally spaced on $[0, 1.2]$. The error term $\boldsymbol{\varepsilon}_{i}=(\varepsilon_{i1},\dots, \varepsilon_{iT})^{{T}}$ is generated from $N (\boldsymbol{0}, \boldsymbol{\Sigma}_E)$, in which $\boldsymbol{\Sigma}_E$ has AR(1) covariance structure with $\rho=0.3$ and $\sigma=0.5$. We consider 4 setups of $\left \{ n, \, T \right \}$: $\left \{ n=100, \, T=20\right \}, \left \{n=100, \, T=50  \right \}, \left \{ n=150,\, T=20 \right \} \text{and} \left \{ n=150, \, T=50 \right \}$. Moreover, to choose $\left \{ \alpha_1(t),\, \alpha_2(t) \right \}$, we also consider three different cases by increasing the distance between the two functions from close to middle, then to far, which are shown below: 
\begin{align*}
\text{Close}\left\{
\begin{array}{ll}
\alpha_1(t)=-0.5t^2+1.25t,
\\
\alpha_2(t)=-t^2+2.5t,
\end{array}\right. 
&&
\text{Middle}\left\{
\begin{array}{ll}
\alpha_1(t)=-0.5t^2+1.25t,
\\
\alpha_2(t)=-1.3t^2+3.25t,
\end{array}\right. 
\\
\text{Far}\left\{
\begin{array}{ll}
\alpha_1(t)=-0.5t^2+1.25t,
\\
\alpha_2(t)=-2.5t^2+6.25t.
\end{array}\right. 
\end{align*}

\begin{figure}[t]
\par
\begin{center}
$
\begin{array}{ccc}
\includegraphics[width=4.3cm]{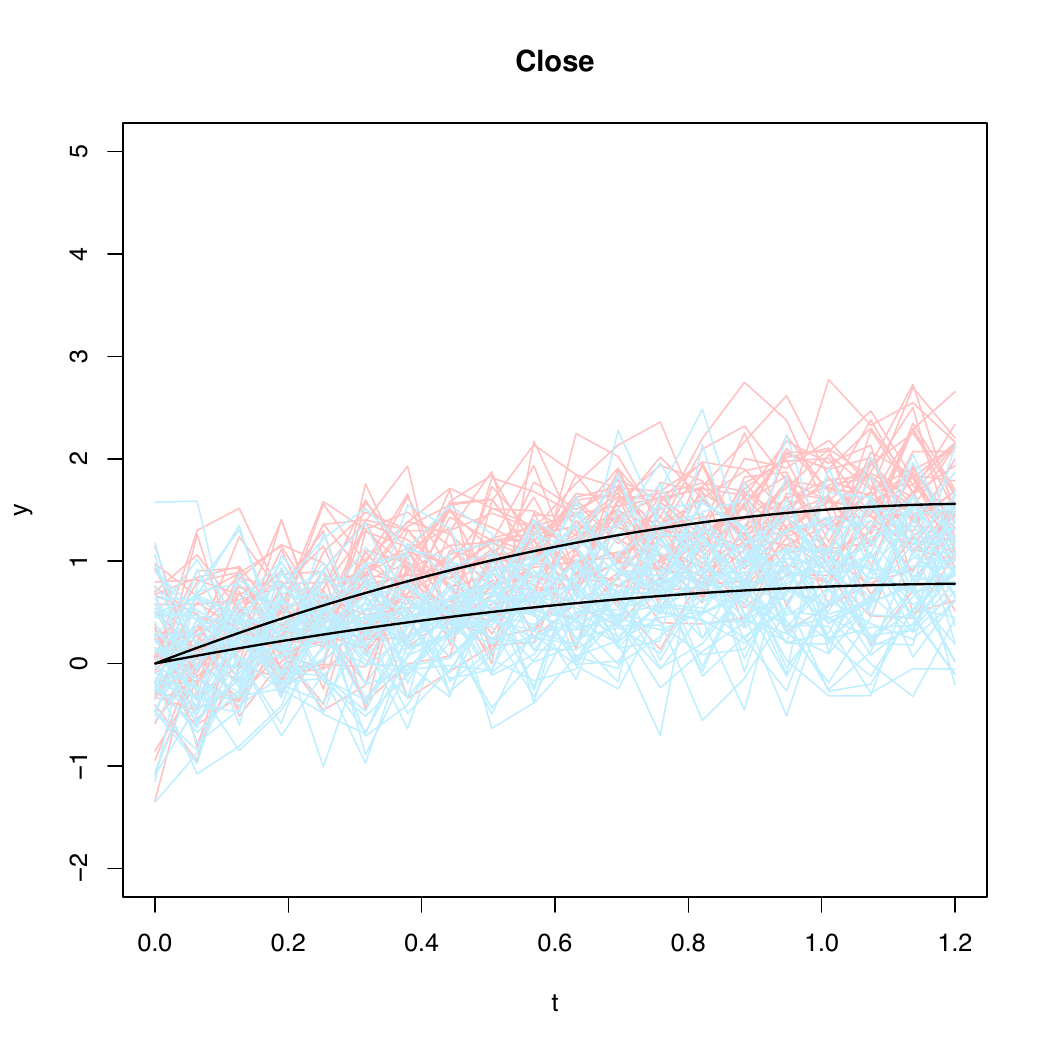}&
\includegraphics[width=4.3cm]{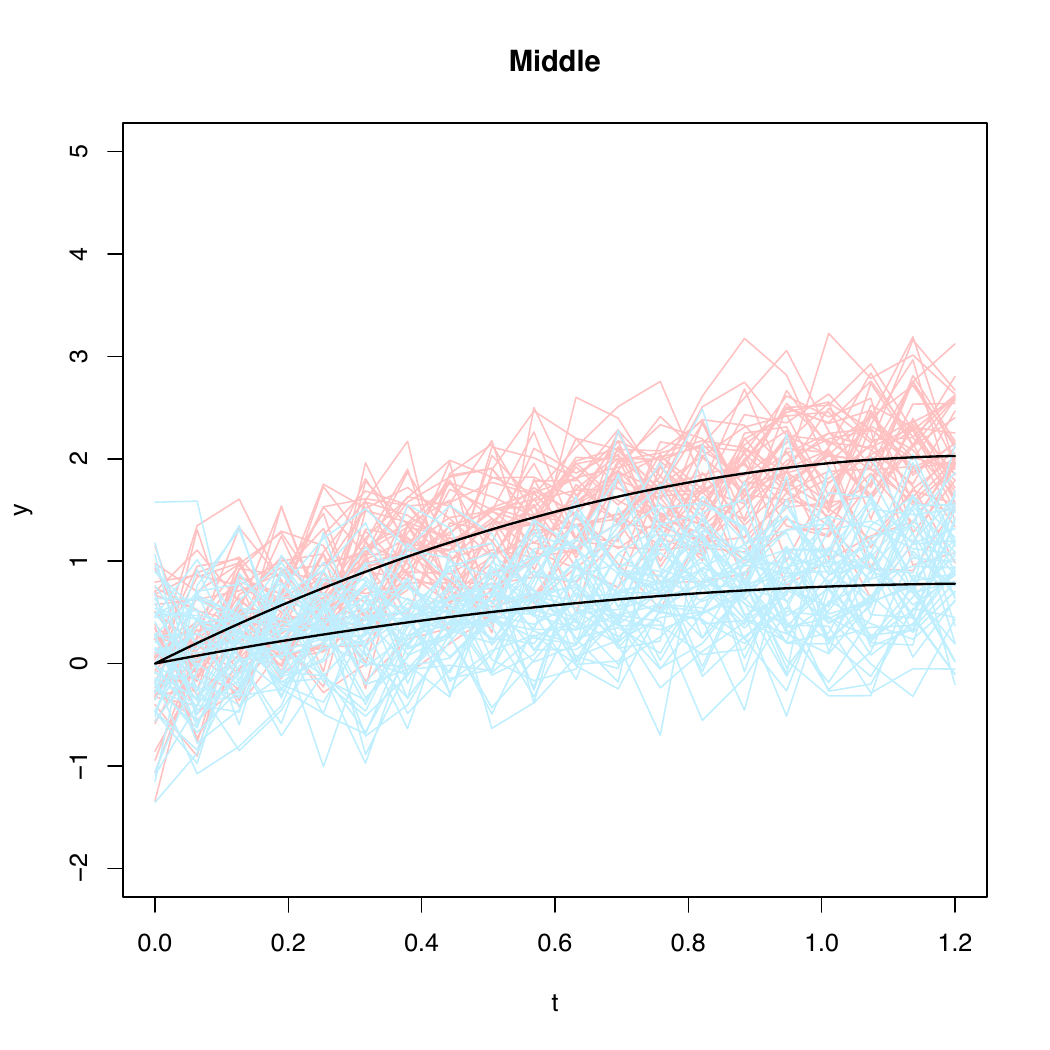}&
\includegraphics[width=4.3cm]{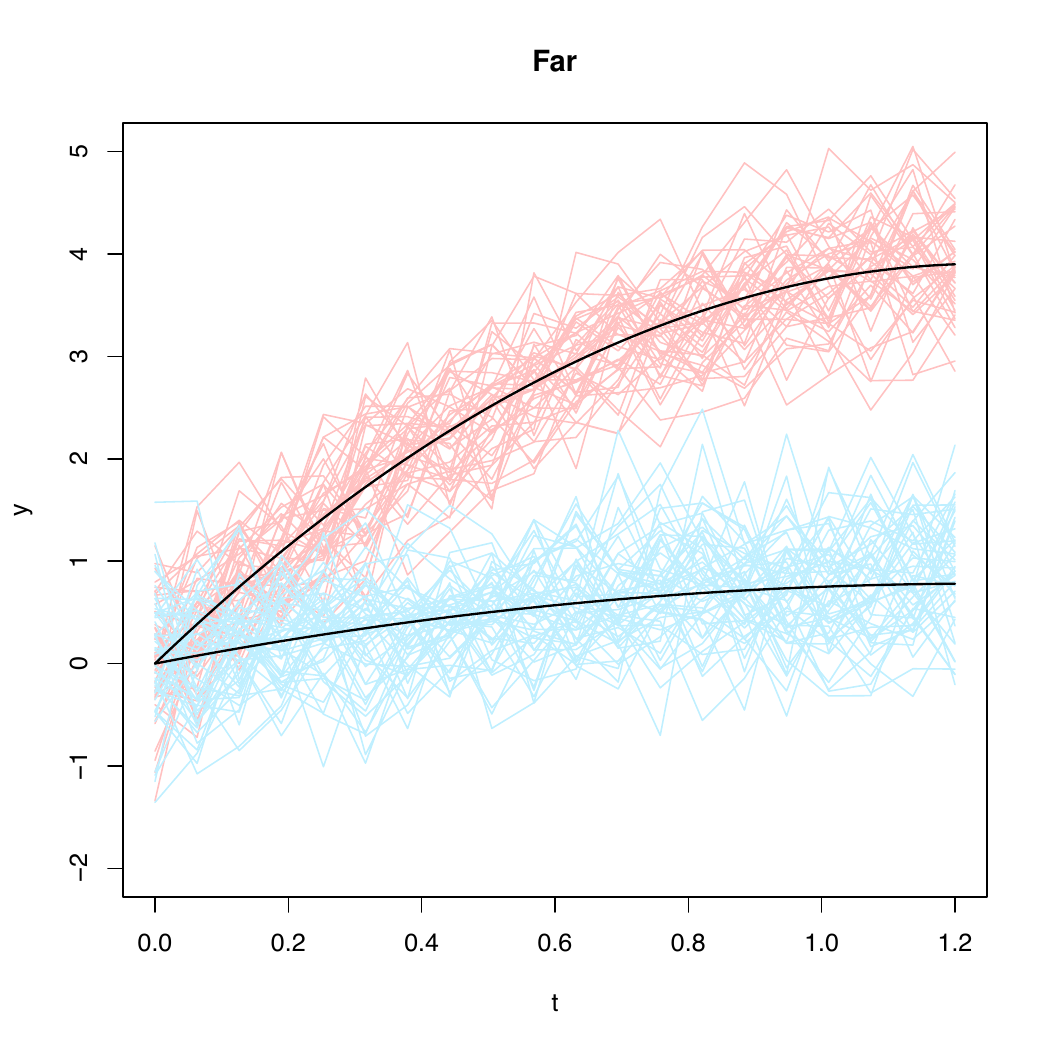}
\end{array}%
$
\end{center}
\caption{The black lines represent the true functions, while the red and blue lines represent the simulated trajectories of the corresponding subgroups under one replication when $n=100, \ T=20$ for balanced data in Two Subgroups Example. The distance between the true functions increases from close, to middle, to far.}
\label{trajectory}
\end{figure}

Figure \ref{trajectory} shows the true functions (black line) and simulated trajectories (blue line and red line) of the three distance cases, respectively, based on one sample with $n=100, \, T=20$ for balanced data. We can see that there are a lot of overlaps in Close and Middle cases, especially in Close case, it looks more like one group.

\indent The unbalanced data is based on the balanced data setting. However, we randomly allow 50\% of the subjects to miss either 30\% or 40\% or 50\% of time points. Next, we conduct simulations to illustrate the performance of our proposed method. 100 replications are taken here. Quadratic splines with one interior knot are used to approximate the nonparametric components. {The quadratic splines are B-splines with order $r=3$. As the order of the B-splines increases, the estimated curve becomes smoother. The quadratic splines can yield smooth enough curves while preventing over-smoothing. Based on the convergence rate given in Theorem 4.2 on page 8, the optimal order of the number of interior knots $J$ is $m^{\frac{1}{2r+1}}_{(n)}$ by letting ${J}/{m_{(n)}}=J^{-2r}$. We choose $J=\left\lfloor
m^{\frac{1}{2r+1}}_{(n)}\right\rfloor =\left\lfloor
m^{\frac{1}{7}}_{(n)}\right\rfloor$, where $\left\lfloor
a\right\rfloor$ denotes the largest integer no bigger than $a$. Then $J=\left\lfloor
m^{\frac{1}{7}}_{(n)}\right\rfloor=1$ when $m_{(n)}=10,20,25, 50$ in our simulation settings.}

\indent Table \ref{k=2} not only reports the summary measurements of the estimated number of subgroups $\hat{K}$ (sample mean, median, per, where per is the percentage of $\hat{K}$ equaling to the true number of subgroups), but also the summary measurements of the clustering accuracy (average values of RI, NMI, \%) by using different model selection criteria (BIC, CH) under different setups of $\left \{ n, \ T \right \}$ when the distance between functions increases (Close, Middle, Far). Balanced and unbalanced data are both included. Note that when calculating RI, NMI and \%, we only include the replications with $\hat{K}$ equaling to the true number of subgroup ($\hat{K}=2$).

\indent From Table \ref{k=2}, we can see that both BIC and CH criteria perform well and give the similar results for most of the cases. When T increases, the summary measurements of $\hat{K}$ (mean, median, per) and accuracy measurements (RI, NMI, \%) both increase. In details, the mean of $\hat{K}$ gets close to 2 and median $\hat{K}$ becomes to 2, where 2 is the true number of subgroups, while the accuracy measurements (RI, NMI, \%) are close to 1 or even become to 1 for both balanced and unbalanced data, which indicates good clustering results. What's more, with the distance between the true functions getting larger, it is much easier to correctly identify the subgroups. Accordingly, we observe that the mean and median of $\hat{K}$ become to 2, while the RI, NMI and \% become to 1 when the distance is sufficiently large (Far case). On the contrary, in Close case, since the trajectories of the two subgroups in Figure \ref{trajectory} show a lot of overlaps, it is more difficult to identify the subgroups, which results in the low percentage (per) of correctly selecting the number of subgroups when $T=20$. Under this case, if we can cluster the subjects into two subgroups, BIC criterion presents higher accuracy performance in group membership. However, if $T$ increases to $50$, all the measurements become much better and it is more likely to correctly identify the subgroups. Compared with unbalanced data, balanced data shows slightly better results.

\indent Furthermore, to study the estimation accuracy, we calculate the square root of the mean squared error (RMSE) of the estimated function in each subgroup only when $\hat{K}$ equals the true number of subgroups $K$. In the $k$th subgroup, we use the formula below to find the corresponding RMSE of the estimated function $\hat{\alpha}_k(t)$ ($\text{RMSE}_k$):
\begin{equation*}
\text{RMSE}_k=\sqrt{\frac{1}{H}\sum^H_{h=1}[\hat{\alpha}_k(t_h)-\alpha_k(t_h)]^2} = \sqrt{ \frac{1}{H}\sum^H_{h=1}[\boldsymbol{B}_{(k)}(t_h)^T\hat{\boldsymbol{\gamma}}_{(k)}-\alpha_k(t_h)]^2}, ~k=1,\dots, {K},
\end{equation*} 
where $\boldsymbol{B}_{(k)}(t)$ is the B-spline basis vector of the $k$th subgroup, $\hat{\boldsymbol{\gamma}}_{(k)}$ is the corresponding estimated B-spline coefficient after refitting model (\ref{mamodel}) and {
$\left \{ t_1, \cdots, t_H\right \}$ is a grid of equally spaced points spanning the original time range $[0, 1.2]$ with $H=50$.}

For oracle (Oracle) method, we use the true group membership to calculate RMSE. As shown in Table \ref{k=2rmse}, the RMSE values under different model selection criteria (BIC, CH) and $\left \{n, \ T \right \}$ setups are comparable to those of the oracle ones for almost all cases. 

\begin{figure}[tbp]
\centering
\scalebox{0.9}{
\includegraphics[width=4.5cm]{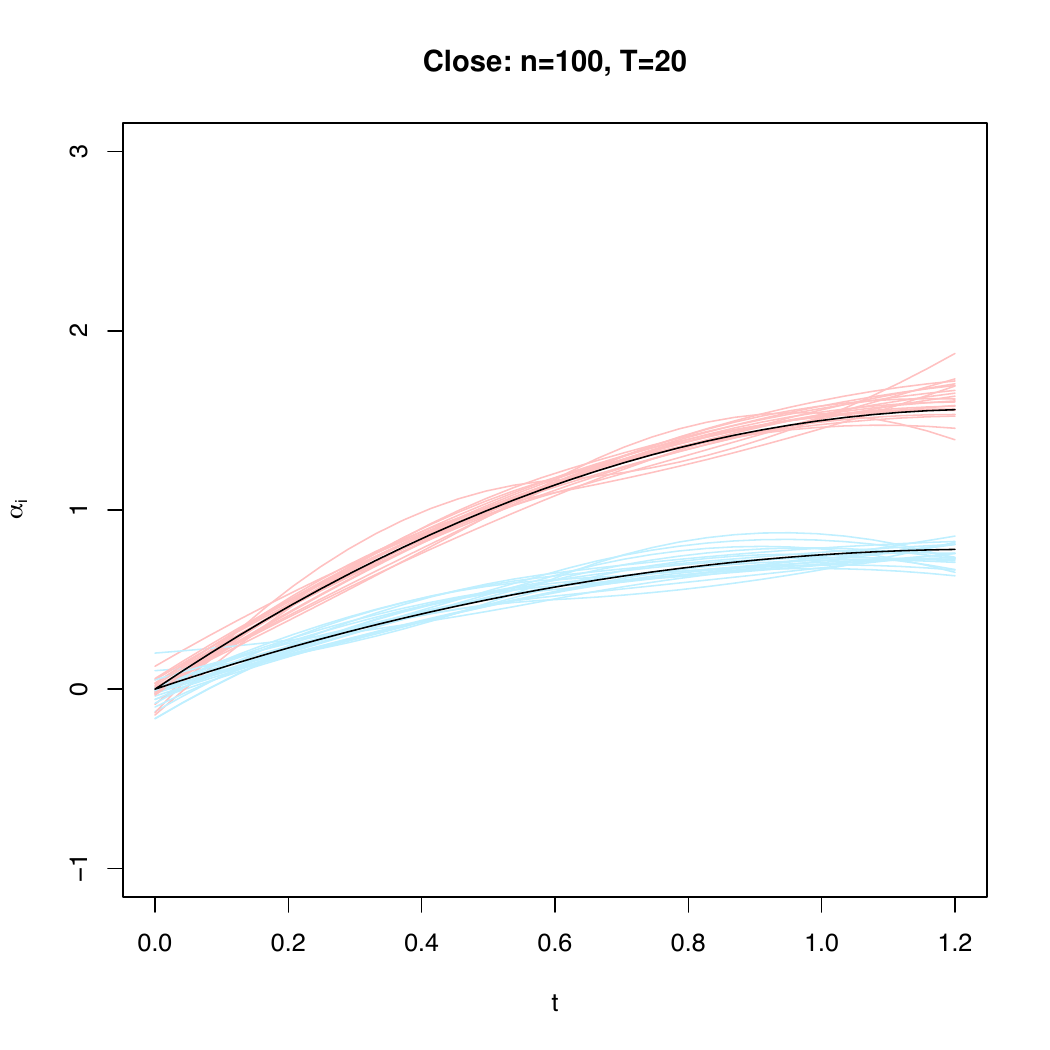}\quad
\includegraphics[width=4.5cm]{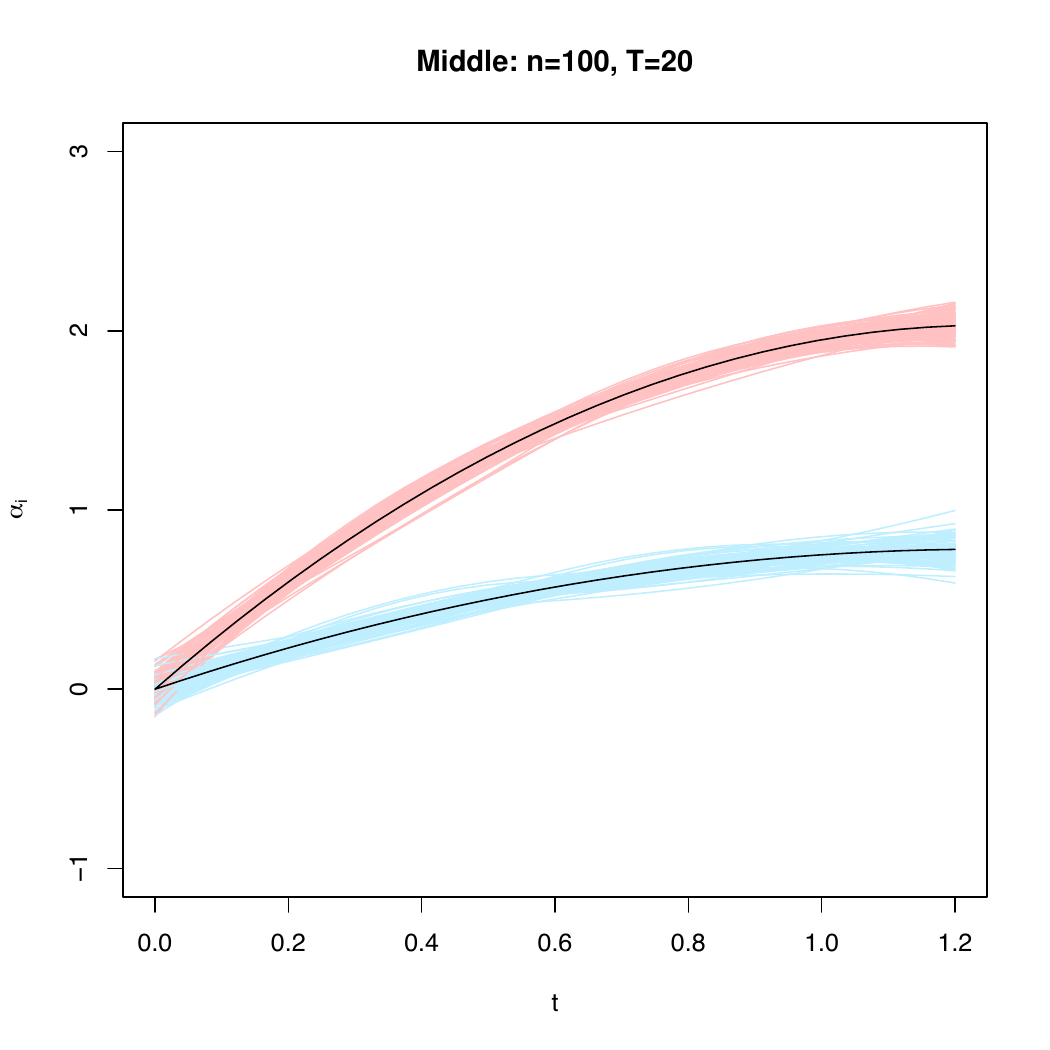}\quad
\includegraphics[width=4.5cm]{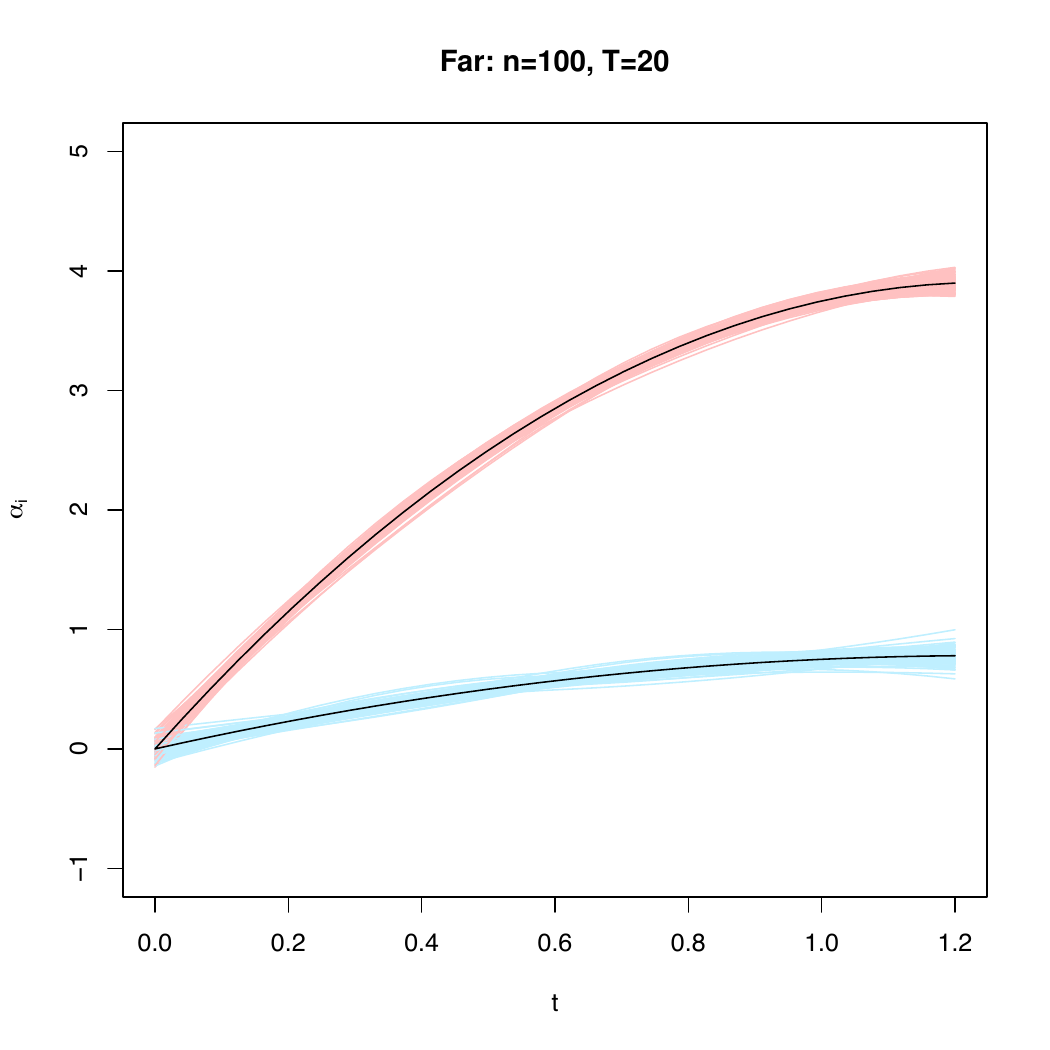}
}

\medskip

\scalebox{0.9}{
\includegraphics[width=4.5cm]{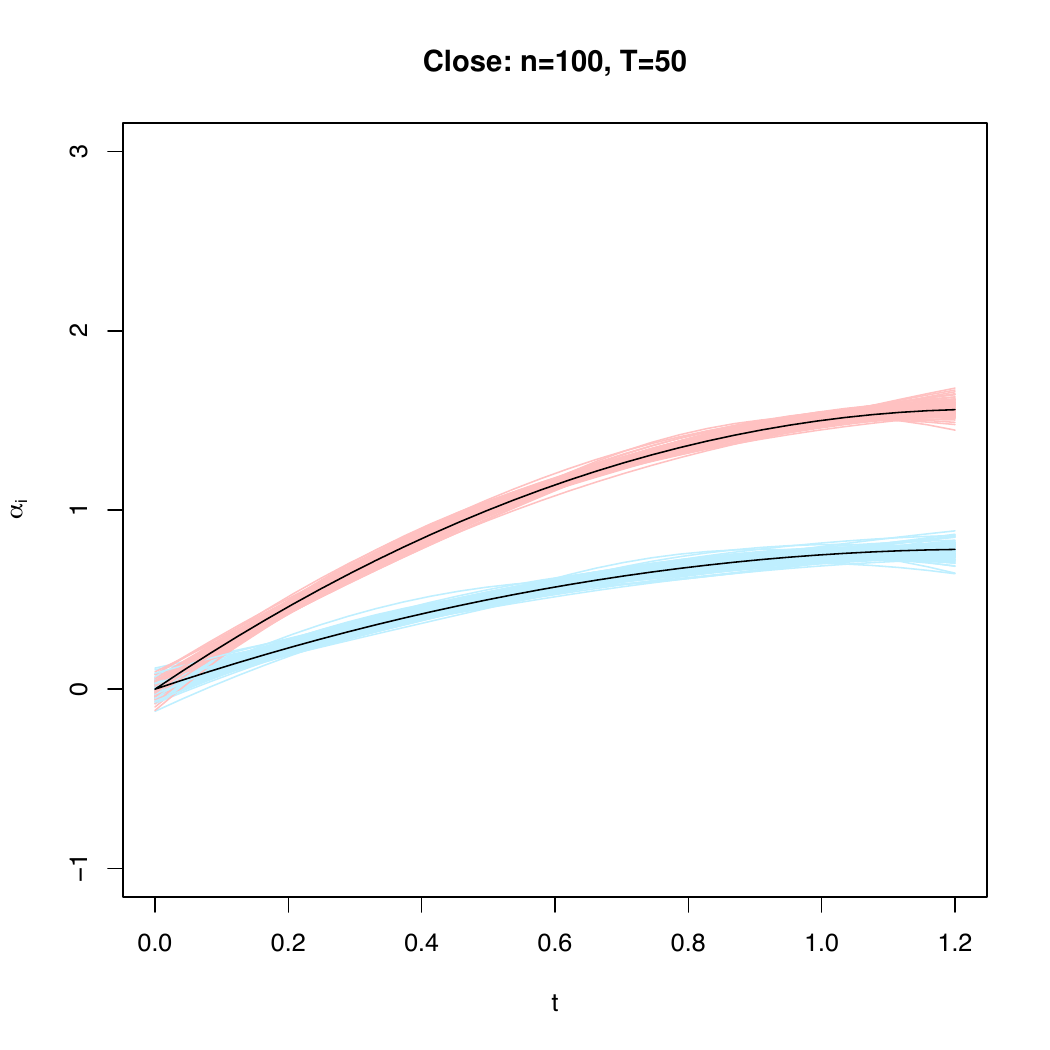}\quad
\includegraphics[width=4.5cm]{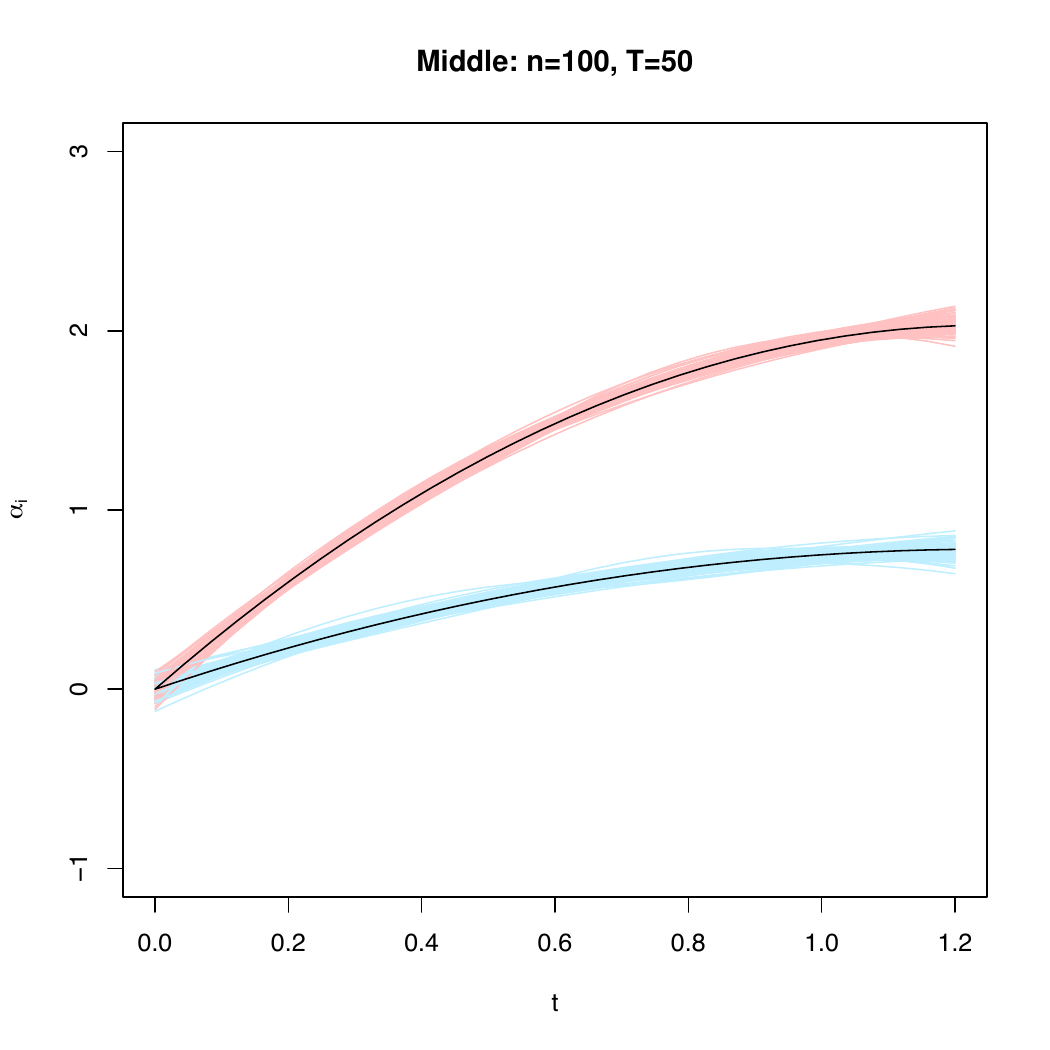}\quad
\includegraphics[width=4.5cm]{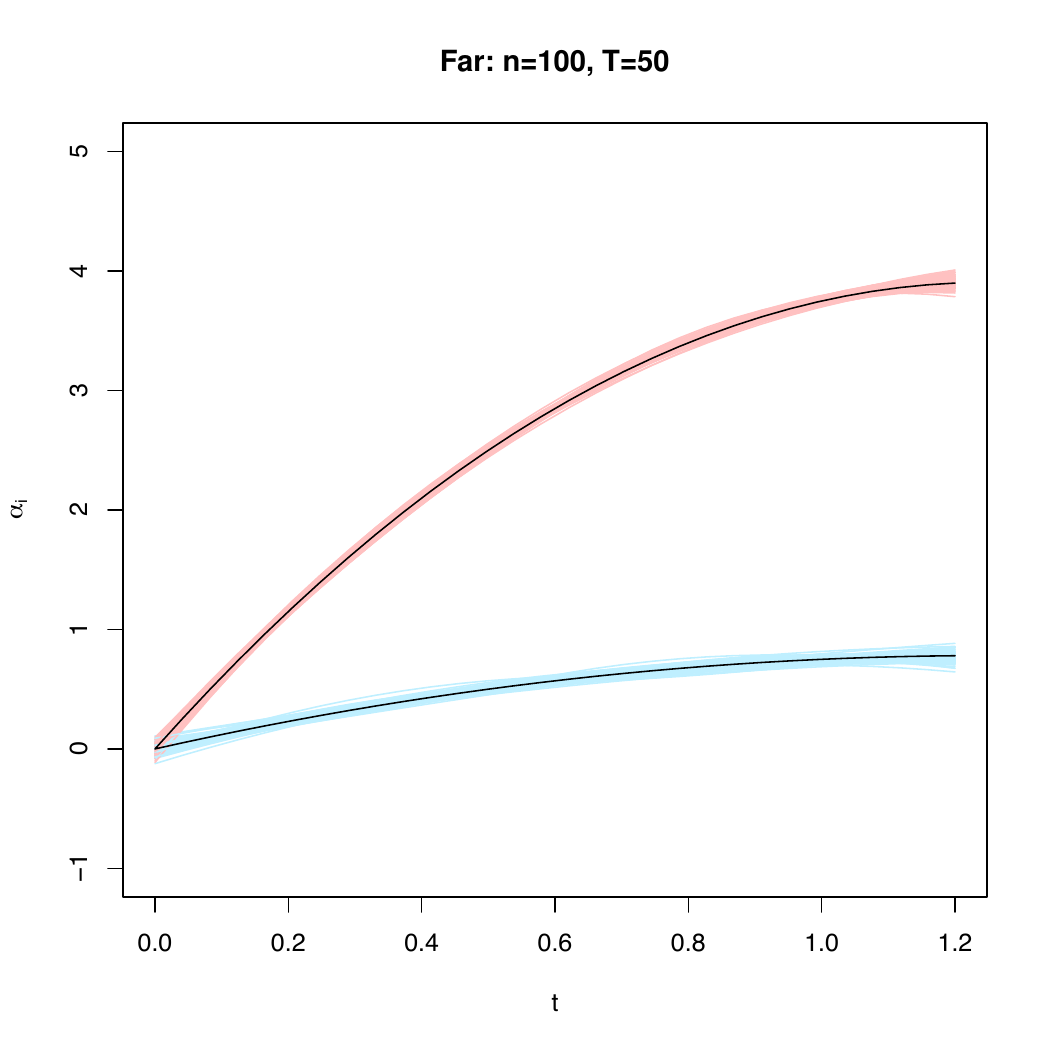}
}

\medskip

\scalebox{0.9}{
\includegraphics[width=4.5cm]{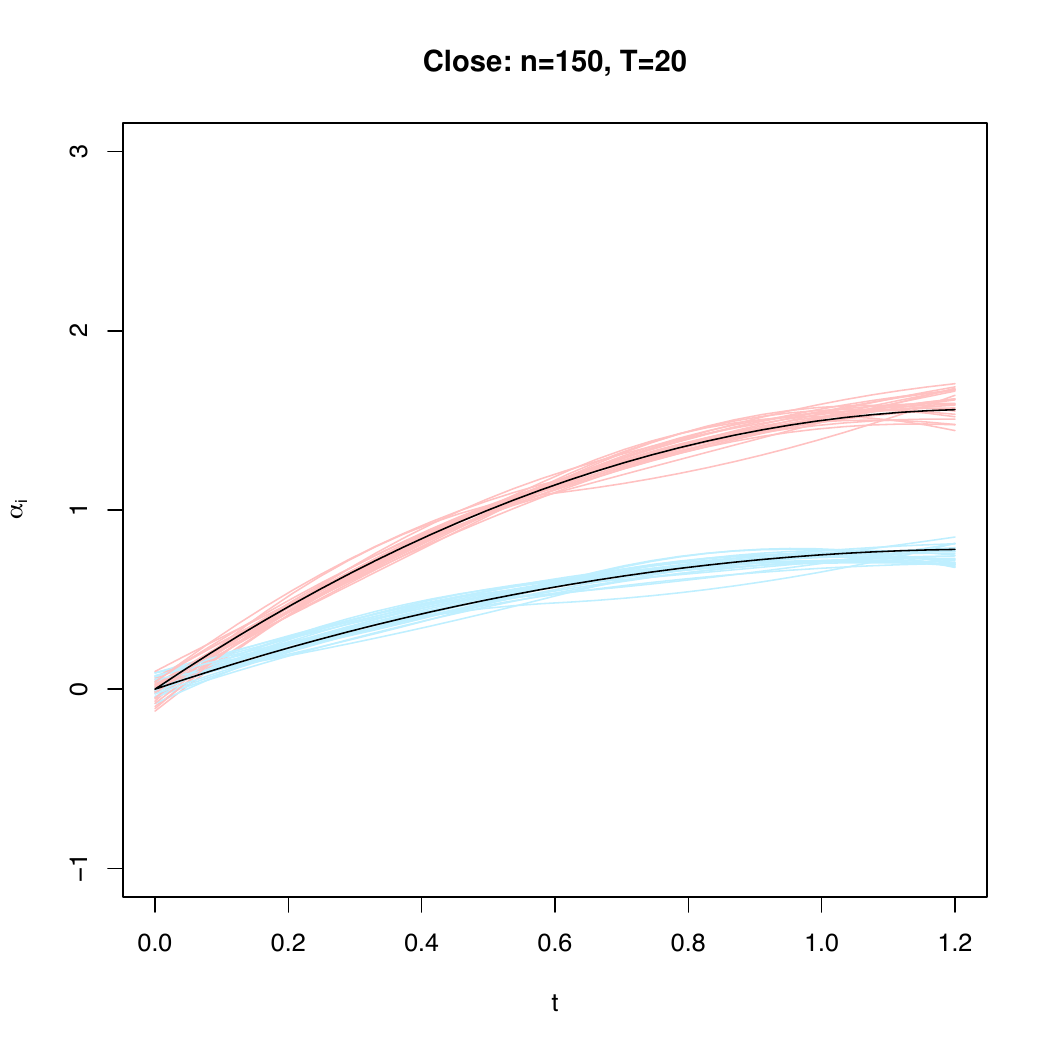}\quad
\includegraphics[width=4.5cm]{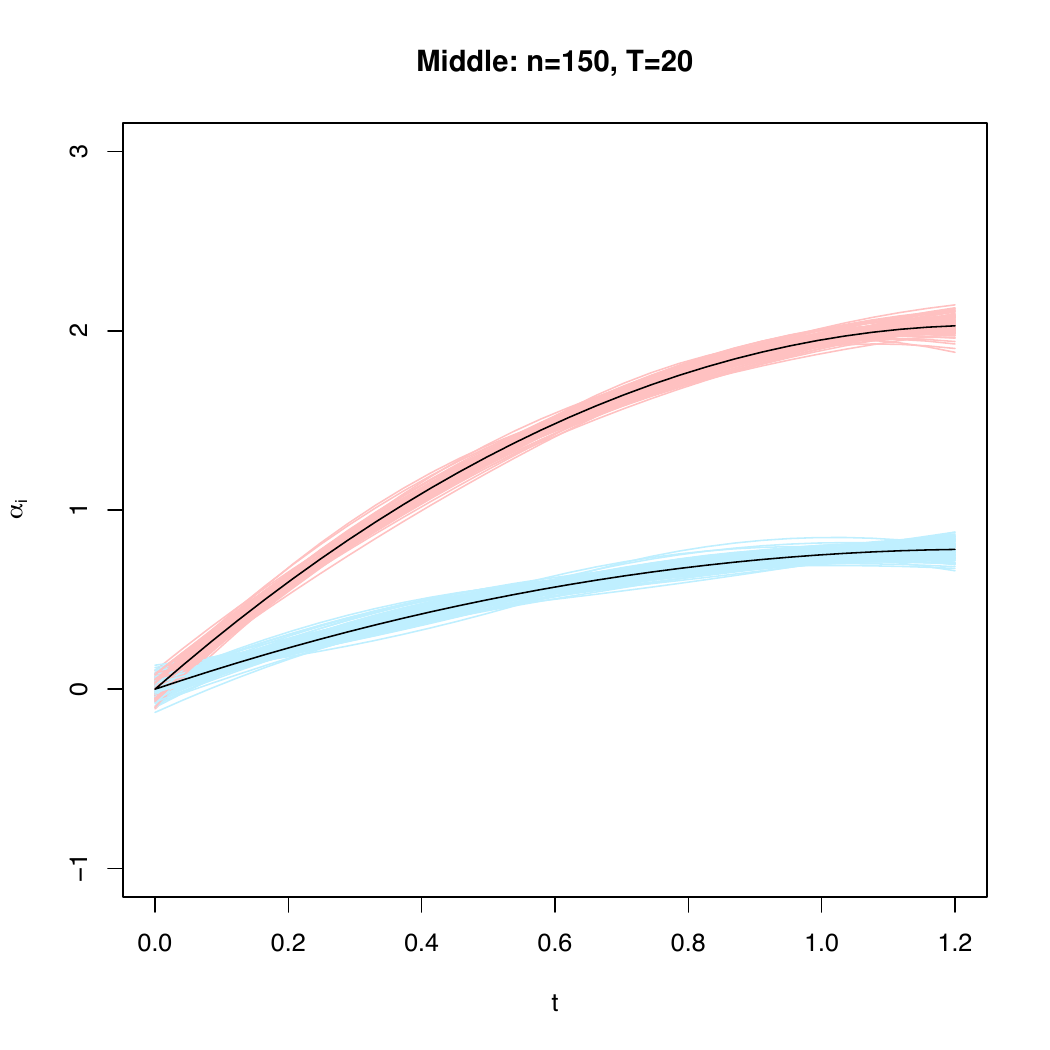}\quad
\includegraphics[width=4.5cm]{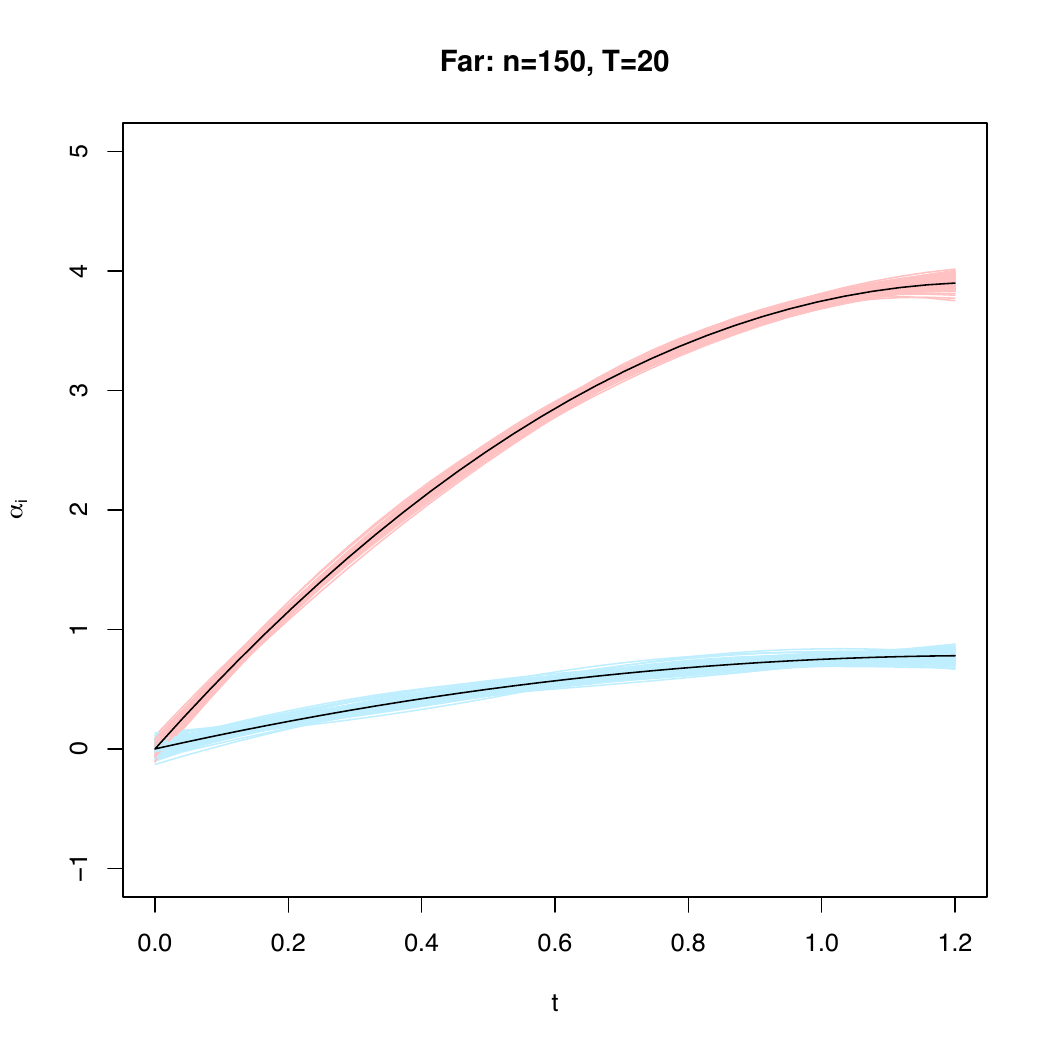}
}

\medskip

\scalebox{0.9}{
\includegraphics[width=4.5cm]{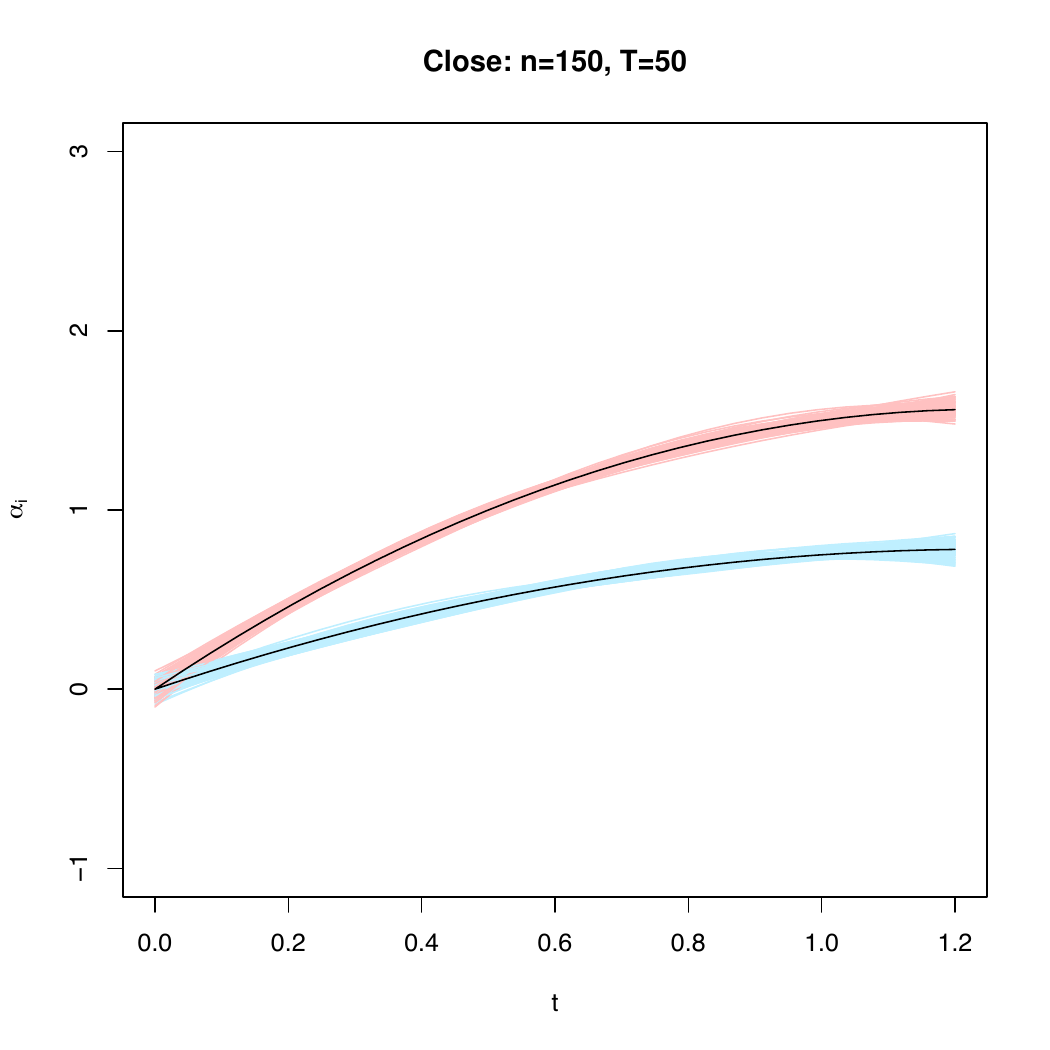}\quad
\includegraphics[width=4.5cm]{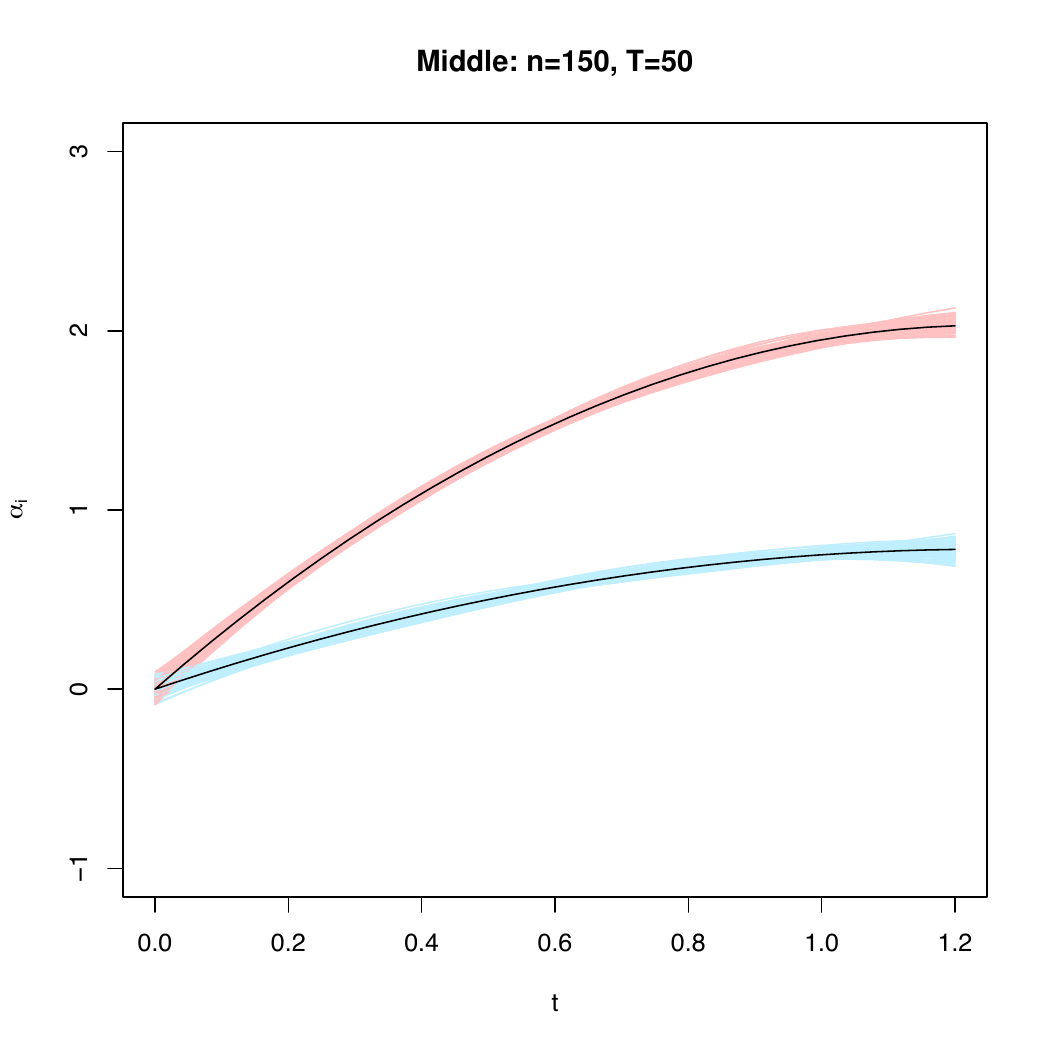}\quad
\includegraphics[width=4.5cm]{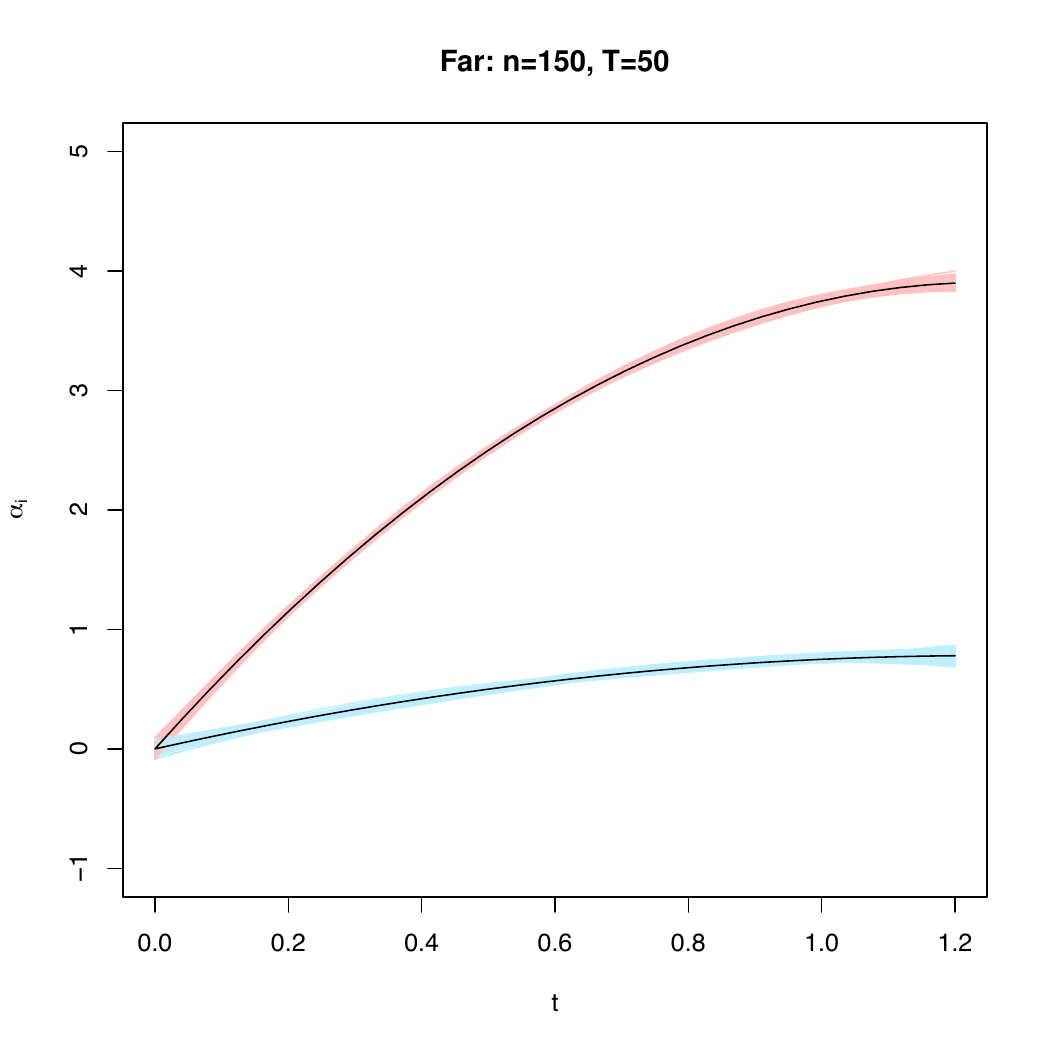}
}

\caption{The black lines represent the true functions, while the red and blue lines are the corresponding fitted curves for the estimated subgroups by using BIC criterion when $\hat{K}=2$ among the 100 replications for balanced data in Two Subgroups Example. On each row, from left to right, it corresponds to close,  middle, and far cases with the same setting of $\left \{ n, \ T \right \}$.}
\label{fitcurve}
\end{figure}

\indent Lastly, the estimated nonparametric curves $\hat{\alpha}_k(t)$ (blue, red lines) and true curves  ${\alpha}_k(t)$ (black line) of the two subgroups for balanced data among the 100 replications are plotted in Figure \ref{fitcurve}. Notice that we only plot the replications when the estimated number of subgroups equals the true number of subgroups. On each row, from left to right, it represents the Close,  Middle, and Far cases with same setting of $\left \{n, \ T \right \}$ respectively. Then either $n$ or $T$ is increased compared to the first row. Given each column, it is obvious that the bands consisting by red or blue lines becomes narrower as $T$ or $n$ increases. Besides, no matter for which setups of $\left\{ n, \ T\right\}$, the estimated curves are very close to the true ones for all distance cases.

{
\indent To further illustrate the performance of our proposed method in unbalanced data, we generate data with $m_i \sim \text{Uniform}\left \{5, 6, ..., 20 \right \}, i = 1,...,n, \ n = 100,1000$, and keep other simulation settings the same as before. We report the numerical results for Middle and Far cases in Table \ref{k=2_new} and \ref{k=2rmse_new}, as the curves from different subgroups in the Close case are too close to be separated based on the previous simulation results; see Table \ref{k=2}. Table \ref{k=2_new} shows that the median value of $\hat{K}$ equals to the true number of subgroups, which is 2. As the mean functions of the subgroups become more separated (from Middle to Far case), the mean value of $\hat{K}$ gets closer to 2, and the average values of RI, NMI and the accuracy percentage (\%) approach to 1. Moreover, Table \ref{k=2rmse_new} shows that the RMSE values of the estimated functions by our method are comparable to those of the oracle ones obtained by assuming that the true memberships are known. These results demonstrate that our proposed method performs well for clustering heterogeneous trajectories from unbalanced data. 

}

\begin{sidewaystable}
\centering
\caption{The sample mean and median of $\hat{K}$, the percentage (per) of $\hat{K}$ equaling to the true number of subgroups, the Rand Index (RI), Normalized mutual information (NMI), and accuracy percentage (\%) equaling the proportion of subjects that are identified correctly under BIC and CH criteria based on 100 realizations in Two Subgroups Example. Balanced and unbalanced data are both included under different $\left \{ n, \ T \right \}$ setups and function distances. }
\scalebox{0.8}{
\begin{tabular}{clccccccccccccc} 
\toprule
\multicolumn{3}{l}{}                                                                             & \multicolumn{6}{c}{\textbf{Balanced} }         & \multicolumn{6}{c}{\textbf{Unbalanced} }         \\ 
\cmidrule(lr){4-9}\cmidrule(lr){10-15}
\textbf{Functions}                & \multicolumn{1}{c}{setting}                      & criterion & mean & median & per  & RI     & NMI    & \%     & mean & median & per  & RI     & NMI    & \%      \\ 
\midrule
\multirow{8}{*}{\textbf{Close} }  & \multicolumn{1}{c}{\multirow{2}{*}{n=100, T=20}} & BIC       & 1.34 & 1.00   & 0.20 & 0.9089 & 0.7459 & 0.9515 & 1.43 & 1.00   & 0.08 & 0.9015 & 0.7289 & 0.9475  \\
                                  & \multicolumn{1}{c}{}                             & CH        & 1.55 & 1.00   & 0.35 & 0.8729 & 0.6818 & 0.9223 & 1.73 & 1.50   & 0.28 & 0.7652 & 0.4861 & 0.8304  \\ 
\cmidrule{2-3}
                                  & \multirow{2}{*}{n=100, T=50}                      & BIC       & 1.98 & 2.00   & 0.98 & 0.9953 & 0.9836 & 0.9977 & 1.97 & 2.00   & 0.97 & 0.9855 & 0.9510 & 0.9927  \\
                                  &                                                   & CH        & 1.98 & 2.00   & 0.98 & 0.9953 & 0.9834 & 0.9977 & 1.97 & 2.00   & 0.97 & 0.9855 & 0.9510 & 0.9927  \\ 
\cmidrule{2-3}
                                  & \multirow{2}{*}{n=150, T=20}                      & BIC       & 1.45 & 1.00   & 0.23 & 0.9271 & 0.7820 & 0.9620 & 1.50 & 1.00   & 0.08 & 0.8868 & 0.6855 & 0.9400  \\
                                  &                                                   & CH        & 1.57 & 1.00   & 0.31 & 0.8746 & 0.6876 & 0.9178 & 1.72 & 1.00   & 0.24 & 0.6563 & 0.2940 & 0.7131  \\ 
\cmidrule{2-3}
                                  & \multirow{2}{*}{n=150, T=50}                      & BIC       & 2.00 & 2.00   & 1.00 & 0.9923 & 0.9719 & 0.9961 & 2.00 & 2.00   & 1.00 & 0.9855 & 0.9484 & 0.9927  \\
                                  &                                                   & CH        & 2.00 & 2.00   & 1.00 & 0.9922 & 0.9717 & 0.9961 & 1.98 & 2.00   & 0.98 & 0.9852 & 0.9472 & 0.9925  \\ 
\midrule

\multirow{8}{*}{\textbf{Middle} } & \multicolumn{1}{c}{\multirow{2}{*}{n=100, T=20}} & BIC       & 2.00 & 2.00   & 1.00 & 0.9960 & 0.9859 & 0.9980 & 2.00 & 2.00   & 1.00 & 0.9903 & 0.9664 & 0.9951  \\
                                  & \multicolumn{1}{c}{}                             & CH        & 2.00 & 2.00   & 1.00 & 0.9952 & 0.9830 & 0.9976 & 2.00 & 2.00   & 1.00 & 0.9901 & 0.9655 & 0.9950  \\ 
\cmidrule{2-3}
                                  & \multirow{2}{*}{n=100, T=50}                      & BIC       & 2.00 & 2.00   & 1.00 & 0.9998 & 0.9993 & 0.9999 & 2.00 & 2.00   & 1.00 & 0.9996 & 0.9985 & 0.9998  \\
                                  &                                                   & CH        & 2.00 & 2.00   & 1.00 & 0.9998 & 0.9993 & 0.9999 & 2.00 & 2.00   & 1.00 & 0.9996 & 0.9985 & 0.9998  \\ 
\cmidrule{2-3}
                                  & \multirow{2}{*}{n=150, T=20}                      & BIC       & 2.00 & 2.00   & 1.00 & 0.9967 & 0.9870 & 0.9983 & 2.00 & 2.00   & 1.00 & 0.9874 & 0.9535 & 0.9937  \\
                                  &                                                   & CH        & 2.00 & 2.00   & 1.00 & 0.9967 & 0.9870 & 0.9983 & 2.00 & 2.00   & 1.00 & 0.9865 & 0.9503 & 0.9932  \\ 
\cmidrule{2-3}
                                  & \multirow{2}{*}{n=150, T=50}                      & BIC       & 2.00 & 2.00   & 1.00 & 1.0000 & 1.0000 & 1.0000 & 2.00 & 2.00   & 1.00 & 0.9999 & 0.9995 & 0.9999  \\
                                  &                                                   & CH        & 2.00 & 2.00   & 1.00 & 0.9999 & 0.9995 & 0.9999 & 2.00 & 2.00   & 1.00 & 0.9997 & 0.9990 & 0.9999  \\ 
\midrule

\multirow{8}{*}{\textbf{Far} }    & \multicolumn{1}{c}{\multirow{2}{*}{n=100, T=20}} & BIC       & 2.00 & 2.00   & 1.00 & 1.0000 & 1.0000 & 1.0000 & 2.00 & 2.00   & 1.00 & 1.0000 & 1.0000 & 1.0000  \\
                                  & \multicolumn{1}{c}{}                             & CH        & 2.00 & 2.00   & 1.00 & 1.0000 & 1.0000 & 1.0000 & 2.00 & 2.00   & 1.00 & 1.0000 & 1.0000 & 1.0000  \\ 
\cmidrule{2-3}
                                  & \multirow{2}{*}{n=100, T=50}                      & BIC       & 2.00 & 2.00   & 1.00 & 1.0000 & 1.0000 & 1.0000 & 2.00 & 2.00   & 1.00 & 1.0000 & 1.0000 & 1.0000  \\
                                  &                                                   & CH        & 2.00 & 2.00   & 1.00 & 1.0000 & 1.0000 & 1.0000 & 2.00 & 2.00   & 1.00 & 1.0000 & 1.0000 & 1.0000  \\ 
\cmidrule{2-3}
                                  & \multirow{2}{*}{n=150, T=20}                      & BIC       & 2.00 & 2.00   & 1.00 & 1.0000 & 1.0000 & 1.0000 & 2.00 & 2.00   & 1.00 & 1.0000 & 1.0000 & 1.0000  \\
                                  &                                                   & CH        & 2.00 & 2.00   & 1.00 & 1.0000 & 1.0000 & 1.0000 & 2.00 & 2.00   & 1.00 & 1.0000 & 1.0000 & 1.0000  \\ 
\cmidrule{2-3}
                                  & \multirow{2}{*}{n=150, T=50}                      & BIC       & 2.00 & 2.00   & 1.00 & 1.0000 & 1.0000 & 1.0000 & 2.00 & 2.00   & 1.00 & 1.0000 & 1.0000 & 1.0000  \\
                                  &                                                   & CH        & 2.00 & 2.00   & 1.00 & 1.0000 & 1.0000 & 1.0000 & 2.00 & 2.00   & 1.00 & 1.0000 & 1.0000 & 1.0000  \\
\bottomrule
\end{tabular}
}
\label{k=2}
\end{sidewaystable}

\begin{sidewaystable}
\centering
\caption{The mean of square root of the MSE (RMSE) for the estimated functions $\hat{\alpha}_1(t), \hat{\alpha}_2(t)$ under BIC, CH and Oracle methods in Two Subgroups Example.}
\scalebox{0.9}{\begin{tabular}{ccccccccccccc} 
\toprule
                      & \multicolumn{4}{c}{\textbf{Close} }                                                      & \multicolumn{4}{c}{\textbf{Middle} }                                                     & \multicolumn{4}{c}{\textbf{Far} }                                                     \\ 
\cmidrule(lr){2-5}\cmidrule(lr){6-9}\cmidrule(lr){10-13}
                    & \multicolumn{2}{c}{Balanced}               & \multicolumn{2}{c}{Unbalanced}             & \multicolumn{2}{c}{Balanced}               & \multicolumn{2}{c}{Unbalanced}             & \multicolumn{2}{c}{Balanced}               & \multicolumn{2}{c}{Unbalanced}             \\ 
\midrule
\textbf{n=100, T=20}  & $\hat{\alpha}_1(t)$  & $\hat{\alpha}_2(t)$  & $\hat{\alpha}_1(t)$  & $\hat{\alpha}_2(t)$  & $\hat{\alpha}_1(t)$  & $\hat{\alpha}_2(t)$  & $\hat{\alpha}_1(t)$  & $\hat{\alpha}_2(t)$  & $\hat{\alpha}_1(t)$  & $\hat{\alpha}_2(t)$  & $\hat{\alpha}_1(t)$  & $\hat{\alpha}_2(t)$   \\ 
\midrule
Oracle                & 0.0363               & 0.0369               & 0.0385               & 0.0400               & 0.0363               & 0.0369               & 0.0385               & 0.0400               & 0.0363               & 0.0369               & 0.0385               & 0.0400                \\
BIC                   & 0.0512               & 0.0467               & 0.0570               & 0.0461               & 0.0365               & 0.0377               & 0.0394               & 0.0408               & 0.0363               & 0.0369               & 0.0385               & 0.0400                \\
CH                    & 0.0626               & 0.0610               & 0.1075               & 0.1128               & 0.0364               & 0.0375               & 0.0395               & 0.0407               & 0.0363               & 0.0369               & 0.0385               & 0.0400                \\ 
\midrule
\textbf{n=100, T=50}  & $\hat{\alpha}_1(t)$  & $\hat{\alpha}_2(t)$  & $\hat{\alpha}_1(t)$  & $\hat{\alpha}_2(t)$  & $\hat{\alpha}_1(t)$  & $\hat{\alpha}_2(t)$  & $\hat{\alpha}_1(t)$  & $\hat{\alpha}_2(t)$  & $\hat{\alpha}_1(t)$  & $\hat{\alpha}_2(t)$  & $\hat{\alpha}_1(t)$  & $\hat{\alpha}_2(t)$   \\ 
\midrule
Oracle                & 0.0247               & 0.0235               & 0.0261               & 0.0243               & 0.0247               & 0.0235               & 0.0261               & 0.0243               & 0.0247               & 0.0235               & 0.0261               & 0.0243                \\
BIC                   & 0.0253               & 0.0236               & 0.0276               & 0.0252               & 0.0248               & 0.0234               & 0.0262               & 0.0243               & 0.0247               & 0.0235               & 0.0261               & 0.0243                \\
CH                    & 0.0253               & 0.0237               & 0.0276               & 0.0253               & 0.0248               & 0.0234               & 0.0262               & 0.0243               & 0.0247               & 0.0235               & 0.0261               & 0.0243                \\ 
\midrule
\textbf{n=150, T=20}  & $\hat{\alpha}_1(t)$  & $\hat{\alpha}_2(t)$  & $\hat{\alpha}_1(t)$  & $\hat{\alpha}_2(t)$  & $\hat{\alpha}_1(t)$  & $\hat{\alpha}_2(t)$  & $\hat{\alpha}_1(t)$  & $\hat{\alpha}_2(t)$  & $\hat{\alpha}_1(t)$  & $\hat{\alpha}_2(t)$  & $\hat{\alpha}_1(t)$  & $\hat{\alpha}_2(t)$   \\ 
\midrule
Oracle                & 0.0314               & 0.0281               & 0.0337               & 0.0302               & 0.0314               & 0.0281               & 0.0337               & 0.0302               & 0.0314               & 0.0281               & 0.0337               & 0.0302                \\
BIC                   & 0.0387               & 0.0403               & 0.0432               & 0.0435               & 0.0317               & 0.0282               & 0.0344               & 0.0302               & 0.0314               & 0.0281               & 0.0337               & 0.0302                \\
CH                    & 0.0602               & 0.0606               & 0.1762               & 0.1698               & 0.0317               & 0.0281               & 0.0347               & 0.0303               & 0.0314               & 0.0281               & 0.0337               & 0.0302                \\ 
\midrule
\textbf{n=150, T=50}  & $\hat{\alpha}_1(t)$  & $\hat{\alpha}_2(t)$  & $\hat{\alpha}_1(t)$  & $\hat{\alpha}_2(t)$  & $\hat{\alpha}_1(t)$  & $\hat{\alpha}_2(t)$  & $\hat{\alpha}_1(t)$  & $\hat{\alpha}_2(t)$  & $\hat{\alpha}_1(t)$  & $\hat{\alpha}_2(t)$  & $\hat{\alpha}_1(t)$  & $\hat{\alpha}_2(t)$   \\ 
\midrule
Oracle                & 0.0199               & 0.0212               & 0.0214               & 0.0221               & 0.0199               & 0.0212               & 0.0214               & 0.0221               & 0.0199               & 0.0212               & 0.0214               & 0.0221                \\
BIC                   & 0.0204               & 0.0217               & 0.0221               & 0.0228               & 0.0199               & 0.0212               & 0.0214               & 0.0221               & 0.0199               & 0.0212               & 0.0214               & 0.0221                \\
CH                    & 0.0204               & 0.0217               & 0.0220               & 0.0227               & 0.0199               & 0.0213               & 0.0214               & 0.0221               & 0.0199               & 0.0212               & 0.0214               & 0.0221                \\
\bottomrule
\end{tabular}}
\label{k=2rmse}
\end{sidewaystable}

\begin{table}[ht]
\centering
\caption{The sample mean and median of $\hat{K}$, the percentage (per) of $\hat{K}$ equaling to the true number of subgroups, the Rand Index (RI), Normalized mutual information (NMI), and accuracy percentage (\%) equaling the proportion of subjects that are identified correctly under BIC and CH criteria based on 100 realizations with $m_i \sim  \text{Uniform} \left \{5, 6, \dots, 20 \right \}$ in Two Subgroups Example.}
\scalebox{0.9}{\begin{tabular}{ccccccccc}
\toprule
Functions               & setting                 & criterion & mean & median & per  & RI     & NMI    & \%     \\ \midrule
\multirow{4}{*}{\textbf{Middle}} & \multirow{2}{*}{n=100}  & BIC       & 2.11 & 2.00   & 0.91 & 0.9526 & 0.8526 & 0.9756 \\
                        &                         & CH        & 2.07 & 2.00   & 0.94 & 0.9542 & 0.8563 & 0.9765 \\ \cmidrule{2-9} 
                        & \multirow{2}{*}{n=1000} & BIC       & 2.02 & 2.00   & 0.98 & 0.9483 & 0.8293 & 0.9734 \\
                        &                         & CH        & 2.01 & 2.00   & 0.99 & 0.9491 & 0.8318 & 0.9738 \\ \midrule
\multirow{4}{*}{\textbf{Far}}    & \multirow{2}{*}{n=100}  & BIC       & 2.00 & 2.00   & 1.00 & 0.9962 & 0.9865 & 0.9981 \\
                        &                         & CH        & 2.00 & 2.00   & 1.00 & 0.9960 & 0.9857 & 0.9980 \\ \cmidrule{2-9} 
                        & \multirow{2}{*}{n=1000} & BIC       & 2.00 & 2.00   & 1.00 & 0.9963 & 0.9830 & 0.9982 \\
                        &                         & CH        & 2.00 & 2.00   & 1.00 & 0.9962 & 0.9828 & 0.9981 \\ 
\bottomrule
\end{tabular}}
\label{k=2_new}
\end{table}

\begin{table}[ht]
\centering
\caption{The mean of square root of the MSE (RMSE) for the estimated functions $\hat{\alpha}_1(t), \hat{\alpha}_2(t)$ under BIC, CH and Oracle methods with  $m_i \sim  \text{Uniform} \left \{5, 6, \dots, 20 \right \}$ in Two Subgroups Example.}
\scalebox{0.9}{\begin{tabular}{ccccccccc}
\toprule
       &                     &                     & \textbf{Middle}              &                     &                     &                     & \textbf{Far}                 &      
\\ \cmidrule(lr){2-5} \cmidrule(lr){6-9}
       & \multicolumn{2}{c}{n=100}                 & \multicolumn{2}{c}{n=1000}                & \multicolumn{2}{c}{n=100}                 & \multicolumn{2}{c}{n=1000}                \\ \midrule
       & $\hat{\alpha}_1(t)$ & $\hat{\alpha}_2(t)$ & $\hat{\alpha}_1(t)$ & $\hat{\alpha}_2(t)$ & $\hat{\alpha}_1(t)$ & $\hat{\alpha}_2(t)$ & $\hat{\alpha}_1(t)$ & $\hat{\alpha}_2(t)$ \\ \midrule
Oracle & 0.0427              & 0.0399              & 0.0127              & 0.0131              & 0.0427              & 0.0399              & 0.0127              & 0.0131              \\
BIC    & 0.0449              & 0.0434              & 0.0165              & 0.0154              & 0.0424              & 0.0401              & 0.0131              & 0.0131              \\
CH     & 0.0446              & 0.0443              & 0.0163              & 0.0155              & 0.0424              & 0.0401              & 0.0131              & 0.0132              \\ \bottomrule
\end{tabular}
}
\label{k=2rmse_new}
\end{table}

\subsection{Three Subgroups Example}
We simulate data from the heterogeneous model with three subgroups
\begin{equation*}
Y_{ij}=\beta _{i}(t_{ij})+\varepsilon
_{ij},~~ i=1,\dots,n, ~j=1,\dots,m_i, 
\end{equation*}%
where $\beta_i(t)=\alpha_1(t)$ if $i \in \mathcal{G}_1$,  $\beta_i(t)=\alpha_2(t)$ if $i \in \mathcal{G}_2$ and $\beta_i(t)=\alpha_3(t)$ if $i \in \mathcal{G}_3$. We generate data in the same way as that in Two Subgroups Example. The three functions for Close, Middle and Far cases are chosen as:
\begin{align*}
\text{Close}\left\{
\begin{array}{lll}
\alpha_1(t)=-0.6t^2+1.5t,
\\
\alpha_2(t)=-1.3t^2+3.25t+0.2,
\\
\alpha_3(t)=-2.2t^2+5.5t+0.1,
\end{array}\right.
&&
\text{Middle}\left\{
\begin{array}{lll}
\alpha_1(t)=-0.4t^2+t,
\\
\alpha_2(t)=-1.3t^2+3.25t+0.2,
\\
\alpha_3(t)=-2.4t^2+6t+0.1,
\end{array}\right.  
\\
\text{Far}\left\{
\begin{array}{lll}
\alpha_1(t)=-0.3t^2+0.75t,
\\
\alpha_2(t)=-4t^2+10t+0.2,
\\
\alpha_3(t)=-8.5t^2+21.25t+0.3.
\end{array}\right. 
\end{align*}

\begin{figure}[H]
\par
\begin{center}
$
\begin{array}{ccc}
\includegraphics[width=4.3cm]{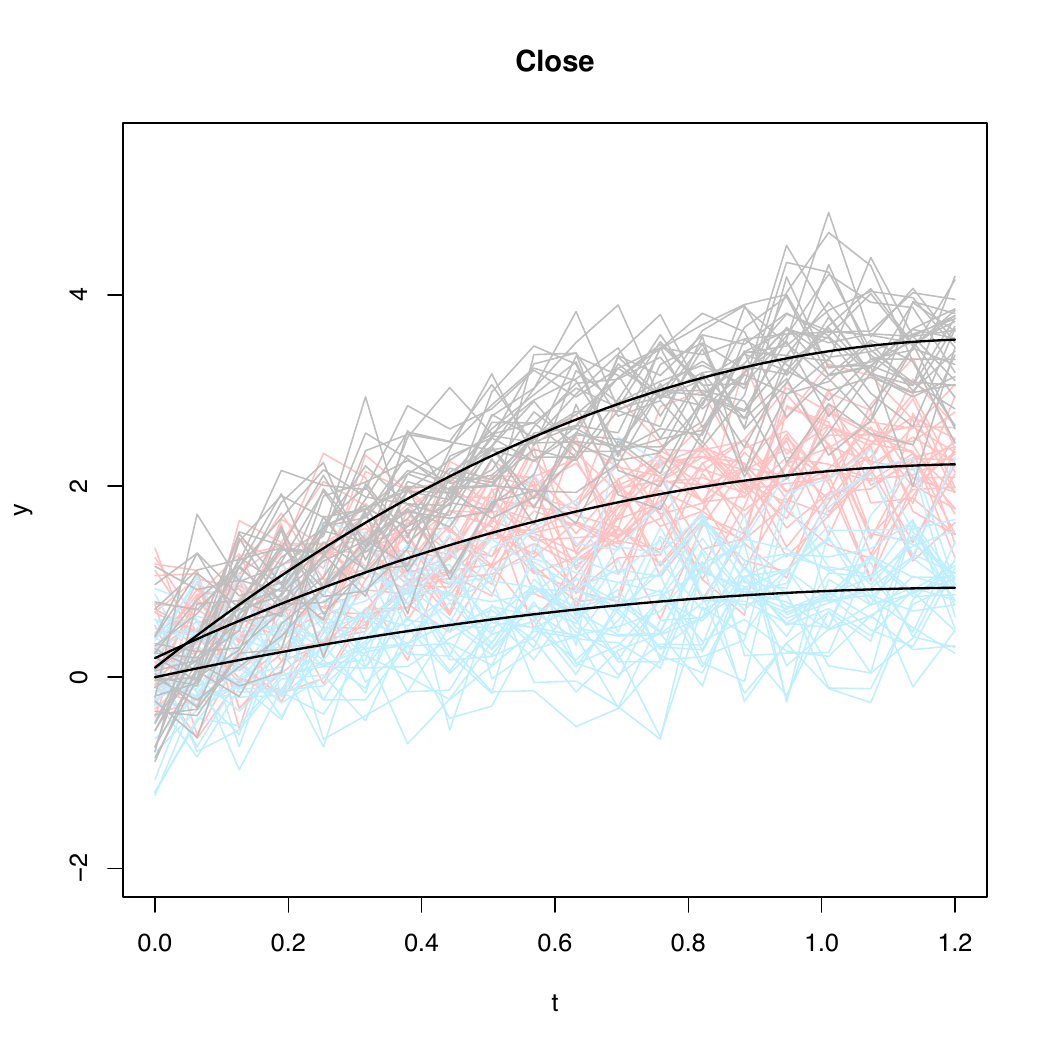}&
\includegraphics[width=4.3cm]{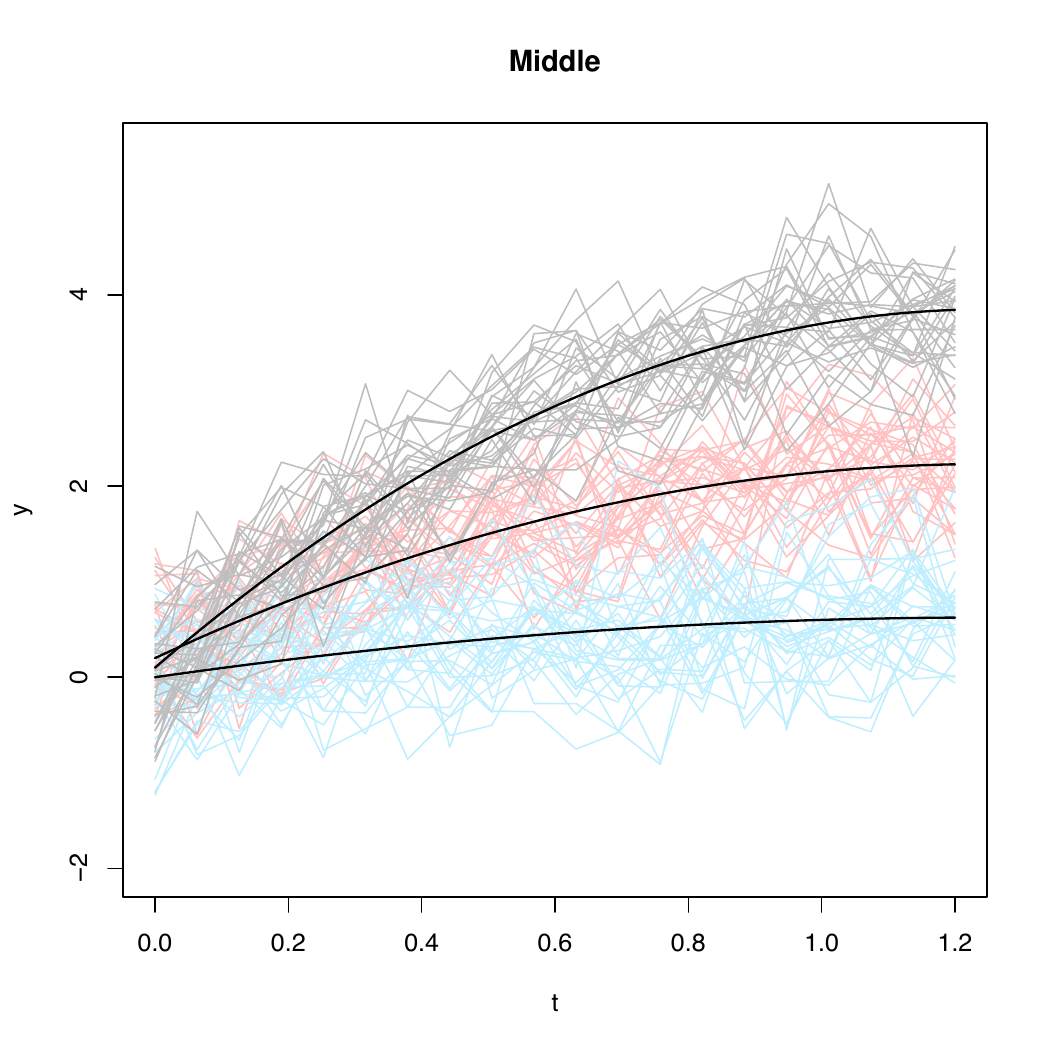}&
\includegraphics[width=4.3cm]{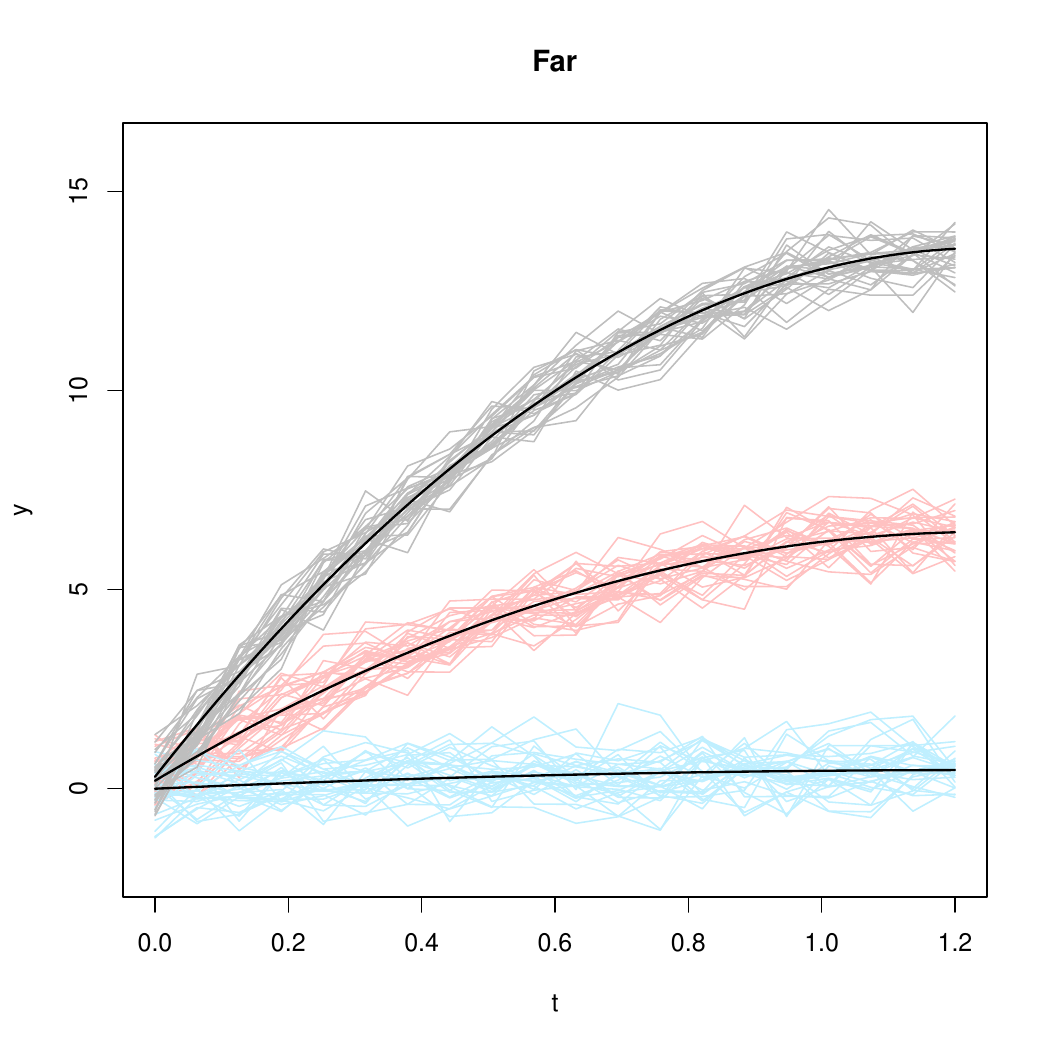}
\end{array}%
$
\end{center}
\caption{The black lines represent the true functions, while the grey, red and blue lines represent the simulated trajectories of the corresponding subgroups under one replication when $n=100, \ T=20$ for balanced data. The distance between the true functions increases from close, to middle, to far.}
\label{k=3trajectory}
\end{figure}

\begin{figure}[tbp]
\centering
\scalebox{0.9}{
\includegraphics[width=4.5cm]{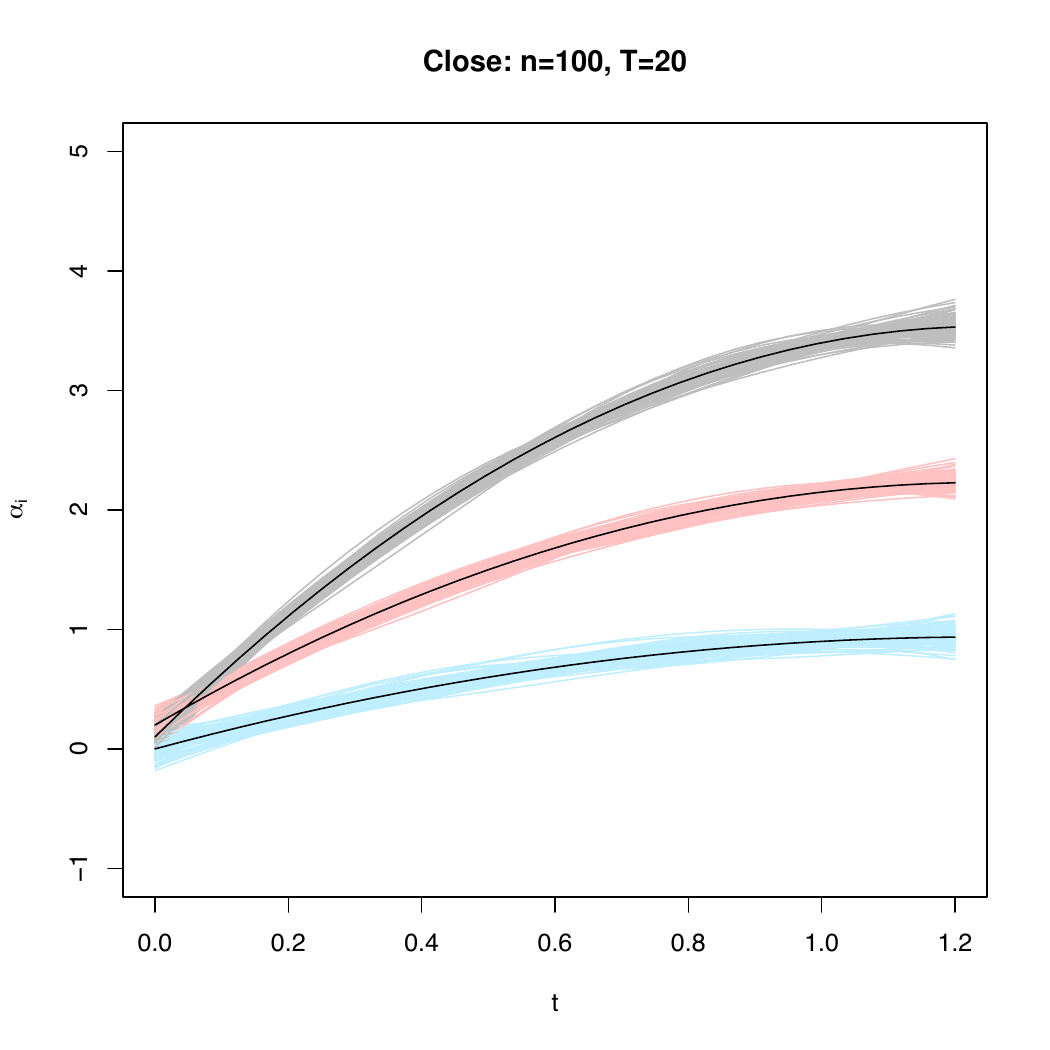}\quad
\includegraphics[width=4.5cm]{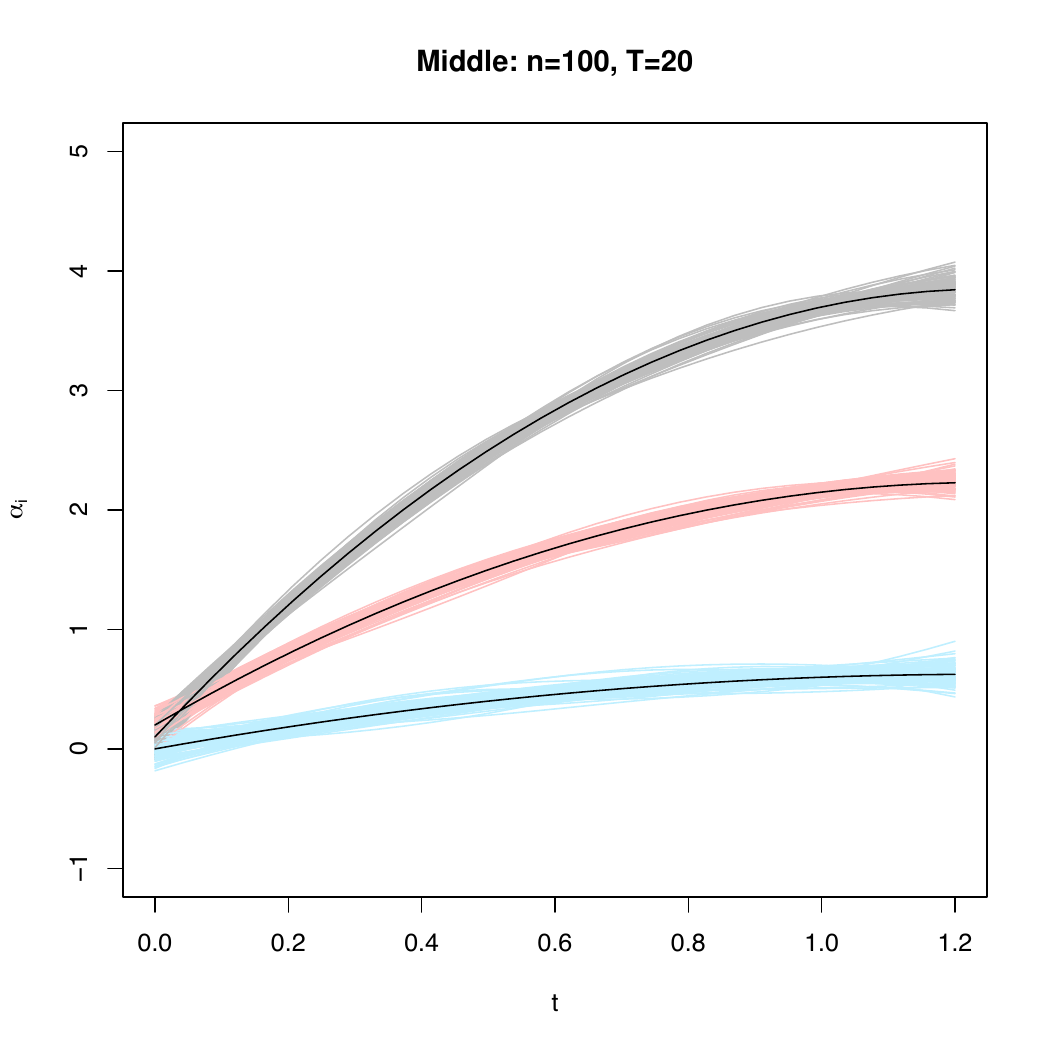}\quad
\includegraphics[width=4.5cm]{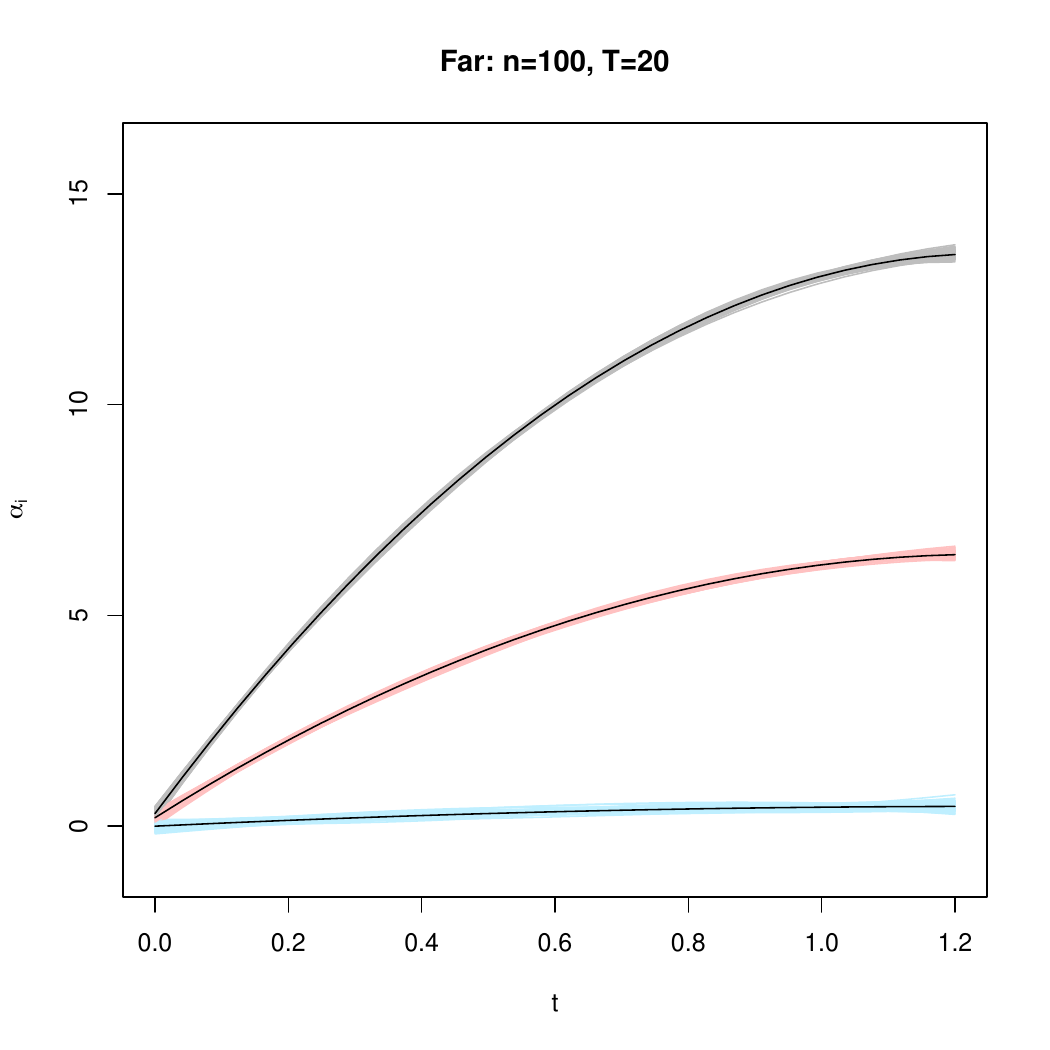}
}

\medskip

\scalebox{0.9}{
\includegraphics[width=4.5cm]{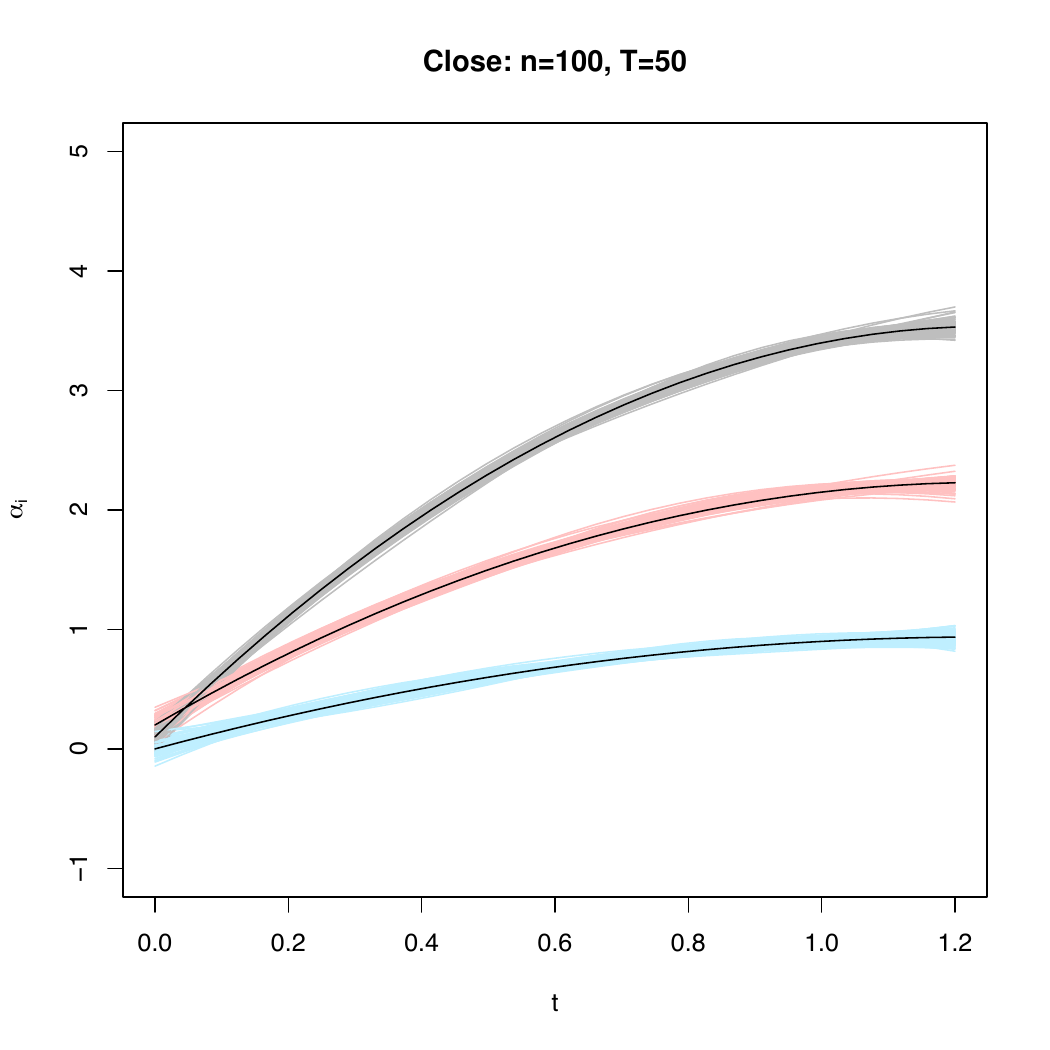}\quad
\includegraphics[width=4.5cm]{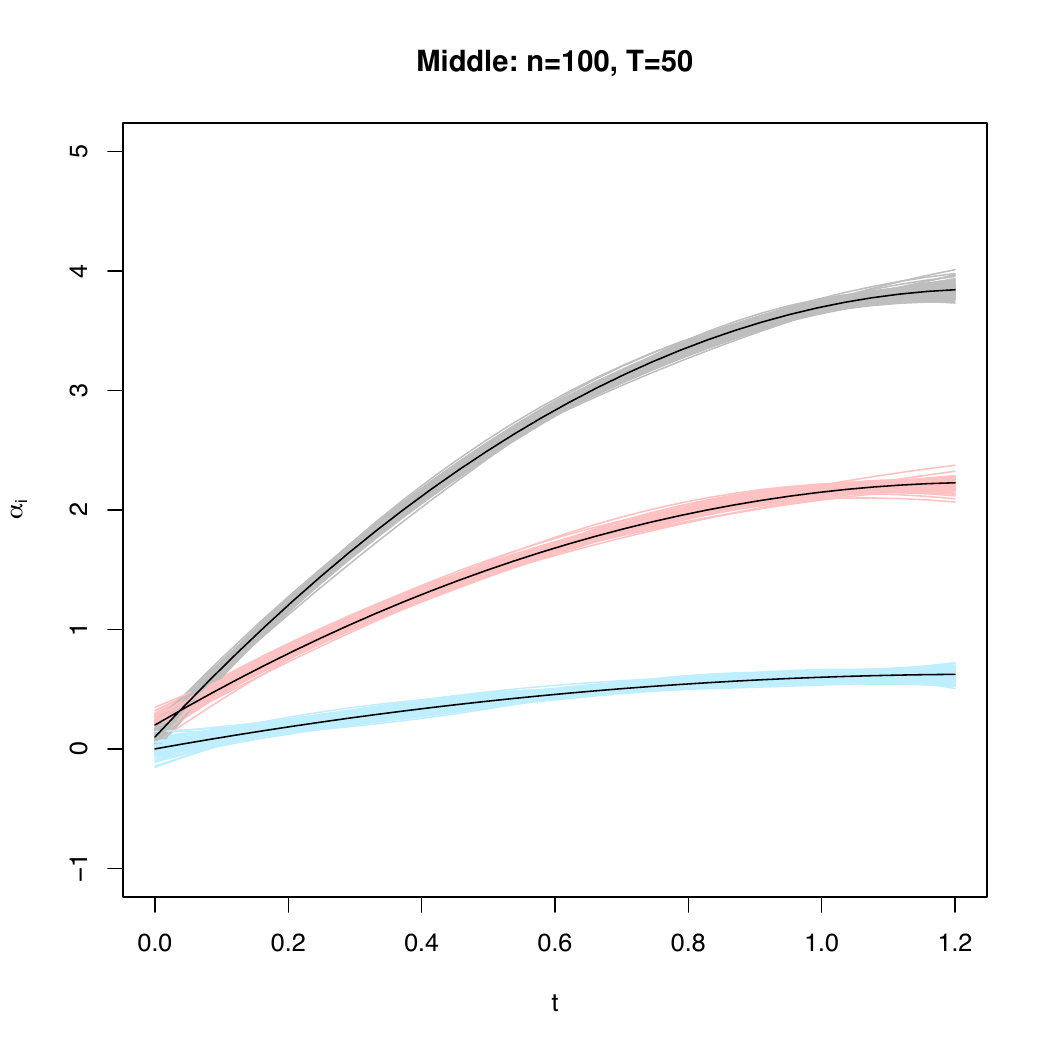}\quad
\includegraphics[width=4.5cm]{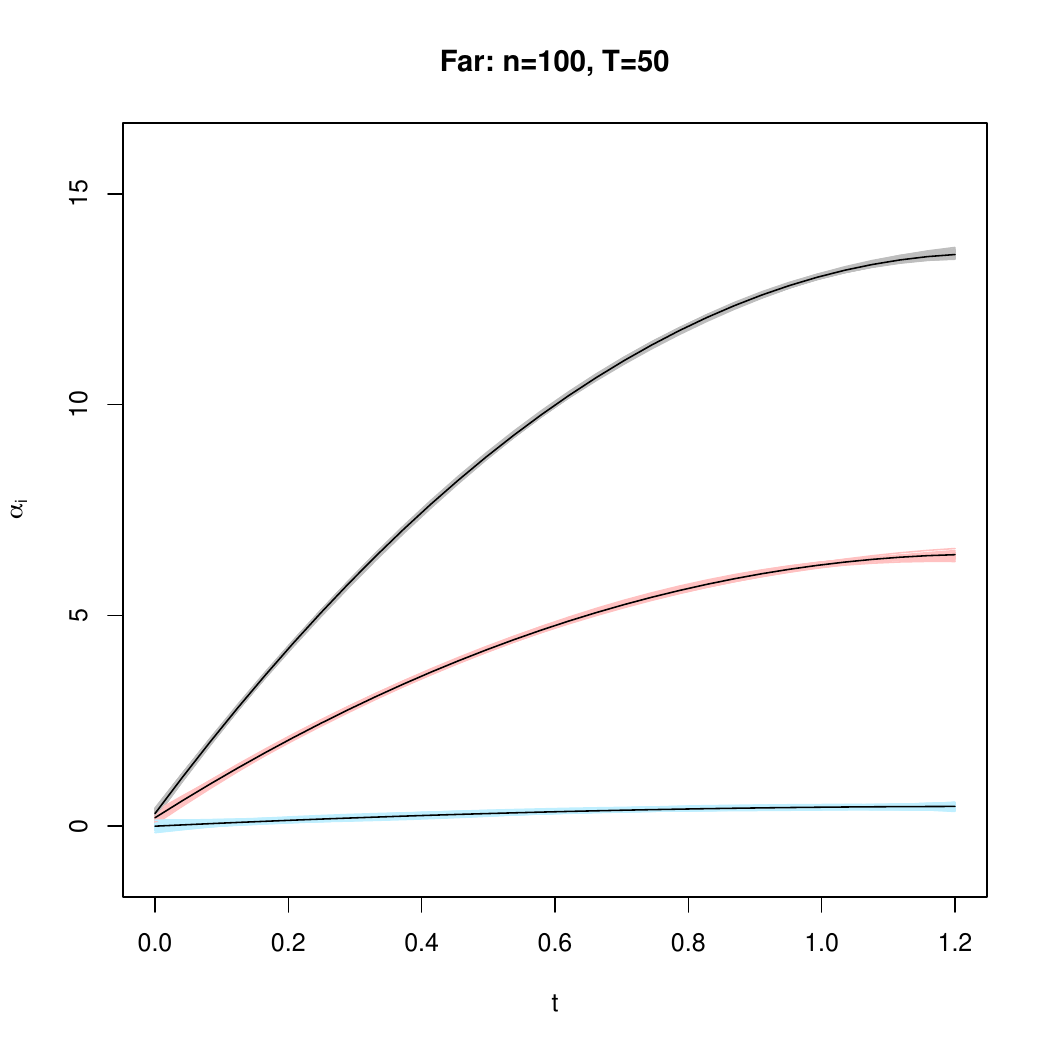}
}

\medskip

\scalebox{0.9}{
\includegraphics[width=4.5cm]{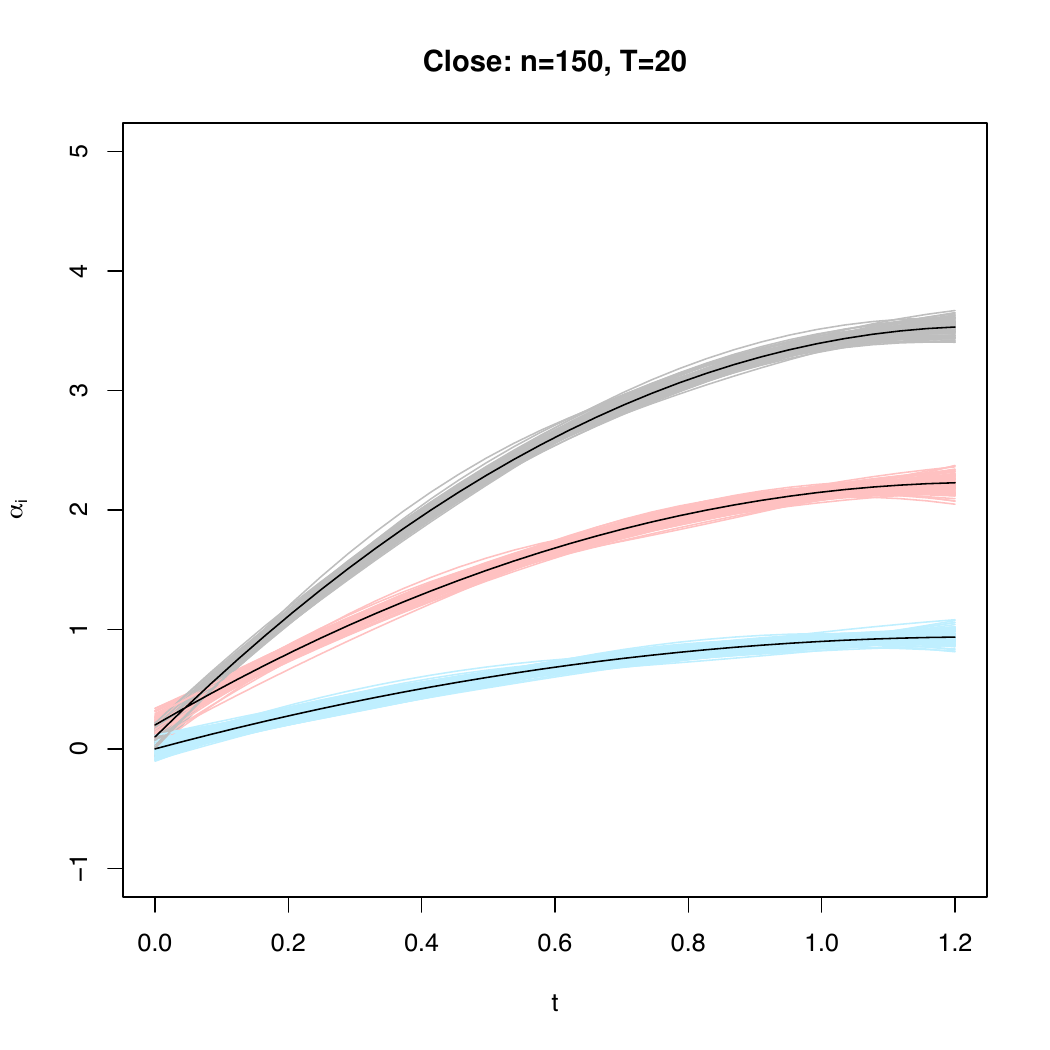}\quad
\includegraphics[width=4.5cm]{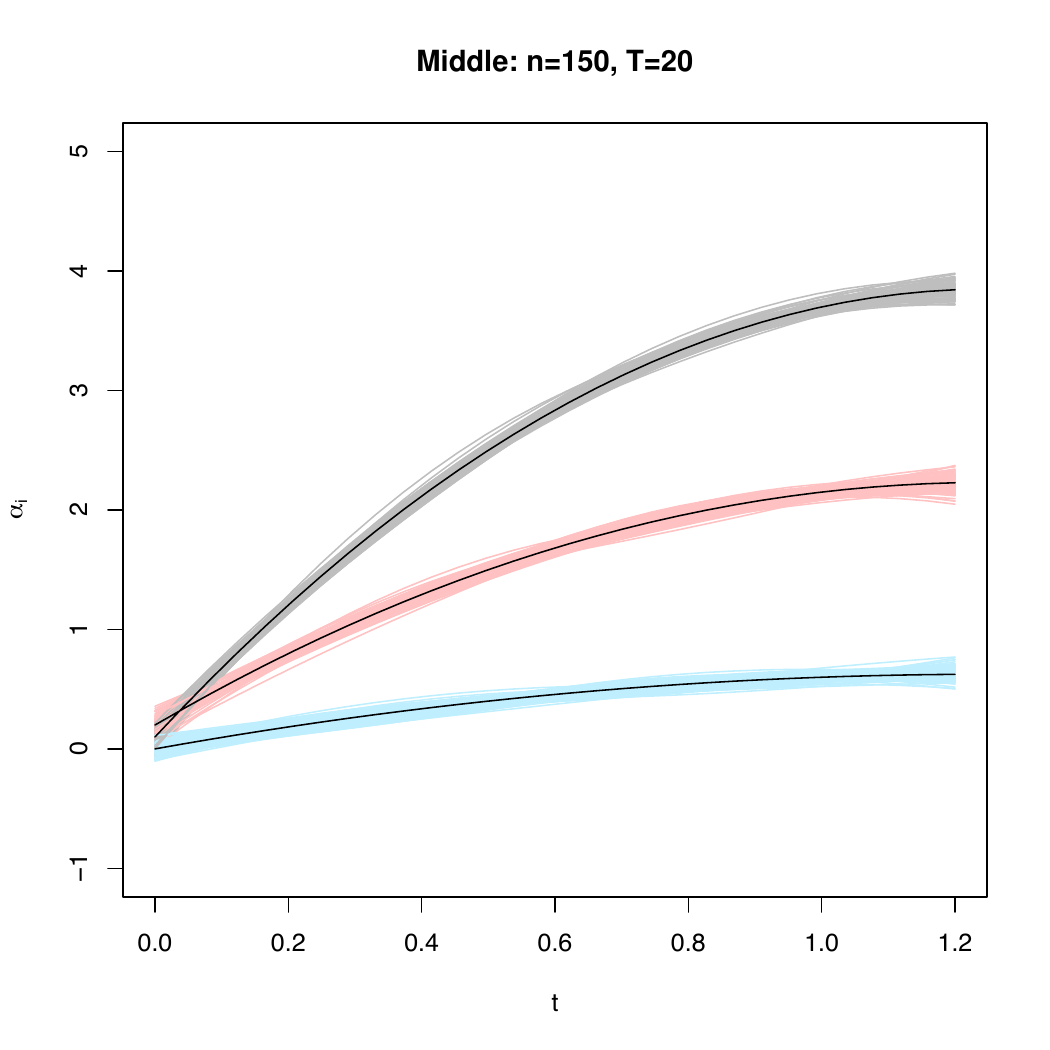}\quad
\includegraphics[width=4.5cm]{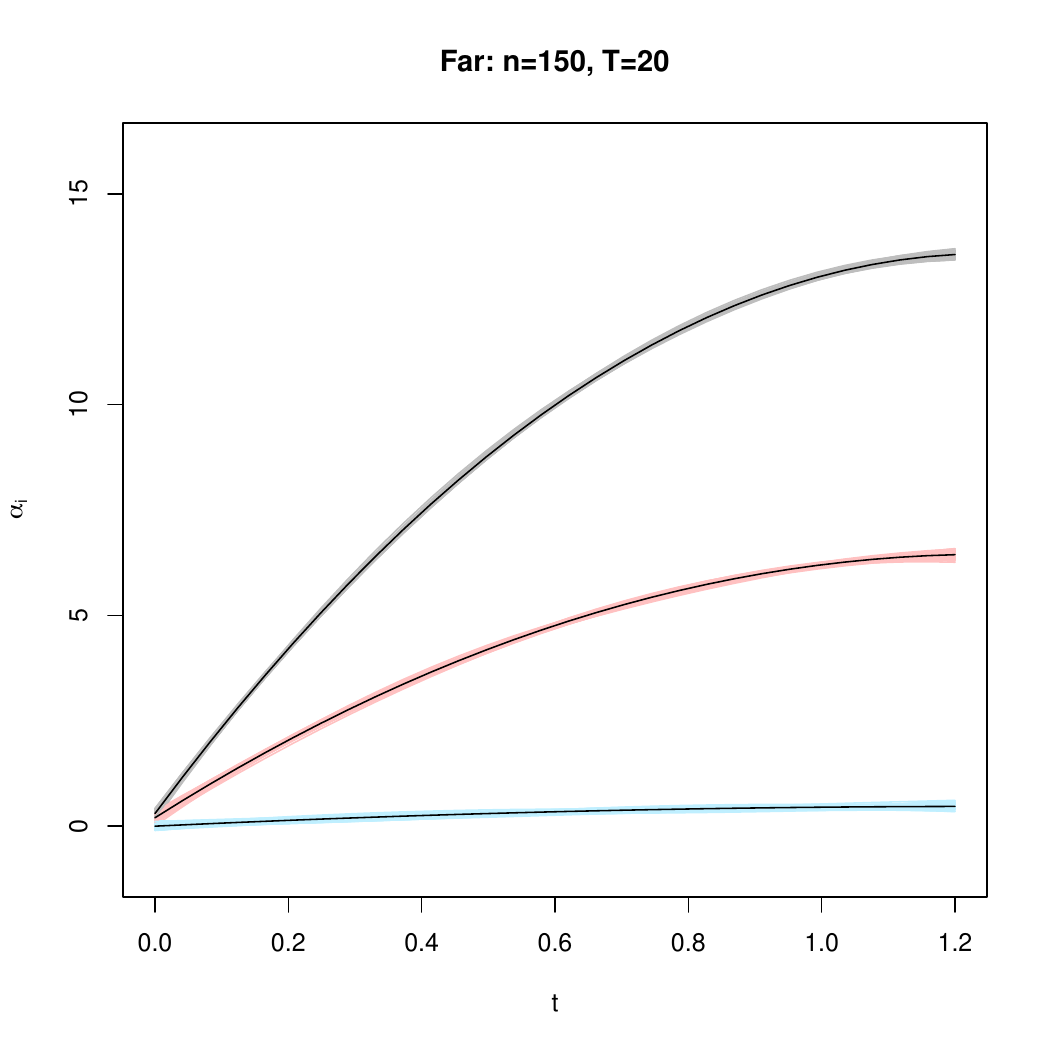}
}

\medskip

\scalebox{0.9}{
\includegraphics[width=4.5cm]{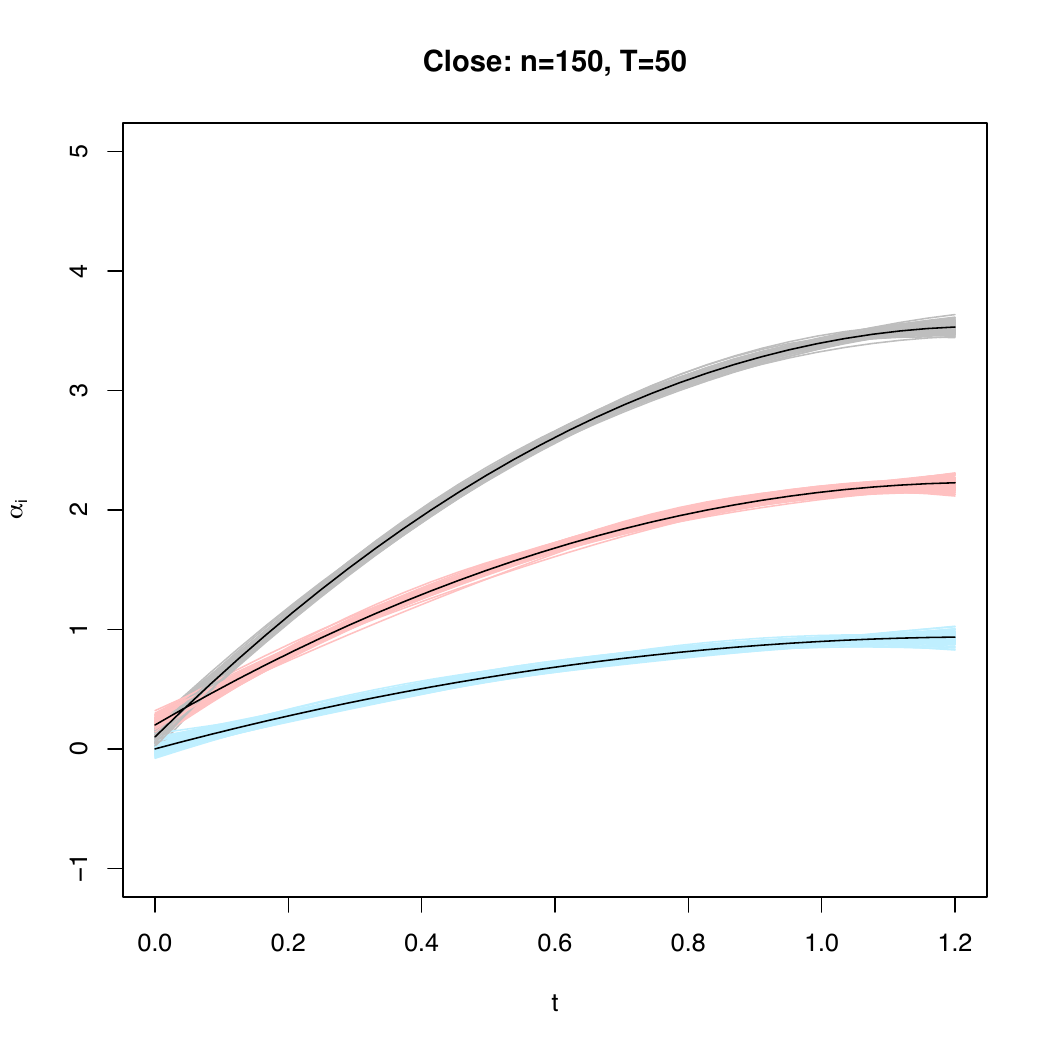}\quad
\includegraphics[width=4.5cm]{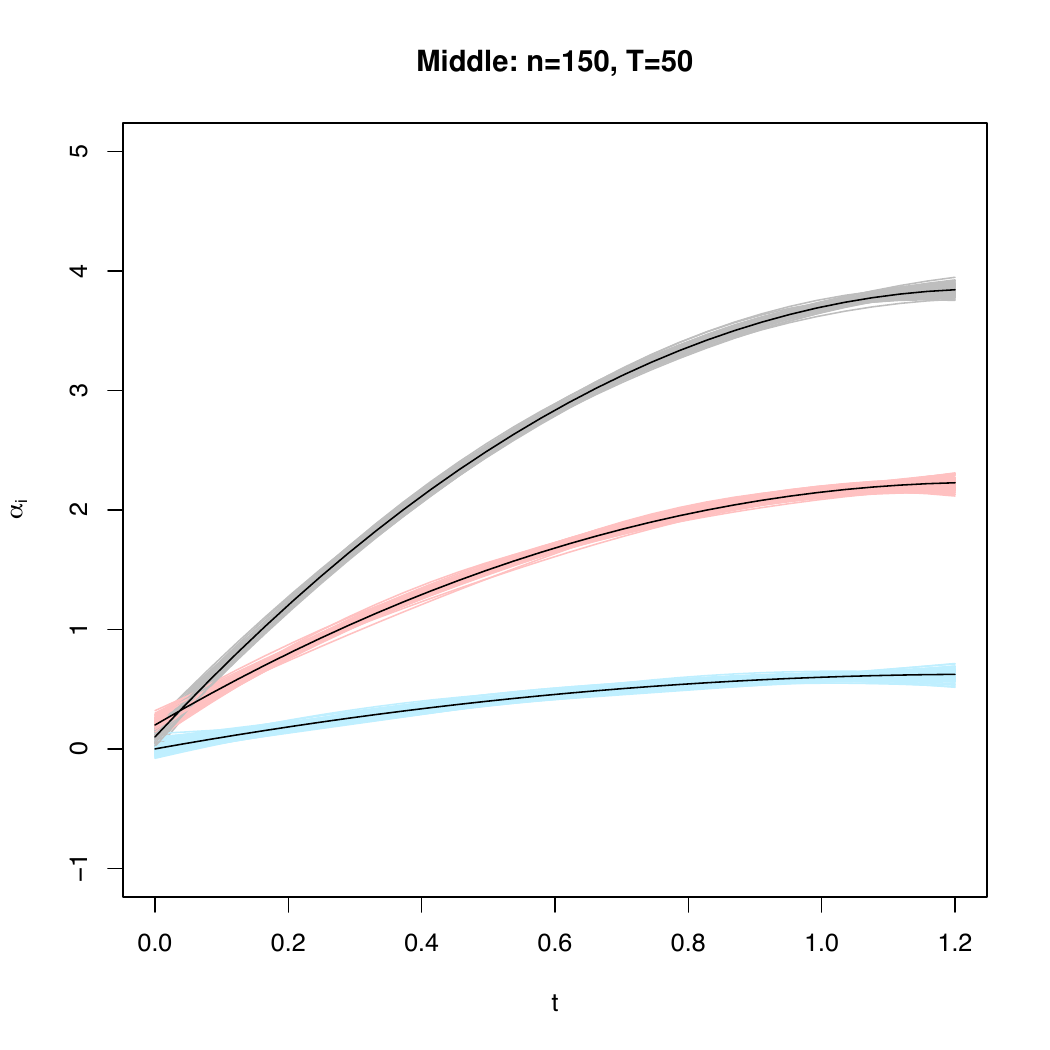}\quad
\includegraphics[width=4.5cm]{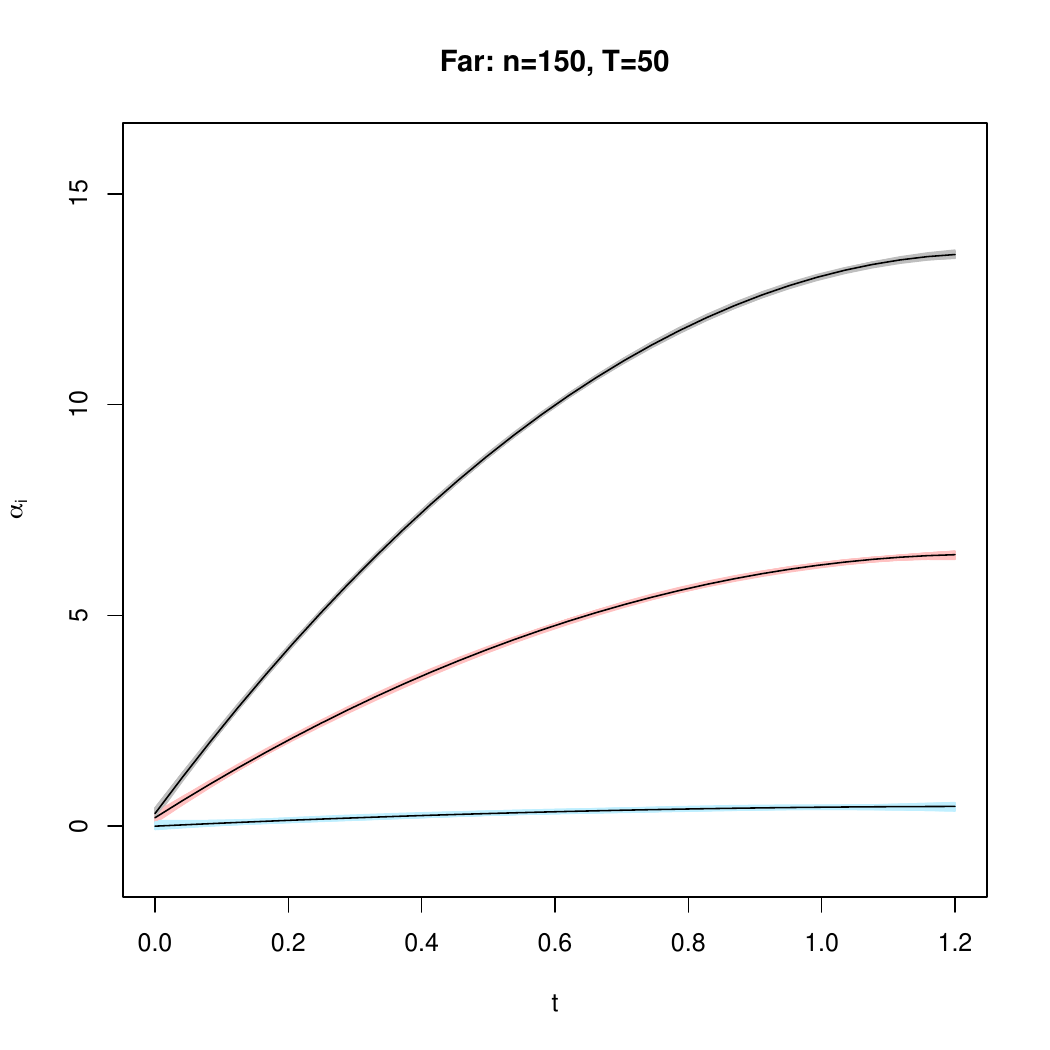}
}

\caption{The black lines represent the true functions, while the grey, red and blue lines are the corresponding fitted curves for the estimated subgroups by using BIC criterion when $\hat{K}=3$ among the 100 replications for balanced data in Three Subgroups Example. On each row, from left to right, it corresponds to close,  middle, and far cases with the same setting of $\left \{ n, \ T \right \}$.}
\label{k=3fitcurve}
\end{figure}

\indent Figure \ref{k=3trajectory} displays the true functions and corresponding trajectories of the three subgroups under one sample with $n=100, \ T=20$ for balanced data. From left to right, the distance between true functions gets larger. We next conduct simulations to do subgroup analysis by using our method. Table \ref{k=3}, based on 100 realizations, presents the mean, median, per of $\hat{K}$ and the average values of RI, NMI, \% for all setups under BIC and CH criteria, respectively. In this table, we observe that the performance for balanced data is better than the corresponding unbalanced data. BIC and CH criteria are consistent due to similar results. When T or the distance between true functions increases, the values of RI, NMI,  and \% become larger. Moreover, to demonstrate the estimation accuracy, Table \ref{k=3rmse} lists the average values of RMSE for the estimated functions $\hat{\alpha}_k(t) \ (k=1,2,3)$ when $\hat{K}$ equals 3, while Figure \ref{k=3fitcurve} shows the estimated nonparametric curves (grey, red, blue lines) and true curves (black lines). From Table \ref{k=3rmse}, it can be seen that the RMSE of $\hat{\alpha}_k(t)$'s are close to those of the oracle estimators. In Figure \ref{k=3fitcurve}, we also observe that the estimated curves are very close to the true curves. And the bands formed by the corresponding estimated curves become narrower as $n$ or $T$ increases.

\begin{sidewaystable}
\centering
\caption{The sample mean and median of $\hat{K}$, the percentage (per) of $\hat{K}$ equaling to the true number of subgroups, the Rand Index (RI), Normalized mutual information (NMI), and accuracy percentage (\%) equaling the proportion of subjects that are identified correctly under BIC and CH criteria based on 100 realizations in Three Subgroups Example. Balanced and unbalanced data are both considered under different $\left \{ n, \  T \right \}$ setups and function distances.} 
\scalebox{0.8}{
\begin{tabular}{clccccccccccccc} 
\toprule
\multicolumn{3}{l}{}                                                                             & \multicolumn{6}{c}{\textbf{Balanced} }         & \multicolumn{6}{c}{\textbf{Unbalanced} }         \\ 
\cmidrule(lr){4-9}\cmidrule(lr){10-15}
\textbf{Functions}                & \multicolumn{1}{c}{setting}                      & criterion & mean & median & per  & RI     & NMI    & \%     & mean & median & per  & RI     & NMI    & \%      \\ 
\midrule
\multirow{8}{*}{\textbf{Close} }  & \multicolumn{1}{c}{\multirow{2}{*}{n=100, T=20}} & BIC       & 3.00 & 3.00   & 1.00 & 0.9962 & 0.9882 & 0.9971 & 3.00 & 3.00   & 1.00 & 0.9870 & 0.9603 & 0.9899  \\
                                  & \multicolumn{1}{c}{}                             & CH        & 2.79 & 3.00   & 0.79 & 0.9965 & 0.9891 & 0.9973 & 2.52 & 3.00   & 0.52 & 0.9887 & 0.9665 & 0.9915  \\ 
\cmidrule{2-3}
                                  & \multirow{2}{*}{n=100, T=50}                      & BIC       & 2.98 & 3.00   & 0.98 & 1.0000 & 1.0000 & 1.0000 & 2.98 & 3.00   & 0.98 & 0.9998 & 0.9993 & 0.9998  \\
                                  &                                                   & CH        & 2.98 & 3.00   & 0.98 & 1.0000 & 1.0000 & 1.0000 & 2.98 & 3.00   & 0.98 & 0.9996 & 0.9988 & 0.9997  \\ 
\cmidrule{2-3}
                                  & \multirow{2}{*}{n=150, T=20}                      & BIC       & 3.00 & 3.00   & 1.00 & 0.9982 & 0.9939 & 0.9987 & 3.00 & 3.00   & 1.00 & 0.9910 & 0.9709 & 0.9931  \\
                                  &                                                   & CH        & 2.91 & 3.00   & 0.91 & 0.9982 & 0.9940 & 0.9987 & 2.69 & 3.00   & 0.69 & 0.9903 & 0.9691 & 0.9928  \\ 
\cmidrule{2-3}
                                  & \multirow{2}{*}{n=150, T=50}                      & BIC       & 3.00 & 3.00   & 1.00 & 1.0000 & 1.0000 & 1.0000 & 3.00 & 3.00   & 1.00 & 0.9999 & 0.9997 & 0.9999  \\
                                  &                                                   & CH        & 3.00 & 3.00   & 1.00 & 1.0000 & 1.0000 & 1.0000 & 3.00 & 3.00   & 1.00 & 0.9999 & 0.9997 & 0.9999  \\ 
\midrule
\multirow{8}{*}{\textbf{Middle} } & \multicolumn{1}{c}{\multirow{2}{*}{n=100, T=20}} & BIC       & 3.00 & 3.00   & 1.00 & 0.9996 & 0.9988 & 0.9997 & 3.00 & 3.00   & 1.00 & 0.9965 & 0.9889 & 0.9973  \\
                                  & \multicolumn{1}{c}{}                             & CH        & 3.00 & 3.00   & 1.00 & 0.9995 & 0.9983 & 0.9996 & 2.98 & 3.00   & 0.98 & 0.9962 & 0.9880 & 0.9970  \\ 
\cmidrule{2-3}
                                  & \multirow{2}{*}{n=100, T=50}                      & BIC       & 3.00 & 3.00   & 1.00 & 1.0000 & 1.0000 & 1.0000 & 3.00 & 3.00   & 1.00 & 1.0000 & 1.0000 & 1.0000  \\
                                  &                                                   & CH        & 3.00 & 3.00   & 1.00 & 1.0000 & 1.0000 & 1.0000 & 3.00 & 3.00   & 1.00 & 1.0000 & 1.0000 & 1.0000  \\ 
\cmidrule{2-3}
                                  & \multirow{2}{*}{n=150, T=20}                      & BIC       & 3.00 & 3.00   & 1.00 & 0.9998 & 0.9994 & 0.9999 & 3.00 & 3.00   & 1.00 & 0.9978 & 0.9925 & 0.9983  \\
                                  &                                                   & CH        & 3.00 & 3.00   & 1.00 & 0.9999 & 0.9997 & 0.9999 & 3.00 & 3.00   & 1.00 & 0.9978 & 0.9929 & 0.9984  \\ 
\cmidrule{2-3}
                                  & \multirow{2}{*}{n=150, T=50}                      & BIC       & 3.00 & 3.00   & 1.00 & 1.0000 & 1.0000 & 1.0000 & 3.00 & 3.00   & 1.00 & 1.0000 & 1.0000 & 1.0000  \\
                                  &                                                   & CH        & 3.00 & 3.00   & 1.00 & 1.0000 & 1.0000 & 1.0000 & 3.00 & 3.00   & 1.00 & 1.0000 & 1.0000 & 1.0000  \\ 
\midrule
\multirow{8}{*}{\textbf{Far} }    & \multicolumn{1}{c}{\multirow{2}{*}{n=100, T=20}} & BIC       & 3.00 & 3.00   & 1.00 & 1.0000 & 1.0000 & 1.0000 & 3.00 & 3.00   & 1.00 & 1.0000 & 1.0000 & 1.0000  \\
                                  & \multicolumn{1}{c}{}                             & CH        & 3.00 & 3.00   & 1.00 & 1.0000 & 1.0000 & 1.0000 & 3.00 & 3.00   & 1.00 & 1.0000 & 1.0000 & 1.0000  \\ 
\cmidrule{2-3}
                                  & \multirow{2}{*}{n=100, T=50}                      & BIC       & 3.00 & 3.00   & 1.00 & 1.0000 & 1.0000 & 1.0000 & 3.00 & 3.00   & 1.00 & 1.0000 & 1.0000 & 1.0000  \\
                                  &                                                   & CH        & 3.00 & 3.00   & 1.00 & 1.0000 & 1.0000 & 1.0000 & 3.00 & 3.00   & 1.00 & 1.0000 & 1.0000 & 1.0000  \\ 
\cmidrule{2-3}
                                  & \multirow{2}{*}{n=150, T=20}                      & BIC       & 3.00 & 3.00   & 1.00 & 1.0000 & 1.0000 & 1.0000 & 3.00 & 3.00   & 1.00 & 1.0000 & 1.0000 & 1.0000  \\
                                  &                                                   & CH        & 3.00 & 3.00   & 1.00 & 1.0000 & 1.0000 & 1.0000 & 3.00 & 3.00   & 1.00 & 1.0000 & 1.0000 & 1.0000  \\ 
\cmidrule{2-3}
                                  & \multirow{2}{*}{n=150, T=50}                      & BIC       & 3.00 & 3.00   & 1.00 & 1.0000 & 1.0000 & 1.0000 & 3.00 & 3.00   & 1.00 & 1.0000 & 1.0000 & 1.0000  \\
                                  &                                                   & CH        & 3.00 & 3.00   & 1.00 & 1.0000 & 1.0000 & 1.0000 & 3.00 & 3.00   & 1.00 & 1.0000 & 1.0000 & 1.0000  \\
\bottomrule
\end{tabular}
}
\label{k=3}
\end{sidewaystable}

\begin{sidewaystable}
\centering
\caption{The mean of square root of the MSE (RMSE) for the estimated functions $\hat{\alpha}_1(t), \hat{\alpha}_2(t), \hat{\alpha}_3(t)$ under BIC, CH and Oracle methods in Three Subgroups Example.}
\scalebox{0.75}{
\begin{tabular}{ccccccccccccccccccc} 
\toprule
                      & \multicolumn{6}{c}{\textbf{Close} }                                                                                                   & \multicolumn{6}{c}{\textbf{Middle} }                                                                                                   & \multicolumn{6}{c}{\textbf{Far} }                                                                                                      \\ 
\cmidrule(lr){2-7}\cmidrule(lr){8-13}\cmidrule(lr){14-19}
                      & \multicolumn{3}{c}{Balanced}                                     & \multicolumn{3}{c}{Unbalanced}                                    & \multicolumn{3}{c}{Balanced}                                      & \multicolumn{3}{c}{Unbalanced}                                    & \multicolumn{3}{c}{Balanced}                                      & \multicolumn{3}{c}{Unbalanced}                                    \\ 
\midrule
\textbf{n=100, T=20}  & $\hat{\alpha}_1(t)$  & $\hat{\alpha}_2(t)$  & $\hat{\alpha}_3(t)$  & $\hat{\alpha}_1(t)$  & $\hat{\alpha}_2(t)$  & $\hat{\alpha}_3(t)$  & $\hat{\alpha}_1(t)$  & $\hat{\alpha}_2(t)$  & $\hat{\alpha}_3(t)$  & $\hat{\alpha}_1(t)$  & $\hat{\alpha}_2(t)$  & $\hat{\alpha}_3(t)$  & $\hat{\alpha}_1(t)$  & $\hat{\alpha}_2(t)$  & $\hat{\alpha}_3(t)$  & $\hat{\alpha}_1(t)$  & $\hat{\alpha}_2(t)$  & $\hat{\alpha}_3(t)$   \\ 
\midrule
Oracle                & 0.0472               & 0.0430              & 0.0465               & 0.0508               & 0.0462               & 0.0498               & 0.0472               & 0.0430               & 0.0465               & 0.0508               & 0.0462               & 0.0498               & 0.0472               & 0.0430               & 0.0465               & 0.0508               & 0.0462             & 0.0498                \\
BIC                   & 0.0466               & 0.0447              & 0.0470               & 0.0513               & 0.0489               & 0.0508               & 0.0469               & 0.0432               & 0.0467               & 0.0505               & 0.0468               & 0.0500               & 0.0472               & 0.0430               & 0.0465               & 0.0508               & 0.0462             & 0.0498                \\
CH                    & 0.0466               & 0.0425              & 0.0474               & 0.0543               & 0.0438               & 0.0506               & 0.0469               & 0.0433               & 0.0467               & 0.0504               & 0.0462               & 0.0501               & 0.0472               & 0.0430               & 0.0465               & 0.0508               & 0.0462             & 0.0498                \\ 
\midrule
\textbf{n=100, T=50}  & $\hat{\alpha}_1(t)$  & $\hat{\alpha}_2(t)$  & $\hat{\alpha}_3(t)$  & $\hat{\alpha}_1(t)$  & $\hat{\alpha}_2(t)$  & $\hat{\alpha}_3(t)$  & $\hat{\alpha}_1(t)$  & $\hat{\alpha}_2(t)$  & $\hat{\alpha}_3(t)$  & $\hat{\alpha}_1(t)$  & $\hat{\alpha}_2(t)$  & $\hat{\alpha}_3(t)$  & $\hat{\alpha}_1(t)$  & $\hat{\alpha}_2(t)$  & $\hat{\alpha}_3(t)$  & $\hat{\alpha}_1(t)$  & $\hat{\alpha}_2(t)$  & $\hat{\alpha}_3(t)$   \\ 
\midrule
Oracle                & 0.0300               & 0.0295              & 0.0316               & 0.0312               & 0.0318               & 0.0332               & 0.0300               & 0.0295               & 0.0316               & 0.0312               & 0.0318               & 0.0332               & 0.0300               & 0.0295               & 0.0316               & 0.0312               & 0.0318             & 0.0332                \\
BIC                   & 0.0299               & 0.0297              & 0.0317               & 0.0310               & 0.0322               & 0.0332               & 0.0300               & 0.0295               & 0.0316               & 0.0312               & 0.0318               & 0.0332               & 0.0300               & 0.0295               & 0.0316               & 0.0312               & 0.0318             & 0.0332                \\
CH                    & 0.0299               & 0.0297              & 0.0317               & 0.0310               & 0.0322               & 0.0332               & 0.0300               & 0.0295               & 0.0316               & 0.0312               & 0.0318               & 0.0332               & 0.0300               & 0.0295               & 0.0316               & 0.0312               & 0.0318             & 0.0332                \\ 
\midrule
\textbf{n=150, T=20}  & $\hat{\alpha}_1(t)$  & $\hat{\alpha}_2(t)$  & $\hat{\alpha}_3(t)$  & $\hat{\alpha}_1(t)$  & $\hat{\alpha}_2(t)$  & $\hat{\alpha}_3(t)$  & $\hat{\alpha}_1(t)$  & $\hat{\alpha}_2(t)$  & $\hat{\alpha}_3(t)$  & $\hat{\alpha}_1(t)$  & $\hat{\alpha}_2(t)$  & $\hat{\alpha}_3(t)$  & $\hat{\alpha}_1(t)$  & $\hat{\alpha}_2(t)$  & $\hat{\alpha}_3(t)$  & $\hat{\alpha}_1(t)$  & $\hat{\alpha}_2(t)$  & $\hat{\alpha}_3(t)$   \\ 
\midrule
Oracle                & 0.0353               & 0.0373              & 0.0355               & 0.0382               & 0.0394               & 0.0377               & 0.0353               & 0.0373               & 0.0355               & 0.0382               & 0.0394               & 0.0377               & 0.0353               & 0.0373               & 0.0355               & 0.0382               & 0.0394             & 0.0377                \\
BIC                   & 0.0353               & 0.0376              & 0.0357               & 0.0393               & 0.0414              & 0.0384               & 0.0353               & 0.0374               & 0.0356               & 0.0383               & 0.0401               & 0.0379               & 0.0353               & 0.0373               & 0.0355               & 0.0382               & 0.0394             & 0.0377                \\
CH                    & 0.0345               & 0.0360              & 0.0356               & 0.0381               & 0.0405               & 0.0389               & 0.0353               & 0.0374               & 0.0355               & 0.0383               & 0.0401               & 0.0379               & 0.0353               & 0.0373               & 0.0355               & 0.0382               & 0.0394             & 0.0377                \\ 
\midrule
\textbf{n=150, T=50}  & $\hat{\alpha}_1(t)$  & $\hat{\alpha}_2(t)$  & $\hat{\alpha}_3(t)$  & $\hat{\alpha}_1(t)$  & $\hat{\alpha}_2(t)$  & $\hat{\alpha}_3(t)$  & $\hat{\alpha}_1(t)$  & $\hat{\alpha}_2(t)$  & $\hat{\alpha}_3(t)$  & $\hat{\alpha}_1(t)$  & $\hat{\alpha}_2(t)$  & $\hat{\alpha}_3(t)$  & $\hat{\alpha}_1(t)$  & $\hat{\alpha}_2(t)$  & $\hat{\alpha}_3(t)$  & $\hat{\alpha}_1(t)$  & $\hat{\alpha}_2(t)$  & $\hat{\alpha}_3(t)$   \\ 
\midrule
Oracle                & 0.0226               & 0.0255              & 0.0250               & 0.0251               & 0.0273               & 0.0266               & 0.0226               & 0.0255               & 0.0250               & 0.0251               & 0.0273               & 0.0266               & 0.0226               & 0.0255               & 0.0250               & 0.0251               & 0.0273             & 0.0266                \\
BIC                   & 0.0226               & 0.0255              & 0.0250               & 0.0251               & 0.0272               & 0.0266               & 0.0226               & 0.0255               & 0.0250               & 0.0251               & 0.0273               & 0.0266               & 0.0226               & 0.0255               & 0.0250               & 0.0251               & 0.0273             & 0.0266                \\
CH                    & 0.0226               & 0.0255              & 0.0250               & 0.0251               & 0.0273               & 0.0266               & 0.0226               & 0.0255               & 0.0250               & 0.0251               & 0.0273               & 0.0266               & 0.0226               & 0.0255               & 0.0250               & 0.0251               & 0.0273             & 0.0266                \\
\bottomrule
\end{tabular}
}

\label{k=3rmse}
\end{sidewaystable}


\section{Real data application}
\label{SEC:real}

 In this section, we apply our method to Alzheimer's disease (AD) data, which can be obtained from the Alzheimer's Disease Neuroimaging Initiative (ADNI) database (adni.loni.usc.edu). The ADNI was launched in 2003 as a public-private partnership, led by Principal Investigator Michael W. Weiner, MD. The primary goal of ADNI is to test whether serial magnetic resonance imaging (MRI), positron emission tomography (PET), other biological markers, and clinical and neuropsychological assessment can be combined to measure the progression of mild cognitive impairment (MCI) and early Alzheimer's disease (AD). For up-to-date information, see \url{www.adni-info.org}.

\indent {We consider two steps in our analytic procedure. The first step is to use the proposed method to identify the latent subgroups and recover the memberships in each subgroup using the observed data. The second step is to use the information from the identified subgroups and the baseline covariates to classify  future patients into the identified subgroups.}

\indent {
In the first step, to conduct latent subgroup analysis, we use the longitudinal data of ADASCOG13 (Alzheimer's Disease Assessment Scale-Cognitive Subscale) for each patient from ADNI1, ADNIGO and ADNI2 at different time points (0, 6, 12, 18, 24, 36, 48, 60, 72, 84, 96, 108, 120 months). The data are unbalanced due to the fact that patients may have missing measurements at some time points. Thus, the number of observed measurements for all patients ranges from 1 to 13.  ADASCOG13 is widely used as a test of cognitive functions, consisting of thirteen tests, with the values ranging from 0 to 85 to assess the severity of the dementia. Higher values indicate more severe of the dementia due to more cognitive errors. To apply our subgroup analysis method, we delete patients with less than 4 measurements. As a result, there are 1253 patients used in our analysis.}

\begin{figure}
\centering
\includegraphics[scale=0.45]{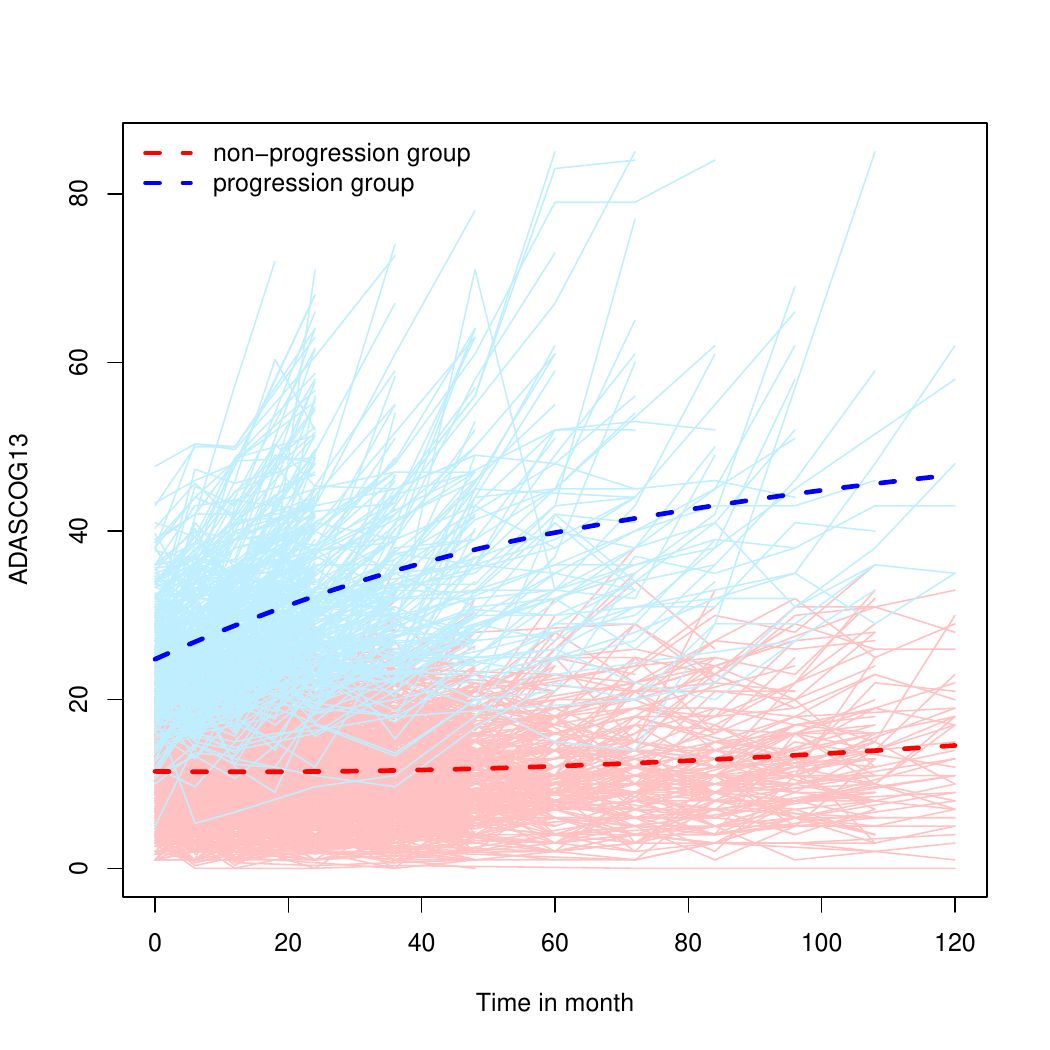}
\caption{The trajectories of individual patients within each identified subgroup (blue, red solid lines) and the estimated mean curve (dashed lines) for each subgroup based on ADASCOG13. The blue group is the progression group, with higher values of ADASCOG13, indicating faster cognition decline.}
\label{realdata}
\end{figure}

{\indent We take ADASCOG13 as the response to fit the heterogeneous model (\ref{model2}). The values of ADASCOG13 are standardized to apply the fusion penalized method. Following the guidance from our simulation studies, we use  quadratic splines with one interior knot to approximate the nonparametric functions. {As a result, we identify two subgroups, one subgroup with 892 patients and the other one with 361 patients. Figure \ref{realdata} displays the trajectories of individual patients within each subgroup and the estimated mean curve for each subgroup. Clearly, the subgroup depicted in red can be viewed as a non-progression group as the values of the estimated mean curve for this subgroup remain constant over time. In contrast, the subgroup shown in blue can be viewed as a progression group, as we can observe a clear increasing trend of the estimated mean curve for this subgroup over time. Note that the increasing value of ADASCOG13 indicates cognitive decline.} Therefore, the progression group is potentially of interest to be recruited in clinical trials when testing whether a drug can slow down the cognitive decline. By our proposed fusion learning method, we can successfully identify two subgroups with their memberships recovered. }

{\indent In the second step, we are interested in classifying future patients into the two identified subgroups using information from baseline covariates. We collect information of several baseline covariates, including ADASCOG13, mmseTOT (Mini-Mental State Examination total score), FAQTOTAL (functional activities questionnaires total score), cdrSB (clinical dementia rating sum of boxes), ApoE4 (Apolipoprotein E4) status and Education. Among them, ADASCOG13, mmseTOT, FAQTOTAL and cdrSB are the baseline measurements of cognition or functional activities. We exclude the 8 patients whose covariates are not observed at the baseline in the classification step. Thus, there are 889 patients in the non-progression group and 356 patients in the progression group. To understand which covariates that contribute to the group difference, we conduct a two-sample test to compare the means between the two subgroups for each covariate. The P-values are reported in Table \ref{difference}, and they are very small for all covariates. Compared with the non-progression group, patients in the progression group clearly have more severe dementia symptoms given that they have higher ADASCOG13, FAQTOTAL and cdrSB and lower mmseTOT at baseline, as well as more AopE4 carriers. Moreover, they also have less education. These findings corroborate the results given in the literature. In general, the cognition tends to decline more quickly if the disease of a patient is more severe at baseline. ApoE4 is known as one important risk factor for AD onset, and ApoE4 carriers tend to show earlier cognitive decline onset than the non-carriers \citep{safieh2019apoe4}. Additionally, some studies have shown that patients with lower education are more likely to develop AD \citep{katzman1993education}. Based on the results in Table \ref{difference}, we include all baseline covariates in the classification step.}

\begin{table}[ht]
\centering
\caption{Mean and standard deviation (SD) for each baseline covariate; P-value shows the significant difference existing in the two subgroups. ApoE4 is tested by two proportion z-test, while other covariates are tested by two sample t-test.} 
\scalebox{0.9}{\begin{tabular}{cccc} \toprule
                             & \textbf{Non-progression group} & \textbf{Progression group} &                  \\ \cmidrule{2-3}
\textbf{Baseline Covariates} & \textbf{Mean (SD)}                  & \textbf{Mean (SD)}              & \textbf{P-value} \\ \midrule
ADASCOG13                    & 11.61 (5.15)                        & 24.91 (6.42)                    & $<0.001$           \\
mmseTOT                      & 28.46 (1.62)                        & 25.31 (2.41)                    & $<0.001$           \\
FAQTOTAL                     & 1.46 (3.05)                         & 7.67 (6.71)                     & $<0.001$           \\
cdrSB                        & 0.81 (0.96)                         & 2.70 (1.66)                     & $<0.001$           \\
ApoE4 carrier (\%)            & 35\% (0.02)                          & 69\% (0.02)                      & $<0.001$           \\
Education                    & 16.21 (2.71)                        & 15.36 (3.05)                    & $<0.001$           \\ \bottomrule
\end{tabular}}
\label{difference}
\end{table}

{\indent Next, we use the two identified subgroups obtained from our fusion learning method, and the six baseline covariates given in Table \ref{difference} to perform classification. Binary variables created from the memberships of the progression group and the  non-progression group  are used as the responses, and the six baseline covariates are used as the predictors in the classification task. We randomly split the dataset into 80\% training data and 20\% test data. The training data is used to fit a predictive model, while the test data is used to examine the prediction performance. We apply four popular supervised methods (predictive models) for classification, including the logistic regression, random forest, boosting (gradient boosting machines) and support vector machine (SVM) with linear kernel. The four methods are implemented using the R packages ``stats", ``randomForest",  ``gbm" and ``e1071", respectively. For the methods involving hyper parameters, we apply 5-fold cross validation (CV) based on a grid search to select the optimal hyper parameters that maximize the accuracy. Table \ref{prediction} reports the accuracy, specificity, precision, recall, F1 score and AUC (area under the ROC curve) obtained from the test data for the four predictive models. They are commonly used metrics to evaluate the classification performance. Accuracy is the percentage of correct predictions. Specificity is the  proportion of true negatives out of the total actual negatives, and it measures how well a method can identify the true negatives. Precision is the ratio of true positives to all positives, while recall refers to the ratio of true positives to the size of the actual positive class. Precision measures the ability of a classification (predictive) model to identify the true positives, and recall assesses its ability to find all the positive cases. F1 score is the weighted average between precision and recall. AUC measures the ability of a classifier to distinguish between classes. From Table \ref{prediction}, we observe that the values of accuracy, specificity and AUC are all above 0.9 for the four classification methods (predictive models). The values of recall and F1 score for the boosting method also exceed 0.9. In general, boosting outperforms the other three methods based on all metrics, and therefore it is recommended for the classification task. In conclusion, our two-step procedure is useful for identifying latent subgroups and then further classifying future patients into the identified subgroups based on their baseline characteristics.}


\begin{table}[tbp]
\centering
\caption{Accuracy, specificity, precision, recall, F1 score and AUC obtained from the test data. The progression group is defined as the positive class.} 
\scalebox{0.9}{\begin{tabular}{ccccccc} 
\toprule
\multirow{2}{*}{\textbf{Predictive model}} & \multicolumn{6}{c}{\textbf{Performance}}    \\          
\cmidrule{2-7} 
                                  & \textbf{accuracy} & \textbf{specificity} & \textbf{precision} & \textbf{recall} & \textbf{F1 score}    & \textbf{AUC} \\ 
\midrule
\textbf{logistic}                          & 0.924    & 0.944       & 0.859     & 0.871  & 0.865 & 0.908 \\
\textbf{random forest}                    & 0.920    & 0.939       & 0.847     & 0.871  & 0.859 & 0.905 \\
\textbf{boosting}                          & 0.944    & 0.955       & 0.889     & 0.914  & 0.901 & 0.935 \\
\textbf{SVM}                               & 0.932    & 0.950       & 0.873     & 0.886  & 0.879 & 0.918  \\
\bottomrule
\end{tabular}
}
\label{prediction}
\end{table}


\section{Conclusions and discussions}
\label{discussion}

In this paper, we consider the subgroup analysis for longitudinal trajectories of the AD data based on a heterogeneous nonparametric regression model. We use B-splines to approximate the nonparametric functional curves, and cluster the subjects into subgroups by applying concave pairwise fusion penalties on the spline coefficients. Our method can automatically identify the latent memberships, and recover the disease trajectory curves of subgroups simultaneously without a prior knowledge of the number of the subgroups. Different from the GMM method that requires to specify an underlying distribution of the data, our method only needs a working correlation matrix of the repeated measures within each subject. Moreover, the resulting estimators of the functional curves are robust to the specification of the working correlation matrix.  Simulation studies indicate promising performance of our proposed method. It has been demonstrated as an effective tool for subgroup analysis of the AD data considered in this paper. As a future work, we plan to extend the proposed method to the joint modeling of survival and longitudinal data, which commonly occur in clinical studies. However, further investigations are needed to develop the computational algorithm and theoretical properties.

\section*{Acknowledgement}

\noindent 
Data used in preparation of this article were obtained from the Alzheimer's Disease Neuroimaging Initiative (ADNI) database (adni.loni.usc.edu). As such, the investigators within the ADNI contributed to the design and implementation of ADNI and/or provided data but did not participate in analysis or writing of this report. A complete listing of ADNI investigators can be found at: http://adni.loni.usc.edu/wp-content/uploads/how\_to\_apply/ADNI\_Acknowledgement\_List.pdf.

\noindent The authors thank Xiang Zhang and Peter F. Castelluccio, former Eli Lilly and Company employees, for their help in data organization. Yushi Liu, Bochao Jia and Luna Sun are stockholders and employees of Eli Lilly and Company.

\noindent The research of Mingming Liu and Shujie Ma is supported in part by the U.S. NSF grants DMS-17-12558 and DMS-20-14221 and the UCR Academic Senate CoR Grant. Jing Yang's research is supported by the National Natural Science Foundation of China (Grant 11801168), the Natural Science Foundation of Hunan Province (Grant 2018JJ3322), the Scientic Research Fund of Hunan Provincial Education Department (Grant 18B024), and the project of China Scholarship Council for his visiting to Professor Shujie Ma at University of California, Riverside.

\noindent Data collection and sharing for this project was funded by the Alzheimer's Disease Neuroimaging Initiative (ADNI) (National Institutes of Health Grant U01 AG024904) and DOD ADNI (Department of Defense award number W81XWH-12-2-0012). ADNI is funded by the National Institute on Aging, the National Institute of Biomedical Imaging and Bioengineering, and through generous contributions from the following: AbbVie, Alzheimer?s Association; Alzheimer?s Drug Discovery Foundation; Araclon Biotech; BioClinica, Inc.; Biogen; Bristol-Myers Squibb Company; CereSpir, Inc.; Cogstate; Eisai Inc.; Elan Pharmaceuticals, Inc.; Eli Lilly and Company; EuroImmun; F. Hoffmann-La Roche Ltd and its affiliated company Genentech, Inc.; Fujirebio; GE Healthcare; IXICO Ltd.; Janssen Alzheimer Immunotherapy Research \& Development, LLC.; Johnson \& Johnson Pharmaceutical Research \& Development LLC.; Lumosity; Lundbeck; Merck \& Co., Inc.; Meso Scale Diagnostics, LLC.; NeuroRx Research; Neurotrack Technologies; Novartis Pharmaceuticals Corporation; Pfizer Inc.; Piramal Imaging; Servier; Takeda Pharmaceutical Company; and Transition Therapeutics. The Canadian Institutes of Health Research is providing funds to support ADNI clinical sites in Canada. Private sector contributions are facilitated by the Foundation for the National Institutes of Health ( http://www.fnih.org). The grantee organization is the Northern California Institute for Research and Education, and the study is coordinated by the Alzheimer?s Therapeutic Research Institute at the University of Southern California. ADNI data are disseminated by the Laboratory for Neuro Imaging at the University of Southern California.

\clearpage

\section*{{\protect\normalsize Supplementary Material\label{SEC:appendix}}}
\label{appendix}

\renewcommand{\thetheorem}{{\sc A.\arabic{theorem}}} \renewcommand{%
\theremark}{{\sc A.\arabic{remark}}}
\renewcommand{\theproposition}{{\sc
A.\arabic{proposition}}} \renewcommand{\thelemma}{{\sc
A.\arabic{lemma}}}  \renewcommand{\theequation}{A.\arabic{equation}} %
\renewcommand{\thesubsection}{{\it A.\arabic{subsection}}} %
\setcounter{equation}{0} \setcounter{lemma}{0} \setcounter{proposition}{0} %
\setcounter{theorem}{0} \setcounter{subsection}{0}%
\setcounter{remark}{0}

\indent This supplement contains the computation procedure using ADMM algorithm, consistency of initial estimator in ADMM, convergence of ADMM and technical proofs of Theorems \ref{Th1}-\ref{Th4}.

\subsection{Computation procedure using ADMM algorithm}

\indent It is worth noting that the penalty function in (\ref{obf}) is not separable in $\boldsymbol{\gamma}_i$'s. Following \cite{Ma2019}, we derive an ADMM algorithm to minimize the objective function (\ref{obf}). By introducing a new set of parameters $\boldsymbol{\delta}_{ij}=\boldsymbol{\gamma}_i-\boldsymbol{\gamma}_{j}$, the problem can be reformulated as the following constrained optimization:  
\begin{gather}
\text{min} ~~ \frac{1}{2}\sum _{i=1}^{n}(\boldsymbol{Y}_{i} -\boldsymbol{X}_{i} \boldsymbol{\gamma}_{i})^{ T}\boldsymbol{V}_{i}^{-1}(\boldsymbol{Y}_{i} -\boldsymbol{X}_{i} \boldsymbol{\gamma}_{i})+\sum _{1\leq i <j\leq n}p_{\tau }\left (\left \Vert \boldsymbol {\delta}_{ij} \right \Vert_2 , \lambda \right ),   \notag \\
\text{subject to }\boldsymbol{\gamma}_i-\boldsymbol{\gamma}_j-\boldsymbol{\delta}_{ij}=\boldsymbol{0}.  \label{EQ:opt}
\end{gather}
Denote by $\left \langle \boldsymbol{a},\boldsymbol{b} \right \rangle=\boldsymbol{a}^{ T}\boldsymbol{b}$ the inner product of two vectors. The above constrained optimization can be transformed into its augmented Lagrangian optimization problem, i.e, minimize:
\begin{align}
L\left ( \boldsymbol {\gamma}, \boldsymbol{\delta},\boldsymbol{\upsilon} \right )&= \frac{1}{2}\sum _{i=1}^{n}(\boldsymbol{Y}_{i} -\boldsymbol{X}_{i} \boldsymbol{\gamma}_{i})^{T }\boldsymbol{V}_{i}^{-1}(\boldsymbol{Y}_{i} -\boldsymbol{X}_{i} \boldsymbol{\gamma}_{i})+\sum _{1\leq i <j\leq n}p_{\tau }\left (\left \Vert \boldsymbol {\delta}_{ij} \right \Vert_2 , \lambda \right ) \notag \\
&+\sum _{i<j}\left \langle \boldsymbol{\upsilon }_{ij} ,\boldsymbol{\gamma}_{i} -\boldsymbol{\gamma }_{j} -\boldsymbol{\delta }_{ij}\right \rangle+\frac{\vartheta }{2}\sum _{i<j}\left \Vert \boldsymbol{\gamma}_{i} -\boldsymbol{\gamma }_{j} -\boldsymbol{\delta }_{ij}\right \Vert_2 ^{2},
\end{align}
where $\boldsymbol{\delta}=\left \{ \boldsymbol{\delta}_{ij}^{T}, i<j \right\}^{T}$, the dual variables $\boldsymbol{\upsilon}=\left\{\boldsymbol{\upsilon}_{ij}^{T}, i<j\right\}^{T}$ are the Lagrange multipliers and $\vartheta$ is the penalty parameter. Then we can compute the estimates of $( \boldsymbol {\gamma}, \boldsymbol{\delta},\boldsymbol{\upsilon})$ through iterations using the ADMM algorithm. 

\indent Given the value of $\boldsymbol{\delta}^{m}, \boldsymbol{\upsilon}^{m}$ at step $m$, we update the estimates at step $m+1$ as follows:
\begin{align}
\boldsymbol{\gamma}^{m+1} &=\mathop{\rm argmin}_{\boldsymbol{\gamma}}L\left(\boldsymbol{\gamma}, \boldsymbol{\delta}^{m},\boldsymbol{\upsilon}^{m} \right), \label{A.3}\\
\boldsymbol{\delta}^{m+1}&=\mathop{\rm argmin}_{\boldsymbol{\delta}}L\left(\boldsymbol{\gamma}^{m+1}, \boldsymbol{\delta},\boldsymbol{\upsilon}^{m} \right), \label{A.4}\\
\boldsymbol{\upsilon}_{ij}^{m+1}&=\boldsymbol{\upsilon}_{ij}^{m}+\vartheta\left(\boldsymbol{\gamma}_{i}^{m+1}-\boldsymbol{\gamma}_{j}^{m+1}-\boldsymbol{\delta}_{ij}^{m+1} \right). \label{A.5}
\end{align}
Notice that the problem in (\ref{A.3}) is equivalent to minimizing the function 
\begin{align*}
f\left ( \boldsymbol {\gamma} \right )
&=\frac{1}{2}\sum _{i=1}^{n}(\boldsymbol{Y}_{i} -\boldsymbol{X}_{i} \boldsymbol{\gamma}_i)^{ T }\boldsymbol{V}_{i}^{-1}(\boldsymbol{Y}_{i} -\boldsymbol{X}_{i} \boldsymbol{\gamma}_{i})+\frac{\vartheta }{2}\sum _{i<j}\left \Vert \boldsymbol{\gamma}_{i} -\boldsymbol{\gamma }_{j} -\boldsymbol{\delta }^{m}_{ij}+\vartheta^{-1}\boldsymbol{\upsilon}^{m}_{ij}\right \Vert_2 ^{2}+C_0\\
&=\frac{1}{2}\left(\boldsymbol{Y}-\boldsymbol{X}\boldsymbol{\gamma} \right)^{T}\boldsymbol{V}^{-1}\left(\boldsymbol{Y}-\boldsymbol{X}\boldsymbol{\gamma}\right)+\frac{\vartheta }{2}\left \Vert \boldsymbol{A}\boldsymbol{\gamma } -\boldsymbol{\delta }^{m}+\vartheta^{-1}\boldsymbol{\upsilon}^m\right \Vert_2^{2}+C_0,
\end{align*}
\noindent where $\boldsymbol{Y}=\left(\boldsymbol{Y}^T_1,...,\boldsymbol{Y}^T_{n} \right)^{ T}$, $\boldsymbol{X}=\text {diag} \left(\boldsymbol{X}_1,...,\boldsymbol{X}_n \right)$, $\boldsymbol{V}=\text {diag}\left({\boldsymbol{V}}_1,...,\boldsymbol{V}_n \right)$, $\boldsymbol{A}=\boldsymbol{D}\otimes \boldsymbol{I}_{S}$ (Kronecker product) and $C_0$ is a constant independent of $\boldsymbol{\gamma}$. Here $\boldsymbol{D}=\left\{\left ( \boldsymbol{e}_i- \boldsymbol{e}_j\right ),i<j \right\}^{ T}$, in which $\boldsymbol{e}_i$ is a $n\times 1$ vector with the $i$ th element being 1 and the remaining ones being 0, and $\boldsymbol{I}_{S}$ is a $S \times S$ identity matrix. Thus, we can update $\boldsymbol{\gamma}^{m+1}$ by
\begin{align}
\boldsymbol{\gamma}^{m+1}&=\left ( \boldsymbol{X}^{T}\boldsymbol{V}^{-1}\boldsymbol{X}+\vartheta\boldsymbol{A}^{ T}\boldsymbol{A} \right)^{-1}\left [\boldsymbol{X}^{T}\boldsymbol{V}^{-1}\boldsymbol{Y}+\vartheta\boldsymbol{A}^{ T}\left (\boldsymbol{\delta}^{m}-\vartheta^{-1}\boldsymbol{\upsilon}^{m} \right ) \right ].
\label{A.6}
\end{align}

\indent In (\ref{A.4}), given $\boldsymbol{\gamma}^{m+1}$ and $\boldsymbol{\upsilon}^m$, the minimization problem is the same as minimizing 
$$\frac{\vartheta}{2}\left \| \boldsymbol{\zeta}^{m}_{ij}-\boldsymbol{\delta}_{ij} \right \|_2^{2}+p_{\tau }\left (\left \Vert \boldsymbol{\delta}_{ij}\right \Vert_2, \lambda \right )$$ with respect to $\boldsymbol{\delta}_{ij}$, where $\boldsymbol{\zeta}^{m}_{ij}=\boldsymbol{\gamma}^{m+1}_i-\boldsymbol{\gamma}^{m+1}_j+\vartheta^{-1}\boldsymbol{\upsilon}^{m}_{ij}$. Consequently, for MCP penalty with $\tau>1/\vartheta$, we have:  
\begin{align}
\boldsymbol{\delta}_{ij}^{m+1}=\left\{
\begin{array}{ll}
\frac{\text{ST}(\boldsymbol{\zeta}_{ij}^m,\lambda /\vartheta )}{1-1/(\tau
\vartheta )} & \text{ if }\|\boldsymbol{\zeta}_{ij}^m\|_2\leq \tau \lambda,
\\
\boldsymbol{\zeta}_{ij}^m & \text{ if }\|\boldsymbol{\zeta}_{ij}^m\|_2>\tau
\lambda,
\end{array}
\right. 
\label{A.7}
\end{align}
where $\text{ST}(\boldsymbol{z},t)=(1-t/\left \|\boldsymbol{z} \right \|_2)_{+}\boldsymbol{z}$ is the groupwise soft thresholding operator. 

\noindent Given the discussion above, we summarize the detailed ADMM algorithm as follows:
\begin{algorithm}[H]
\caption{ADMM algorithm}\label{admm}
\begin{algorithmic}[1]
\State Initialize $\boldsymbol{\delta}^0$, $\boldsymbol{\upsilon}^0$.
\For{$m=0, 1, 2, \cdots$}
    \State Update $\boldsymbol{\gamma}^{m+1}$ using (\ref{A.6})
    \State Update $\boldsymbol{\delta}^{m+1}$ using (\ref{A.7})
    \State Update $\boldsymbol{\upsilon}^{m+1}$ using (\ref{A.5})
    \If{the convergence criterion is met, }
       \State Stop and denote the last iteration by $\hat{\boldsymbol{\gamma}}(\lambda)$, 
    \Else
    \State $m=m+1$.
    \EndIf
\EndFor
\Ensure Output
\end{algorithmic}

\end{algorithm}

\indent We stop the ADMM algorithm when the primal residual $\boldsymbol{r}^{m+1}=\boldsymbol{A} \boldsymbol{\gamma}^{m+1}-\boldsymbol{\delta}^{m+1}$ is close to zero such that $\left \|  \boldsymbol{r}^{m+1}\right \|_2 < \varepsilon$ for some small value $\varepsilon>0$.

\subsubsection{Initial value for starting ADMM algorithm and computation of solution path}
\label{initial_value}

\noindent To start ADMM algorithm, an appropriate initial value is very important. First, given model (\ref{mamodel}), we use the ordinary least squares estimate of each subject as the initial estimate $\boldsymbol{\gamma}^0$, i.e. $\boldsymbol{\gamma}^0_i=(\boldsymbol{X}^{T}_i\boldsymbol{X}_i)^{-1}\boldsymbol{X}^{T}_i\boldsymbol{Y}_i, i=1, \cdots, n$, which is a consistent estimate. Then, let initial estimates $\boldsymbol{\delta}^0_{ij}=\boldsymbol{\gamma}^0_i-\boldsymbol{\gamma}^0_j$ in  $\boldsymbol{\delta}^{0}$ and $\boldsymbol{\upsilon}^{0}=\boldsymbol{0}$.

\indent In order to compute the solution path of $\boldsymbol{\gamma}$ against $\lambda$, we consider a grid of $\lambda$ values with $\lambda_\text{min}=\lambda_0<\lambda_1< \cdots <\lambda_K=\lambda_\text{max}$, where $0 \leq \lambda_\text{min}<\lambda_\text{max}<\infty$. Given a $\lambda$ value in $[\lambda_\text{min}, \lambda_\text{max}]$, we can compute $\hat{\boldsymbol{\gamma}}(\lambda)$ given in (\ref{3.4}) by using ADMM algorithm. Referring to \cite{Ma2019}, a warm start and continuation strategy is used for updating the solutions. Specifically, we compute $\hat{\boldsymbol{\gamma}}(\lambda_0)$ by using $\boldsymbol{\gamma}^0$ as the initial value, then $\hat{\boldsymbol{\gamma}}(\lambda_k)$ by using $\hat{\boldsymbol{\gamma}}(\lambda_{k-1})$ as the initial value $(k=1, \cdots, K)$.

\subsubsection{Computational complexity of ADMM algorithm}
{
The computational complexity can be expressed in terms of floating point operations per second (flops) required to find the solution \citep{boyd2011distributed}. Our ADMM algorithm involves updating the estimates of $\boldsymbol{\delta}$, $\boldsymbol{\upsilon}$ and $\boldsymbol{\gamma}$ given in (\ref{A.4}), (\ref{A.5}) and (\ref{A.6}) through iterations. It costs $O((n-1)nS)=O(n^2S)$ flops for computing the updates of $\boldsymbol{\delta}$ and $\boldsymbol{\upsilon}$, given that $\boldsymbol{\delta}$ and $\boldsymbol{\upsilon}$ are $0.5n(n-1)S \times 1$ vectors, where $S$ is the number of B-spline basis functions. Following \citep{boyd2011distributed} (see pages 27-29), the computational cost for updating $\boldsymbol{\gamma}$ given in (\ref{A.6}) is dominated by $\boldsymbol{V}^{-1}$ and  $(\boldsymbol{X}^T\boldsymbol{V}^{-1}\boldsymbol{X}+\vartheta\boldsymbol{A}^T\boldsymbol{A})^{-1}$, which require $O(\sum^n_{i=1} m^3_i)$ and $O(n^3S^3)$ flops, respectively. Therefore, it takes $O(\sum^n_{i=1} m^3_i+n^3S^3)$ flops to update $\boldsymbol{\gamma}$ in the first iteration. Since the values  $(\boldsymbol{X}^T\boldsymbol{V}^{-1}\boldsymbol{X}+\vartheta\boldsymbol{A}^T\boldsymbol{A})^{-1}$ and $\boldsymbol{X}^T\boldsymbol{V}^{-1}\boldsymbol{Y}$ remain the same through iterations, and we only need to compute them once, in the subsequent iterations, the computational cost for updating $\boldsymbol{\gamma}$ is reduced to $O(nS \times 0.5n(n-1)S)=O(n^3S^2)$, which is cost of computing $\boldsymbol{A}^{ T}\left (\boldsymbol{\delta}-\vartheta^{-1}\boldsymbol{\upsilon} \right )$ given in (\ref{A.6}). In sum, the overall cost of updating $(\boldsymbol{\gamma}, \boldsymbol{\delta}, \boldsymbol{\upsilon})$ is $O(n^2S+\sum^{n}_{i=1} m^3_i + n^3S^3)=O(\sum ^n_{i=1}m^3_i+n^3S^3)$ in the first iteration, and it is $O(n^2S+n^3S^2)=O(n^3S^2)$ in the subsequent iterations. 
}

\subsection{ Consistency and convergence}

Let $C$ denotes a generic constant that might assume different values at different places. Without loss of generality, we consider the following B-spline basis functions that span $G$, that is, $B_l=S^{1/2}B^{*}_l$, $l=1,\ldots,S$, where $\{B^{*}_l\}_{l=1}^{S}$ are the B-splines defined in Chapter 5 of \cite{DeVore1993}. {It follows from Theorem 4.2 of \cite{DeVore1993} that
\begin{equation}\label{A1}
M_1 \|\boldsymbol{\gamma}\|_2^2 \leq \int\left\{ \sum_{l=1}^{S}{B_l(t)\gamma_l}\right\}^2 dt \leq M_2 \|\boldsymbol{\gamma}\|_2^2
\end{equation}
for some constants $0<M_{1}<M_{2}<\infty$, where $\boldsymbol{\gamma}=(\gamma_1,\ldots,\gamma_{S})^T$.}

\begin{lemma}\label{Lemma1}
For each $i$, there exist some constants $0<M_{1}<M_{2}<\infty$ such that, except on an event whose probability tends to zero, all the eigenvalues of $\boldsymbol{X}_i^T\boldsymbol{X}_i/m_i$ fall between $M_1$ and $M_2$.
\end{lemma}

\begin{lemma}\label{Lemma2}
Assume the random variables $\xi$ and $\eta$ be $F_1^k$-measurable and $F_{k+s}^{\infty}$-measurable, respectively. If $E(|\xi|^p)<\infty$, $E(|\eta|^q)<\infty$ for some $p,q>1$ and $1/p+1/q<1$. Then, under $\alpha$-mixing,
\[
|E(\xi\eta)-E(\xi)E(\eta)|\leq 10 \alpha(s)^{1-\frac{1}{p}-\frac{1}{q}}\|\xi\|_p\|\eta\|_q,
\]
where $\|\xi\|_p=E^{1/p}(|\xi|^p)$ denotes the $L_p$-norm of $\xi$.
\end{lemma}

The proof of Lemma \ref{Lemma1} and Lemma \ref{Lemma2} can be respectively referred to Lemma 2 of \cite{Huang2004} and Theorem 7.3 of \cite{Roussas1987}.

{Define $\beta_i^*(t)=\boldsymbol{B}(t)^T\boldsymbol{\gamma}^* \in G$ such that $\|\beta_i^*-\beta_i\|_2=\inf_{g\in G}\|g-\beta_i\|_2 \triangleq \varpi_i$, it follows from the result on page 149 of \cite{de2001practical} that $\varpi_i=J^{-r}$ if $\beta_i(\cdot)\in  C^{(r)}$.}

\subsubsection{Consistency of initial estimator}

\begin{proposition}\label{Pro1}
Under conditions (C1)-(C4), the initial estimators {$\hat{{\beta}}_i^{(0)}(t)$, $i=1,\ldots,n$, satisfy $\|\hat{{\beta}}_i^{(0)}-{\beta}_i\|_2^2=O_p(J/m_i+J^{-2r})$.}
\end{proposition}

Proof. Recall that
\[
\hat{\boldsymbol{\gamma}}_i^{(0)}=\arg\min_{\boldsymbol{\gamma}_i}(\boldsymbol{Y}_i-\boldsymbol{X}_i\boldsymbol{\gamma}_i)^T(\boldsymbol{Y}_i-\boldsymbol{X}_i\boldsymbol{\gamma}_i)=(\boldsymbol{X}_i^T\boldsymbol{X}_i)^{-1}\boldsymbol{X}_i^T\boldsymbol{Y}_i,
\]
{and $\hat{{\beta}}_i^{(0)}(t)=\boldsymbol{B}(t)^T\hat{\boldsymbol{\gamma}}_i^{(0)}$.} Now, define $\tilde{\boldsymbol{\gamma}}_i^{(0)}=(\boldsymbol{X}_i^T\boldsymbol{X}_i)^{-1}\boldsymbol{X}_i^T\tilde{\boldsymbol{Y}}_i$, $\tilde{{\beta}}_i^{(0)}(t)=\boldsymbol{B}(t)^T\tilde{\boldsymbol{\gamma}}_i^{(0)}$, where $\tilde{\boldsymbol{Y}}_i=(\beta_i(t_{i1}),\ldots,\beta_i(t_{im_i}))^T$. Obviously,
\[
\hat{\boldsymbol{\gamma}}_i^{(0)}-\tilde{\boldsymbol{\gamma}}_i^{(0)}=(\boldsymbol{X}_i^T\boldsymbol{X}_i)^{-1}\boldsymbol{X}_i^T(\boldsymbol{Y}_i-\tilde{\boldsymbol{Y}}_i)=(\boldsymbol{X}_i^T\boldsymbol{X}_i)^{-1}\boldsymbol{X}_i^T\boldsymbol{\varepsilon}_i,
\]
where $\boldsymbol{\varepsilon}_i=(\varepsilon_{i1},\ldots,\varepsilon_{im_i})^T$. It follows from Lemma \ref{Lemma1} that there exists a constant $C>0$, such that
\[
E\left( \boldsymbol{\varepsilon}_i^T \boldsymbol{X}_i (\boldsymbol{X}_i^T\boldsymbol{X}_i)^{-1}(\boldsymbol{X}_i^T\boldsymbol{X}_i)^{-1}\boldsymbol{X}_i^T\boldsymbol{\varepsilon}_i \right) \leq C \frac{1}{m_i^2}E\left( \boldsymbol{\varepsilon}_i^T \boldsymbol{X}_i \boldsymbol{X}_i^T\boldsymbol{\varepsilon}_i \right).
\]
{
Taking $p=q=4$ in Lemma \ref{Lemma2} and by the properties of B-splines, we can obtain
\begin{eqnarray*}
E\left( \boldsymbol{\varepsilon}_i^T \boldsymbol{X}_i \boldsymbol{X}_i^T\boldsymbol{\varepsilon}_i \right)&=&E\left\{ \sum_{l=1}^{S} {\left( \sum_{j=1}^{m_i}{B_l(t_{ij})}\varepsilon_{ij} \right)^2} \right\}= E\left\{ \sum_{j,j^{\prime}=1}^{m_i} {  \sum_{l=1}^{S} { \varepsilon_{ij}\varepsilon_{ij^{\prime}} B_l(t_{ij})B_l(t_{ij^{\prime}})} }\right\} \\
&\leq& S \sum_{j,j^{\prime}=1}^{m_i} { |E(\varepsilon_{ij}\varepsilon_{ij^{\prime}})| } \leq 10 S \sum_{1\leq j,j^{\prime} \leq m_i} {\alpha(|j-j^{\prime}|)^{1/2}[E(|\varepsilon_{ij}|^4)]^{1/4} [E(|\varepsilon_{ij^{\prime}}|^4)]^{1/4} } \\
&=&O_p(Jm_i),
\end{eqnarray*}
where the last equality holds because $\sum_{s=1}^{\infty} { \alpha(s)^{1/2} [E(|\varepsilon_{ij}|^4)]^{1/4} [E(|\varepsilon_{ij^{\prime}}|^4)]^{1/4} }$ is bounded from condition (C2). Thus, $\|\hat{\boldsymbol{\gamma}}_i^{(0)}-\tilde{\boldsymbol{\gamma}}_i^{(0)}\|_2^2=O_p(J/m_i)$.} This together with expression (\ref{A1}) leads to
{
\begin{equation}\label{PPro:1}
\|\hat{{\beta}}_i^{(0)}-\tilde{{\beta}}_i^{(0)}\|_2^2=O\left( \|\hat{\boldsymbol{\gamma}}_i^{(0)}-\tilde{\boldsymbol{\gamma}}_i^{(0)}\|_2^2 \right)=O_p(J/m_i).
\end{equation}
}

On the other hand, by Lemma \ref{Lemma1}, we have
\[
\|\tilde{\boldsymbol{\gamma}}_i^{(0)}-\boldsymbol{\gamma}_i^*\|_2^2 =O_p\left( \frac{1}{m_i} (\tilde{\boldsymbol{\gamma}}_i^{(0)}-\boldsymbol{\gamma}_i^*)^T\boldsymbol{X}_i^T\boldsymbol{X}_i(\tilde{\boldsymbol{\gamma}}_i^{(0)}-\boldsymbol{\gamma}_i^*) \right).
\]
Noting that $\boldsymbol{X}_i\tilde{\boldsymbol{\gamma}}_i^{(0)}=\boldsymbol{X}_i(\boldsymbol{X}_i^T\boldsymbol{X}_i)^{-1}\boldsymbol{X}_i^T\tilde{\boldsymbol{Y}}_i$ is an orthogonal projection of $\tilde{\boldsymbol{Y}}_i$. Hence,
\begin{eqnarray*}
\frac{1}{m_i}(\tilde{\boldsymbol{\gamma}}_i^{(0)}-\boldsymbol{\gamma}_i^*)^T\boldsymbol{X}_i^T\boldsymbol{X}_i(\tilde{\boldsymbol{\gamma}}_i^{(0)}-\boldsymbol{\gamma}_i^*) &\leq& \frac{1}{m_i} (\tilde{\boldsymbol{Y}}_i-\boldsymbol{X}_i\boldsymbol{\gamma}_i^*)^T(\tilde{\boldsymbol{Y}}_i-\boldsymbol{X}_i\boldsymbol{\gamma}_i^*) \\
& =& \frac{1}{m_i} \sum_{j=1}^{m_i} { \left(\beta_i(t_{ij})-\beta_i^*(t_{ij})\right)^2 } \\
&=& O(\varpi_i^2),
\end{eqnarray*}
{It follows from expression (\ref{A1}) that
\begin{equation}\label{PPro:2}
\|\tilde{{\beta}}_i^{(0)}-{\beta}_i^*\|_2^2=O\left( \|\tilde{\boldsymbol{\gamma}}_i^{(0)}-\boldsymbol{\gamma}_i^*\|_2^2 \right)=O_p(\varpi_i^2).
\end{equation} 
Therefore, by the definition of $\varpi_i$, equations (\ref{PPro:1})-(\ref{PPro:2}) and the triangle inequality, we have
\begin{eqnarray*}
&& \|\hat{{\beta}}_i^{(0)}-{\beta}_i\|_2^2 \\
&\leq& \|\hat{{\beta}}_i^{(0)}-\tilde{{\beta}}_i^{(0)}\|_2^2+\|\tilde{{\beta}}_i^{(0)}-{\beta}_i^*\|_2^2+\|{\beta}_i^*-{\beta}_i\|_2^2 \\
&=&O_p(J/m_i) +O_p(\varpi_i^2)+O_p(\varpi_i^2)=O_p(J/m_i+J^{-2r}).
\end{eqnarray*} }
This completes the proof.

\subsubsection{Convergence of ADMM}

\begin{proposition}\label{Pro2}
Let $\boldsymbol{r}^{m+1}=\boldsymbol{A}\boldsymbol{\gamma}^{m+1}-\boldsymbol{\delta}^{m+1}$ and $\boldsymbol{s}^{m+1}=\vartheta \boldsymbol{A}^T(\boldsymbol{\delta}^{m+1}-\boldsymbol{\delta}^{m})$ respectively be the primal residual and dual residual in the ADMM. Then, $\lim_{m\rightarrow \infty}\|\boldsymbol{r}^{m+1}\|_2^2=0$ and $\lim_{m\rightarrow \infty}\|\boldsymbol{s}^{m+1}\|_2^2=0$ hold for MCP penalty.
\end{proposition}
Proof. Taking a careful examination of our constructed objective function $L(\boldsymbol{\gamma},\boldsymbol{\delta},\boldsymbol{\nu})$ with that of \cite{Ma2019}, the conclusion $\lim_{m\rightarrow \infty}\|\boldsymbol{r}^{m+1}\|_2^2=0$ can be directly derived by a similar proof of proposition 1 in \cite{Ma2019}. Recall that $\boldsymbol{\gamma}^{m+1}$ minimize $L(\boldsymbol{\gamma},\boldsymbol{\delta}^{m},\boldsymbol{\nu}^{m})$ by definition, thus
\begin{eqnarray*}
0&=&\frac{\partial L(\boldsymbol{\gamma},\boldsymbol{\delta}^{m},\boldsymbol{\nu}^{m})}{\partial\boldsymbol{\gamma}}\Big|_{\boldsymbol{\gamma}=\boldsymbol{\gamma}^{m+1}}=\boldsymbol{X}^T\boldsymbol{V}^{-1}(\boldsymbol{X}\boldsymbol{\gamma}^{m+1}-\boldsymbol{Y})+ \boldsymbol{A}^T\{\boldsymbol{\nu}^{m}+\vartheta(\boldsymbol{A\gamma}^{m+1}-\boldsymbol{\delta}^{m})\} \\
&=& \boldsymbol{X}^T\boldsymbol{V}^{-1}(\boldsymbol{X\gamma}^{m+1}-\boldsymbol{Y})+\boldsymbol{A}^T\{[\boldsymbol{\nu}^{m+1}-\vartheta(\boldsymbol{A\gamma}^{m+1}-\boldsymbol{\delta}^{m+1})]+\vartheta(\boldsymbol{A\gamma}^{m+1}-\boldsymbol{\delta}^{m}) \} \\
&=& \boldsymbol{X}^T\boldsymbol{V}^{-1}(\boldsymbol{X\gamma}^{m+1}-\boldsymbol{Y})+\boldsymbol{A}^T\boldsymbol{\nu}^{m+1}+\vartheta \boldsymbol{A}^T(\boldsymbol{\delta}^{m+1}-\boldsymbol{\delta}^{m}),
\end{eqnarray*}
which implies
\[
\boldsymbol{s}^{m+1}=\vartheta \boldsymbol{A}^T(\boldsymbol{\delta}^{m+1}-\boldsymbol{\delta}^{m}) =-\{ \boldsymbol{X}^T\boldsymbol{V}^{-1}(\boldsymbol{X\gamma}^{m+1}-\boldsymbol{Y})+\boldsymbol{A}^T\boldsymbol{\nu}^{m+1}\}.
\]
In view of $\lim_{m\rightarrow \infty}\|\boldsymbol{r}^{m}\|_2^2=\lim_{m\rightarrow \infty}\|\boldsymbol{A\gamma}^{m}-\boldsymbol{\delta}^{m}\|_2^2=0$, we have
\[
0=\lim_{m\rightarrow \infty} \frac{\partial L(\boldsymbol{\gamma},\boldsymbol{\delta}^{m},\boldsymbol{\nu}^{m})}{\partial\boldsymbol{\gamma}}\Big|_{\boldsymbol{\gamma}=\boldsymbol{\gamma}^{m+1}}=\lim_{m\rightarrow \infty} \{ \boldsymbol{X}^T\boldsymbol{V}^{-1}(\boldsymbol{X\gamma}^{m+1}-\boldsymbol{Y})+ \boldsymbol{A}^T\boldsymbol{\nu}^{m+1} \}=\lim_{m\rightarrow \infty}-\boldsymbol{s}^{m+1}.
\]
Therefore, we obtain $\lim_{m\rightarrow \infty}\|\boldsymbol{s}^{m+1}\|_2^2=0$, this completes the proof.

\subsection{Proof of the Theorems}
To prove the main theoretical results in this article, we first present the following lemma which will be frequently used in the sequel. {Let $\bar{\beta}_i(t)=\boldsymbol{B}^T(t)\bar{\boldsymbol{\gamma}}_i$, where}
\begin{equation}\label{PLe3:0}
\bar{\boldsymbol{\gamma}}_i=\arg\min_{\boldsymbol{\gamma}_i} (\boldsymbol{Y}_i-\boldsymbol{X}_i\boldsymbol{\gamma}_i)^T\boldsymbol{V}_i^{-1}(\boldsymbol{Y}_i-\boldsymbol{X}_i\boldsymbol{\gamma}_i)=(\boldsymbol{X}_i^T\boldsymbol{V}_i^{-1}\boldsymbol{X}_i)^{-1}\boldsymbol{X}_i^T\boldsymbol{V}_i^{-1}\boldsymbol{Y}_i.
\end{equation}

\begin{lemma}\label{Lemma3}
Under conditions (C1)-(C5), { we have  $\|\bar{{\beta}}_i-{\beta}_i\|_2^2=O_p(J/m_i+J^{-2r})$, $i=1,\ldots,n$.}
\end{lemma}

Proof. For $i=1,\ldots,n$, {let $\tilde{{\beta}}_i(t)=\boldsymbol{B}(t)^T\tilde{\boldsymbol{\gamma}}_i$ and $\tilde{\boldsymbol{\gamma}}_i=(\boldsymbol{X}_i^T\boldsymbol{V}_i^{-1}\boldsymbol{X}_i)^{-1}\boldsymbol{X}_i^T\boldsymbol{V}_i^{-1}\tilde{\boldsymbol{Y}}_i$, where $\tilde{\boldsymbol{Y}}_i$ is defined as above. Obviously, $\bar{{\beta}}_i(t)-\tilde{{\beta}}_i(t)=\boldsymbol{B}(t)^T(\boldsymbol{X}_i^T\boldsymbol{V}_i^{-1}\boldsymbol{X}_i)^{-1}\boldsymbol{X}_i^T\boldsymbol{V}_i^{-1}\boldsymbol{\varepsilon}_i$.} By Lemma \ref{Lemma1} and the bounded assumption on the eigenvalues of $V$, it is easy to verify that there exist two constants $0<C_1\leq C_2<\infty$, such that
\[
C_1 \frac{1}{m_i^2} E(\boldsymbol{\varepsilon}_i^T\boldsymbol{V}_i^{-1}\boldsymbol{X}_i\boldsymbol{X}_i^T\boldsymbol{V}_i^{-1}\boldsymbol{\varepsilon}_i) \leq E\left( \|\bar{{\beta}}_i(t)-\tilde{{\beta}}_i(t)\|_2^2 \right) \leq C_2 \frac{1}{m_i^2} E(\boldsymbol{\varepsilon}_i^T\boldsymbol{V}_i^{-1}\boldsymbol{X}_i\boldsymbol{X}_i^T\boldsymbol{V}_i^{-1}\boldsymbol{\varepsilon}_i).
\]
According to the operation properties of the trace and expectation, we have
\begin{eqnarray*}
E(\boldsymbol{\varepsilon}_i^T\boldsymbol{\Sigma}_i^{-1}\boldsymbol{X}_i\boldsymbol{X}_i^T\boldsymbol{\Sigma}_i^{-1}\boldsymbol{\varepsilon}_i)&=&\mbox{trace}\left\{ E(\boldsymbol{\varepsilon}_i^T\boldsymbol{\Sigma}_i^{-1}\boldsymbol{X}_i\boldsymbol{X}_i^T\boldsymbol{\Sigma}_i^{-1}\boldsymbol{\varepsilon}_i) \right\}=E\left( \mbox{trace}\{\boldsymbol{X}_i^T\boldsymbol{\Sigma}_i^{-1}\boldsymbol{\varepsilon}_i\boldsymbol{\varepsilon}_i^T\boldsymbol{\Sigma}_i^{-1}\boldsymbol{X}_i\} \right) \\
&=& E\left( \mbox{trace}\{\boldsymbol{X}_i^T\boldsymbol{V}_i^{-1}\boldsymbol{\Sigma}_i\boldsymbol{V}_i^{-1}\boldsymbol{X}_i\} \right)= \mbox{trace}\left\{ E(\boldsymbol{X}_i^T\boldsymbol{V}_i^{-1}\boldsymbol{\Sigma}_i\boldsymbol{V}_i^{-1}\boldsymbol{X}_i) \right\} \\
&=& O_p(m_iJ),
\end{eqnarray*}
where the last equality holds due to condition (C5) and Lemma \ref{Lemma1}. Hence, we obtain
\begin{equation}\label{PLe3:1}
\|\bar{{\beta}}_i-\tilde{{\beta}}_i\|_2^2=O_p(J/m_i).
\end{equation}
Furthermore, as $\boldsymbol{V}_i^{-1/2}\boldsymbol{X}_i(\boldsymbol{X}_i^T\boldsymbol{V}_i^{-1}\boldsymbol{X}_i)^{-1}\boldsymbol{X}_i^T\boldsymbol{V}_i^{-1}\tilde{\boldsymbol{Y}}_i$ is an orthogonal projection of $\boldsymbol{V}_i^{-1/2}\tilde{\boldsymbol{Y}}_i$, we have
\begin{eqnarray*}
&&\frac{1}{m_i} (\tilde{\boldsymbol{\gamma}}_i-\boldsymbol{\gamma}_i^*)^T\boldsymbol{X}_i^T\boldsymbol{V}_i^{-1}\boldsymbol{X}_i(\tilde{\boldsymbol{\gamma}}_i-\tilde{\boldsymbol{\gamma}}_i^*) \\
&=& \frac{1}{m_i} \left\{(\boldsymbol{X}_i^T\boldsymbol{V}_i^{-1}\boldsymbol{X}_i)^{-1}\boldsymbol{X}_i^T\boldsymbol{V}_i^{-1}\tilde{\boldsymbol{Y}}_i-\tilde{\boldsymbol{\gamma}}_i^*\right\}^T \boldsymbol{X}_i^T\boldsymbol{V}_i^{-1/2}\boldsymbol{V}_i^{-1/2}\boldsymbol{X}_i\left\{(\boldsymbol{X}_i^T\boldsymbol{V}_i^{-1}\boldsymbol{X}_i)^{-1}\boldsymbol{X}_i^T\boldsymbol{V}_i^{-1}\tilde{\boldsymbol{Y}}_i-\tilde{\boldsymbol{\gamma}}_i^*\right\} \\
&=& \frac{1}{m_i} \|\boldsymbol{V}_i^{-1/2}\boldsymbol{X}_i(\boldsymbol{X}_i^T\boldsymbol{V}_i^{-1}\boldsymbol{X}_i)^{-1}\boldsymbol{X}_i^T\boldsymbol{V}_i^{-1}\tilde{\boldsymbol{Y}}_i- \boldsymbol{V}_i^{-1/2}\boldsymbol{X}_i \boldsymbol{\gamma}_i^*\|_2^2 \\
&\leq& \frac{1}{m_i} \|\boldsymbol{V}_i^{-1/2}\tilde{\boldsymbol{Y}}_i-\boldsymbol{V}_i^{-1/2}\boldsymbol{X}_i\boldsymbol{\gamma}_i^*\|_2^2=\frac{1}{m_i} (\tilde{\boldsymbol{Y}}_i-\boldsymbol{X}_i\boldsymbol{\gamma}_i^*)^T\boldsymbol{V}_i^{-1}(\tilde{\boldsymbol{Y}}_i-\boldsymbol{X}_i\boldsymbol{\gamma}_i^*) \\
&=& O_p\left( \frac{1}{m_i}\|\tilde{\boldsymbol{Y}}_i-\boldsymbol{X}_i\boldsymbol{\gamma}_i^*\|_2^2 \right) = O_p\left( \|{\beta}_i-{\beta}_i^*\|_2^2 \right)=O_p(\varpi_i^2),
\end{eqnarray*}
where the antepenult equality holds by the bounded assumption on the eigenvalues of $V$. Combining the expression (\ref{A1}), condition (C5) as well as Lemma \ref{Lemma1} leads to
\begin{equation}\label{PLe3:2}
\|\tilde{{\beta}}_i-{\beta}_i^*\|_2^2=O\left( \|\tilde{\boldsymbol{\gamma}}_i-\boldsymbol{\gamma}_i^*\|_2^2 \right)=O_p\left( \frac{1}{m_i} (\tilde{\boldsymbol{\gamma}}_i-\boldsymbol{\gamma}_i^*)^T\boldsymbol{X}_i^T\boldsymbol{\Sigma}_i^{-1}\boldsymbol{X}_i(\tilde{\boldsymbol{\gamma}}_i-\tilde{\boldsymbol{\gamma}}_i^*) \right)= O_p(\varpi_i^2).
\end{equation}
Consequently, it follows from equations (\ref{PLe3:1}), (\ref{PLe3:2}) and the definition of $\varpi_i$ that
\[
\|\bar{{\beta}}_i-{\beta}_i\|_2^2 \leq \|\bar{{\beta}}_i-\tilde{{\beta}}_i\|_2^2+\|\tilde{{\beta}}_i-{\beta}_i^*\|_2^2 +\|{\beta}_i^*-{\beta}_i\|_2^2=O_p(J/m_i+\varpi_i^2)=O_p(J/m_i+J^{-2r}).
\]
This finishes the proof.\\


\noindent \emph{Proof of Theorem \ref{Th1}.} Notice that, it is equivalent to individually obtaining $\hat{\boldsymbol{\theta}}_{k}=\arg\min_{ \boldsymbol{\theta}_k } (\boldsymbol{Y}_{(k)}-\boldsymbol{X}_{(k)}\boldsymbol{\theta}_k)^T\boldsymbol{V}^{-1}_{(k)}(\boldsymbol{Y}_{(k)}-\boldsymbol{X}_{(k)}\boldsymbol{\theta}_k)=\left( \boldsymbol{X}_{(k)}^T\boldsymbol{V}_{(k)}^{-1}\boldsymbol{X}_{(k)} \right)^{-1} \boldsymbol{X}_{(k)}^T\boldsymbol{V}_{(k)}^{-1}\boldsymbol{Y}_{(k)}$ for $k=1,\ldots,K$, where $\boldsymbol{Y}_{(k)}=\left \{ \boldsymbol{Y}^T_i: i \in \mathcal{G}_k \right \}^T, \boldsymbol{X}_{(k)}=\left \{ \boldsymbol{X}^T_i: i \in \mathcal{G}_k \right \}^T$ and $\boldsymbol{V}_{(k)}=\text{diag}\left\{\boldsymbol{V}_i: i \in \mathcal{G}_k \right \}$. Then  $\hat{{\alpha}}_k(t)=\boldsymbol{B}(t)^T\hat{\boldsymbol{\theta}}_k$ for any $t \in \mathbb{T}$. 
{According to Lemma \ref{Lemma3}, we have
\[
\|\hat{{\alpha}}_k- {\alpha}_k\|_2^2=O_p(J/N_k+J^{-2r})\leq O_p(J/N_0+J^{-2r}),
\]
where $\alpha_k(t)$ is the true function in the $k$th group. As a result,
\[
\|\hat{\boldsymbol{{\alpha}}}^{or}-\boldsymbol{\alpha}\|_2^2=\sum_{k=1}^K  { \|\hat{{\alpha}}^{or}_k- {\alpha}_k\|_2^2 } = O_p(J/N_0+J^{-2r})
\]
for any fixed $K$.} This completes the proof.\\

\noindent 
\emph{Proof of Theorem \ref{Th2}.} Let $\hat{\beta}_i^{or}(t)$ and $\hat{\boldsymbol{\gamma}}^{or}_i$ be the estimated function and estimated B-spline coefficient for subject $i$ given the true membership, respectively. We first prove $\|\hat{{\beta}}_i-\hat{{\beta}}_i^{or}\|_2^2=O_p(J/m_{(n)}+J^{-2r})$ for each $i$, where . Let $\delta_n=J/m_{(n)}+J^{-2r}$, if one can show that for any $\omega>0$, there exists a large enough constant $M>0$ satisfying
\begin{equation}\label{PTH2:1}
P \left\{ \inf_{\|\boldsymbol{X}_i(\boldsymbol{\gamma}_i-\hat{\boldsymbol{\gamma}}_i^{or})\|_2^2=M\delta_n} L_n(\boldsymbol{\gamma}) > L_n(\hat{\boldsymbol{\gamma}}^{or}) \right\} \geq 1-\omega,
\end{equation}
which means a local minimizer of $L_n(\boldsymbol{\gamma})$ existed in the region $\mathbb{B}_0=\{\boldsymbol{\gamma}:\|\boldsymbol{X}_i(\boldsymbol{\gamma}_i-\hat{\boldsymbol{\gamma}}_i^{or})\|_2^2 \leq M\delta_n\}$. Then, $\|\hat{{\beta}}_i-\hat{{\beta}}_i^{or}\|_2^2=O_p(J/m_{(n)}+J^{-2r})$ can be proved.

Let $L_{n1}=\frac{1}{2}(\boldsymbol{Y}-\boldsymbol{X\gamma})^T\boldsymbol{V}^{-1}(\boldsymbol{Y}-\boldsymbol{X\gamma})$, thus $\bar{\boldsymbol{\gamma}}=(\bar{\boldsymbol{\gamma}}_1^T,\ldots,\bar{\boldsymbol{\gamma}}_n^T)^T$ minimize $L_{n1}$, where $\bar{\boldsymbol{\gamma}}_i$, $i=1,\ldots,n$ are defined in (\ref{PLe3:0}). It follows from Lemma \ref{Lemma3} that $\|{\|\bar{{\beta}}_i-{\beta}_i\|_2^2}=O_p(J/m_{(n)}+J^{-2r})$ for each $i$. Combining this result with Theorem \ref{Th1} leads to
\[
\|\bar{{\beta}}_i-\hat{{\beta}}_i^{or}\|_2^2 \leq \|\bar{{\beta}}_i-{\beta}_i\|_2^2 + \|\hat{{\beta}}_i^{or}-{\beta}_i\|_2^2 = O_p(J/m_{(n)}+J^{-2r}),
\]
which is equivalent to
\begin{equation}\label{PTH2:2}
\|\boldsymbol{X}_i(\bar{\boldsymbol{\gamma}}_i-\hat{\boldsymbol{\gamma}}_i^{or})\|_2^2 \leq C_0\delta_n
\end{equation}
for some constant $C_0$ from expression (\ref{A1}). {Moreover, for any $k\neq k^{\prime}$,
\begin{eqnarray*}
\|\hat{{\alpha}}_k-\hat{{\alpha}}_{k^{\prime}}\|_2 &=& \|\hat{{\alpha}}_k-{\alpha}_k+{\alpha}_k-\hat{{\alpha}}_{k^{\prime}}+ {\alpha}_{k^{\prime}}-{\alpha}_{k^{\prime}}\|_2 \\
&\geq& \|{\alpha}_k-{\alpha}_{k^{\prime}}\|_2-\|\hat{{\alpha}}_k-{\alpha}_k\|_2-\|\hat{{\alpha}}_{k^{\prime}}-{\alpha}_{k^{\prime}}\|_2\\
&\geq& b -\|\hat{{\alpha}}_k-{\alpha}_k\|_2-\|\hat{{\alpha}}_{k^{\prime}}-{\alpha}_{k^{\prime}}\|_2.
\end{eqnarray*} 
}
Thus, we have $\|\hat{{\alpha}}_k-\hat{{\alpha}}_{k^{\prime}}\|_2\geq b$ for sufficiently large $N_0$ from Theorem \ref{Th1}. Accordingly, $\|\hat{\boldsymbol{\theta}}_k-\hat{\boldsymbol{\theta}}_{k^{\prime}}\|_2\geq Cb$ for some constant $C>0$. Similarly, for any $i\in \mathcal{G}_k$, $j\in \mathcal{G}_{k^{\prime}}$, $k\neq k^{\prime}$, we can derive that $\|\boldsymbol{\gamma}_i-\boldsymbol{\gamma}_j\|_2\geq Cb$ for any $\boldsymbol{\gamma}$ lies in the constraint $\mathbb{B}_0$ and sufficiently large $m_n$.

In addition, as $P_{\tau}(\cdot,\lambda)\geq0$ and $P_{\tau}(0,\lambda)=0$, then
\begin{eqnarray*}
&& L_n(\boldsymbol{\gamma}) - L_n(\hat{\boldsymbol{\gamma}}^{or}) \\
&\geq& L_{n1}(\boldsymbol{\gamma}) - L_{n1}(\hat{\boldsymbol{\gamma}}^{or}) +  \sum_{\begin{array}{c}
                                                             i\in \mathcal{G}_k, j\in \mathcal{G}_{k^{\prime}} \\
                                                             k\neq k^{\prime}
                                                            \end{array}}
{ \left\{ P_{\tau}(\|\boldsymbol{\gamma}_i-\boldsymbol{\gamma}_j\|_2,\lambda)\right\} } - \sum_{ k\neq k^{\prime}} {P_{\tau}(\|\hat{\boldsymbol{\theta}}_k-\hat{\boldsymbol{\theta}}_{k^{\prime}}\|_2,\lambda) }.
\end{eqnarray*}
As $\|\hat{\boldsymbol{\theta}}_k-\hat{\boldsymbol{\theta}}_{k^{\prime}}\|_2\geq Cb$ and $\|\boldsymbol{\gamma}_i-\boldsymbol{\gamma}_j\|_2\geq Cb$ for any $i\in \mathcal{G}_k$, $j\in \mathcal{G}_{k^{\prime}}$, $k\neq k^{\prime}$ from previous arguments, it follows from the condition $Cb \geq \tau\lambda$ that
\[
\sum_{\begin{array}{c}
                                                             i\in \mathcal{G}_k, j\in \mathcal{G}_{k^{\prime}} \\
                                                             k\neq k^{\prime}
                                                            \end{array}}
{ \left\{ P_{\tau}(\|\boldsymbol{\gamma}_i-\boldsymbol{\gamma}_j\|_2,\lambda)\right\} }=0
\]
and
\[
\sum_{ k\neq k^{\prime}} {P_{\tau}(\|\hat{\boldsymbol{\theta}}_k-\hat{\boldsymbol{\theta}}_{k^{\prime}}\|_2,\lambda) }=0
\]
Thus,
\[
L_n(\boldsymbol{\gamma}) - L_n(\hat{\boldsymbol{\gamma}}^{or}) \geq L_{n1}(\boldsymbol{\gamma}) - L_{n1}(\hat{\boldsymbol{\gamma}}^{or}).
\]

Further by the definition of $\bar{\boldsymbol{\gamma}}$ and (\ref{PTH2:2}), we have $L_{n1}(\boldsymbol{\gamma})\geq L_{n1}(\hat{\boldsymbol{\gamma}}^{or})$ for any $\boldsymbol{\gamma}$ satisfying $\|\boldsymbol{X}_i(\boldsymbol{\gamma}_i-\hat{\boldsymbol{\gamma}}_i^{or})\|_2^2=M\delta_n $ with sufficiently large $M$. Therefore, (\ref{PTH2:1}) is proved, which means
\[
\|\hat{{\beta}}_i-\hat{{\beta}}_i^{or}\|_2^2=O_p(J/m_{(n)}+J^{-2r}).
\]
Combing this result with Theorem \ref{Th1} yields
\[
\| \hat{{\beta}}_i-{\beta}_i \|_2^2 \leq \| \hat{{\beta}}_i-\hat{{\beta}}_i^{or} \|_2^2+\| \hat{{\beta}}_i^{or}-{\beta}_i \|_2^2 = O_p(J/m_{(n)}+J^{-2r}).
\]
This completes the proof of Theorem \ref{Th2}.\\

\noindent \emph{Proof of Theorem \ref{Th3}.} (i) For $i,j\in \mathcal{G}_k$, we have $\beta_i=\beta_j$. Then
\begin{eqnarray*}
\|\hat{\beta}_i-\hat{\beta}_j\|_2^2 &\leq& \|\hat{\beta}_i-\beta_i\|_2^2 + \|\beta_i-\beta_j\|_2^2 + \|\hat{\beta}_j-\beta_j\|_2^2 \\
&\leq &2\max_{i}\|\hat{\beta}_i-\beta_i\|_2^2 +0=O_p(J/m_{(n)}+J^{-2r}) \rightarrow 0
\end{eqnarray*}
as $m_{(n)}\rightarrow \infty$, where the last equality holds from Theorem \ref{Th2}. This means $\hat{\beta}_i$ and $\hat{\beta}_j$ will fall into the same group with probability approaching to 1.\\

(ii) For any $i \in \mathcal{G}_k$, $j\in \mathcal{G}_{k^{\prime}}$, $k\neq k^{\prime}$, it follows from Theorem \ref{Th2} that
\begin{eqnarray*}
\|\hat{\beta}_i-\hat{\beta}_j\|_2^2 &=& \|\hat{\beta}_i-\beta_i+ \beta_i-\beta_j+\beta_j-\hat{\beta}_j\|_2^2 \\
& \geq & \min_{\begin{array}{c}
                 i\in \mathcal{G}_k, j\in \mathcal{G}_{k^{\prime}} \\
                 k\neq k^{\prime}
               \end{array}} \|\beta_i-\beta_j\|_2^2-2\max_{1\leq i \leq n}\|\hat{\beta}_i-\beta_i\|_2^2 \\
&=& b^2-O_p(J/m_{(n)}+J^{-2r}) \rightarrow b^2>0,
\end{eqnarray*}
which implies that $\hat{\beta}_i$ and $\hat{\beta}_j$ will fall into the different groups with probability approaching to 1. Therefore, the proof is completed by the combinations of conclusions (i) and (ii).\\

In what follows, let $a_n\asymp b_n$ mean that $a_n/b_n$ and $b_n/a_n$ are bounded for given sequences of positive numbers $a_n$ and $b_n$. For a square integrable function $g$ on $\mathbb{T}$, we define the norms $\|g\|^2=E(g(t)^2)$ and $\|g\|_{\infty}=\sup_{t\in \mathbb{T}}|g(t)|$.\\

{\noindent \emph{Proof of Theorem \ref{Th4}.} We can conclude from the results of Theorems \ref{Th1}-\ref{Th3} that the proposed penalized
estimators performs asymptotically equivalent to the oracle ones as $m_n$ approaching to infinite. Thus, we only need to prove the asymptotic normalities of the oracle estimators $\hat{\boldsymbol{\alpha}}^{or}(t)=(\hat{\alpha}_1(t),\ldots,\hat{\alpha}_K(t))^T=\mathbb{B}(t)\hat{\boldsymbol{\theta}}$, where $\mathbb{B}(t)=\boldsymbol{I}_K \otimes \boldsymbol{B}(t)^T$ (Kronecker product) and $\hat{\boldsymbol{\theta}}=(\hat{\boldsymbol{\theta}}_1^T,\ldots,\hat{\boldsymbol{\theta}}_K^T)^T$.} To this end, we first show that
\begin{equation}\label{PTH4:0}
\text{Var}\left(\hat{\alpha}_k(t)\right)^{-1/2}\left\{\hat{\alpha}_k(t)-\text{E}(\hat{\alpha}_k(t))\right\} \mathop \to \limits^d N(0,1),~~~~k=1,\ldots,K.
\end{equation}

Recall that $\hat{\boldsymbol{\theta}}_{k}=\arg\min_{ \boldsymbol{\theta}_k } (\boldsymbol{Y}_{(k)}-\boldsymbol{X}_{(k)}\boldsymbol{\theta}_k)^T\boldsymbol{V}^{-1}_{(k)}(\boldsymbol{Y}_{(k)}-\boldsymbol{X}_{(k)}\boldsymbol{\theta}_k)=\left( \boldsymbol{X}_{(k)}^T\boldsymbol{V}_{(k)}^{-1}\boldsymbol{X}_{(k)} \right)^{-1} \boldsymbol{X}_{(k)}^T\boldsymbol{V}_{(k)}^{-1}\boldsymbol{Y}_{(k)}$ for $k=1,\ldots,K$, where $\boldsymbol{Y}_{(k)}=\left \{ \boldsymbol{Y}^T_i: i \in \mathcal{G}_k \right \}^T, \boldsymbol{X}_{(k)}=\left \{ \boldsymbol{X}^T_i: i \in \mathcal{G}_k \right \}^T$ and $\boldsymbol{V}_{(k)}=\text{diag}\left\{\boldsymbol{V}_i: i \in \mathcal{G}_k \right \}$. We only consider an fixed $k$ here since other cases can be proved in the same way. For any $t\in \mathbb{T}$, $\hat{\alpha}_k(t)=\boldsymbol{B}(t)^T\hat{\boldsymbol{\theta}}_{k}=\boldsymbol{B}(t)^T\left( \boldsymbol{X}_{(k)}^T\boldsymbol{V}_{(k)}^{-1}\boldsymbol{X}_{(k)} \right)^{-1} \boldsymbol{X}_{(k)}^T\boldsymbol{V}_{(k)}^{-1}\boldsymbol{Y}_{(k)}$, and $\text{E}\left( \hat{\alpha}_k(t) \right)=\boldsymbol{B}(t)^T\left( \boldsymbol{X}_{(k)}^T\boldsymbol{V}_{(k)}^{-1}\boldsymbol{X}_{(k)} \right)^{-1} \boldsymbol{X}_{(k)}^T\boldsymbol{V}_{(k)}^{-1}\alpha_k(t)$. Thus,
\begin{eqnarray*}
\hat{\alpha}_k(t)-\text{E}(\hat{\alpha}_k(t))&=&\boldsymbol{B}(t)^T\left( \boldsymbol{X}_{(k)}^T\boldsymbol{V}_{(k)}^{-1}\boldsymbol{X}_{(k)} \right)^{-1} \boldsymbol{X}_{(k)}^T\boldsymbol{V}_{(k)}^{-1}\boldsymbol{\varepsilon}_{(k)}\\
&=& \boldsymbol{B}(t)^T\left( \boldsymbol{X}_{(k)}^T\boldsymbol{V}_{(k)}^{-1}\boldsymbol{X}_{(k)} \right)^{-1} \boldsymbol{X}_{(k)}^T \boldsymbol{V}_{(k)}^{-1} \boldsymbol{\Sigma}_{(k)}^{1/2} \boldsymbol{\Sigma}_{(k)}^{-1/2}\boldsymbol{\varepsilon}_{(k)} \\
&\triangleq & \boldsymbol{B}(t)^T \left( \boldsymbol{X}_{(k)}^T\boldsymbol{V}_{(k)}^{-1}\boldsymbol{X}_{(k)} \right)^{-1} \boldsymbol{Z}_{(k)}^T\boldsymbol{e}_{(k)},
\end{eqnarray*}
where $\boldsymbol{Z}_{(k)}=\boldsymbol{\Sigma}_{(k)}^{1/2}\boldsymbol{V}_{(k)}^{-1}\boldsymbol{X}_{(k)}$ and $\boldsymbol{e}_{(k)}=\boldsymbol{\Sigma}_{(k)}^{-1/2}\boldsymbol{\varepsilon}_{(k)}$. Obviously, we have $\mbox{E}(\boldsymbol{e}_{(k)})=0$ and $\mbox{Var}(\boldsymbol{e}_{(k)})=I$, which means that the elements $\{e_{(k)\iota}\}_{\iota=1}^{N_k}$ can be seen as independent random variables with zero mean and unit variance.

Denote $\boldsymbol{Z}_{(k)}^{\iota}$ as a $S$-dimensional column vector comprised by the $\iota$th row of $\boldsymbol{Z}_{(k)}$, it follows that
\[
\hat{\alpha}_k(t)-\text{E}\left( \hat{\alpha}_k(t) \right)=\sum_{\iota=1}^{N_k}{ \boldsymbol{B}(t)^T\left( \boldsymbol{X}_{(k)}^T\boldsymbol{V}_{(k)}^{-1}\boldsymbol{X}_{(k)} \right)^{-1} \boldsymbol{Z}_{(k)}^{\iota} e_{(k)\iota} }=\sum_{\iota=1}^{N_k}{ \phi_{(k)\iota}e_{(k)\iota} }
\] and
\[
\mbox{Var}(\hat{\alpha}_k(t)) =\sum_{\iota=1}^{N_k}{\phi_{(k)\iota}^2 },
\]
where $\phi_{(k)\iota}=\boldsymbol{B}(t)^T\left( \boldsymbol{X}_{(k)}^T\boldsymbol{V}_{(k)}^{-1}\boldsymbol{X}_{(k)} \right)^{-1}\boldsymbol{Z}_{(k)}^{\iota}$. Therefore, if the Lindeberg condition holds, that is, $\max_{\iota}{\phi_{(k)\iota}^2}/\sum_{\iota=1}^{N_k}{\phi_{(k)\iota}^2 } \rightarrow 0 $, we can obtain that
\begin{equation}\label{PTH4:*}
\frac{\sum_{\iota=1}^{N_k}{ \phi_{(k)\iota}e_{(k)\iota} }}{\sqrt{\sum_{\iota=1}^{N_k}{\phi_{(k)\iota}^2 }}} ~\mathop \to \limits^d~ N(0,1),
\end{equation}
which indicates the result (\ref{PTH4:0}).

In fact, by the definition of $\boldsymbol{Z}_{(k)}$, condition (C5) and Lemma \ref{Lemma1}, we have
\begin{eqnarray}\label{PTH4:1}
\phi_{(k)\iota}^2 &=& \left( \boldsymbol{B}(t)^T\left( \boldsymbol{X}_{(k)}^T\boldsymbol{V}_{(k)}^{-1}\boldsymbol{X}_{(k)} \right)^{-1} \boldsymbol{Z}_{(k)}^{\iota} \right)^2 \asymp \frac{1}{N_k^2} \left( \boldsymbol{B}(t)^T\boldsymbol{Z}_{(k)}^{\iota} \right)^2 \nonumber \\
& \leq & \frac{C}{N_k^2} \sum_{l=1}^{S}{ B_l^2(t) } \sum_{l=1}^{S}{ B_l^2(t_{(k)\iota}) },
\end{eqnarray}
where the last step holds by the Cauchy-Schwarz inequality and condition (C5). Moreover, based on the same rationale as above, it follows that
\begin{eqnarray}\label{PTH4:2}
\sum_{\iota=1}^{N_k}{\phi_{(k)\iota}^2 } &=& \boldsymbol{B}(t)^T\left( \boldsymbol{X}_{(k)}^T\boldsymbol{V}_{(k)}^{-1}\boldsymbol{X}_{(k)} \right)^{-1}\sum_{\iota=1}^{N_k}{ \boldsymbol{Z}_{(k)}^{\iota}\boldsymbol{Z}_{(k)}^{\iota^T}}\left( \boldsymbol{X}_{(k)}^T\boldsymbol{V}_{(k)}^{-1}\boldsymbol{X}_{(k)} \right)^{-1} \boldsymbol{B}(t) \nonumber \\
&=& \boldsymbol{B}(t)^T\left( \boldsymbol{X}_{(k)}^T\boldsymbol{V}_{(k)}^{-1}\boldsymbol{X}_{(k)} \right)^{-1}\left( \boldsymbol{X}_{(k)}^T\boldsymbol{V}_{(k)}^{-1}\boldsymbol{\Sigma}_{(k)}\boldsymbol{V}_{(k)}^{-1}\boldsymbol{X}_{(k)} \right)\left( \boldsymbol{X}_{(k)}^T\boldsymbol{V}_{(k)}^{-1}\boldsymbol{X}_{(k)} \right)^{-1} \boldsymbol{B}(t) \nonumber \\
&\asymp & \boldsymbol{B}(t)^T\left( \boldsymbol{X}_{(k)}^T\boldsymbol{V}_{(k)}^{-1}\boldsymbol{X}_{(k)} \right)^{-1}\boldsymbol{B}(t) \asymp \frac{1}{N_k} \sum_{l=1}^{S}{ B_l^2(t) }.
\end{eqnarray}
Combining the expressions (\ref{PTH4:1}) and (\ref{PTH4:2}) leads to
\[
\frac{\phi_{(k)\iota}^2}{\sum_{\iota=1}^{N_k}{\phi_{(k)\iota}^2 }} \leq \frac{C}{N_k} \sum_{l=1}^{S}{ B_l^2(t_{(k)\iota}) } \leq \frac{C}{N_k} \sup_{t\in \mathbb{T} }\sum_{l=1}^{S}{ B_l^2(t) }.
\]

Observing that
\[
\sup_{t\in \mathbb{T}}\sqrt{\sum_{l=1}^{S}{ B_l^2(t) }} = \sup_{t\in \mathbb{T} } \sup_{b_l} \frac{\sum_{l=1}^{S}{|B_l(t)b_l|}}{\sqrt{\sum_{l=1}^{S}{b_l^2}}}  \leq \sup_{b_l} \frac{\sup_{t\in \mathbb{T} } \sum_{l=1}^{S}{|B_l(t)b_l|}}{\sqrt{\sum_{l=1}^{S}{b_l^2}}} = \sup_{g \in G}\frac{\|g\|_{\infty}}{\|g\|_2}, \\
\]
where the last step due to expression (\ref{A1}). Based on the definitions of norms and condition (C1) that the density function $f(t)$ is uniformly bounded away from 0 and infinity on $\mathbb{T}$, {it is easy to verify $\|g\|_2 \asymp \|g\|$. }Let $A_n=\sup_{g \in G}\|g\|_{\infty}/\|g\|$, hence
\[
\max_{\iota} \frac{\phi_{(k)\iota}^2}{\sum_{\iota=1}^{N_k}{\phi_{(k)\iota}^2 }} \leq \frac{C}{N_k}\sup_{g \in G}\frac{\|g\|_{\infty}^2}{\|g\|_2^2} \leq \frac{C}{N_k}\sup_{g \in G}\frac{\|g\|_{\infty}^2}{\|g\|^2}(1+o_p(1))=\frac{C A_n^2}{N_k}(1+o_p(1)).
\] 
Based on conditions (C1) and (C4), we have $A_n^2 \asymp S$. Therefore,
\[
\max_{\iota} \frac{\phi_{(k)\iota}^2}{\sum_{\iota=1}^{N_k}{\phi_{(k)\iota}^2 }} \leq \frac{C A_n^2}{N_k}(1+o_p(1)) \asymp \frac{S}{N_k}(1+o_p(1)) \rightarrow 0,
\]
which implies the validation of Lindeberg condition. Consequently, (\ref{PTH4:*}) is proved, and then (\ref{PTH4:0}) holds.

On the other hand, according to conditions (C3) and (C4), it is easy to verity that all the conditions assumed in Theorem 5.1 of \cite{Huang2003} hold. This implies, by virtue of condition (C5), that
\[
\left|\text{E}\left( \hat{\alpha}_k(t)\right)-\alpha_k(t)\right|=O_p(J^{-r}).
\]
Moreover, it follows from (\ref{PTH4:2}), condition (C5), Lemma \ref{Lemma1} and the properties of B-spline that
\begin{eqnarray*}
\text{Var}(\hat{\alpha}_k(t))&=&\boldsymbol{B}(t)^T\left( \boldsymbol{X}_{(k)}^T\boldsymbol{V}_{(k)}^{-1}\boldsymbol{X}_{(k)} \right)^{-1}\left( \boldsymbol{X}_{(k)}^T\boldsymbol{V}_{(k)}^{-1}\boldsymbol{\Sigma}_{(k)}\boldsymbol{V}_{(k)}^{-1}\boldsymbol{X}_{(k)} \right)\left( \boldsymbol{X}_{(k)}^T\boldsymbol{V}_{(k)}^{-1}\boldsymbol{X}_{(k)} \right)^{-1} \boldsymbol{B}(t) \\
&\asymp& S/N_k,
\end{eqnarray*}
where $\boldsymbol{\Sigma}_{(k)}=\text{diag} \left\{\boldsymbol{\Sigma}_i: i \in \mathcal{G}_k \right \}$.
Taking into account of $J/m_{(n)}^{1/(2r+1)}\rightarrow \infty$, we have
\begin{equation}\label{PTH4:3}
\sup_{t\in \mathbb{T}}\left| \frac{\text{E}\left( \hat{\alpha}_k(t)\right)-\hat{\alpha}_k(t)}{\sqrt{\text{Var}(\hat{\alpha}_k(t))}} \right|=o_p(1).
\end{equation}
Combining the results of (\ref{PTH4:0}) and (\ref{PTH4:3}) leads to
\begin{equation}\label{PTH4:4}
\text{Var}\left(\hat{\alpha}_k(t)\right)^{-1/2}\left(\hat{\alpha}_k(t)-\alpha_k(t)\right) \mathop \to \limits^d N(0,1).
\end{equation}

Let $\boldsymbol{X}_0=\text {diag} \left(\boldsymbol{X}_{(1)},...,\boldsymbol{X}_{(K)} \right), \boldsymbol{V}_0=\text {diag}\left({\boldsymbol{V}}_{(1)},...,\boldsymbol{V}_{(K)} \right)$ and  $\boldsymbol{\Sigma}_0=\text {diag} \left(\boldsymbol{\Sigma}_{(1)},...,\boldsymbol{\Sigma}_{(K)} \right)$, where $\boldsymbol{Y}_{(k)}=\left \{ \boldsymbol{Y}^T_i: i \in \mathcal{G}_k \right \}^T, \boldsymbol{X}_{(k)}=\left \{ \boldsymbol{X}^T_i: i \in \mathcal{G}_k \right \}^T, \boldsymbol{V}_{(k)}=\text{diag}\left\{\boldsymbol{V}_i: i \in \mathcal{G}_k \right \}$ and $\boldsymbol{\Sigma}_{(k)}=\text{diag} \left\{\boldsymbol{\Sigma}_i: i \in \mathcal{G}_k \right \}$. Finally, by the expression of $\hat{\boldsymbol{\alpha}}^{or}(t)$ and the independence assumption of different subgroup, we can obtain
\[
\text{Var}\left(\hat{\boldsymbol{\alpha}}^{or}(t)\right)^{-1/2}\left(\hat{\boldsymbol{\alpha}}^{or}(t)-\boldsymbol{\alpha}(t)\right) \mathop \to \limits^d N(\boldsymbol{0},\boldsymbol{I}_K),
\]
where
\begin{align}
\text{Var}\left(\hat{\boldsymbol{\alpha}}^{or}(t)\right)=\mathbb{B}(t)\left( \boldsymbol{X}_0^T\boldsymbol{V}_0^{-1}\boldsymbol{X}_0 \right)^{-1}\left( \boldsymbol{X}_0^T\boldsymbol{V}_0^{-1}\boldsymbol{\Sigma}_0 \boldsymbol{V}_0^{-1}\boldsymbol{X}_0 \right)\left( \boldsymbol{X}_0^T\boldsymbol{V}_0^{-1}\boldsymbol{X}_0 \right)^{-1} \mathbb{B}(t)^T
\label{var}
\end{align}
with $\mathbb{B}(t)=\boldsymbol{I}_K \otimes \boldsymbol{B}(t)^T$ (Kronecker product).
Therefore, we complete the proof of Theorem \ref{Th4} based on (\ref{PTH4:4}) and the conclusions of Theorems \ref{Th1}-\ref{Th3}.\\

\clearpage


\bibliographystyle{dcu}
\bibliography{my}

\end{document}